\documentclass[aps,prd,preprint,draft,showpacs,eqsecnum,nofootinbib]{revtex4}
\usepackage{graphicx,epsf,amssymb}
\newcommand{\be}{\begin{equation}}
\newcommand{\ee}{\end{equation}}
\newcommand{\ba}{\begin{eqnarray}}
\newcommand{\ea}{\end{eqnarray}}

\newcommand{\bea}{\begin{eqnarray}}
\newcommand{\eea}{\end{eqnarray}}
\newcommand{\Tr}{{\rm Tr}}

\newcommand{\nn}{\nonumber}

\newcommand{\NeqFour}{{\cal N} =4}

\def\outdent{\hskip -22mm}
\def\outdenta{\hskip -42mm}

\newif\ifdraft
\drafttrue
\newif\ifpreprint
\preprinttrue

\def\sect#1{section~{\ref{#1}}}

\def\fig#1{fig.~{\ref{#1}}}
\def\figs#1#2{figs.~{\ref{#1}} and {\ref{#2}}}

\def\pol{\varepsilon}
\def\Tr{\, {\rm Tr}}

\def\NeqFour{{\cal N}=4}
\def\NeqOne{{\cal N}=1}
\def\NeqZero{{\cal N}=0}
\def\ns{n_{\mskip-2mu s}}
\def\nf{n_{\mskip-2mu f}}
\def\Nc{N_{c}}

\def\Shift#1#2{{[#1,#2\rangle}}
\def\Lzz{\mathop{\hbox{\rm L}}\nolimits_2}
\def\Lz{\mathop{\hbox{\rm L}}\nolimits_0}
\def\Kz{\mathop{\hbox{\rm K}}\nolimits_0}

\def\sandp#1.#2.#3{%
\left\langle\smash{#1}{\vphantom1}^{-}\right|{#2}%
\left|\smash{#3}{\vphantom1}^{+}\right\rangle}
\def\sandpp#1.#2.#3{%
\left\langle\smash{#1}{\vphantom1}^{+}\right|{#2}%
\left|\smash{#3}{\vphantom1}^{+}\right\rangle}
\def\sandmm#1.#2.#3{%
\left\langle\smash{#1}{\vphantom1}^{-}\right|{#2}%
\left|\smash{#3}{\vphantom1}^{-}\right\rangle}
\def\spab#1.#2.#3{\sandmm#1.#2.#3}
\def\spba#1.#2.#3{\sandpp#1.#2.#3}
\def\spaa#1.#2.#3.#4{\sandmp#1.{#2#3}.#4}
\def\spbb#1.#2.#3.#4{\sandpm#1.{#2#3}.#4}
\def\spa#1.#2{\left\langle#1\,#2\right\rangle}
\def\spb#1.#2{\left[#1\,#2\right]}
\def\spash#1.#2{\vphantom{\hat K}\spa{\smash{#1}}.{\smash{#2}}}
\def\spbsh#1.#2{\vphantom{\hat K}\spb{\smash{#1}}.{\smash{#2}}}
\def\lor#1.#2{\left(#1\,#2\right)}
\def\sand#1.#2.#3{%
\left\langle\smash{#1}{\vphantom1}^{-}\right|{#2}%
\left|\smash{#3}{\vphantom1}^{-}\right\rangle}
\def\sandpp#1.#2.#3{%
\left\langle\smash{#1}{\vphantom1}^{+}\right|{#2}%
\left|\smash{#3}{\vphantom1}^{+}\right\rangle}
\def\sandpm#1.#2.#3{%
\left\langle\smash{#1}{\vphantom1}^{+}\right|{#2}%
\left|\smash{#3}{\vphantom1}^{-}\right\rangle}
\def\sandmp#1.#2.#3{%
\left\langle\smash{#1}{\vphantom1}^{-}\right|{#2}%
\left|\smash{#3}{\vphantom1}^{+}\right\rangle}

\def\MSbar{$\overline{\rm MS}$}
\def\spseq#1.#2{\prod\limits_{k=#1}^{#2} \spa{k}.{(k+1)}}

\newbox\SlashedBox
\def\slashed#1{\setbox\SlashedBox=\hbox{#1}
\hbox to 0pt{\hbox to 1\wd\SlashedBox{\hfil/\hfil}\hss}#1}
\def\hboxtosizeof#1#2{\setbox\SlashedBox=\hbox{#1}
\hbox to 1\wd\SlashedBox{#2}}

\newbox\charbox
\newbox\slabox
\def\s#1{{      
        \setbox\charbox=\hbox{$#1$}
        \setbox\slabox=\hbox{$/$}
        \dimen\charbox=\ht\slabox
        \advance\dimen\charbox by -\dp\slabox
        \advance\dimen\charbox by -\ht\charbox
        \advance\dimen\charbox by \dp\charbox
        \divide\dimen\charbox by 2
        \raise-\dimen\charbox\hbox to \wd\charbox{\hss/\hss}
        \llap{$#1$}
}}

\def\eqn#1{eq.~(\ref{#1})}
\def\Eqn#1{Equation~(\ref{#1})}
\def\eqns#1#2{eqs.~(\ref{#1}) and~(\ref{#2})}

\def\e{\epsilon}
\def\eps{\epsilon}

\def\Gr{{\rm Gr}}

\def\sign{{\mathop{\rm sign}\nolimits}}
\def\lr{\leftrightarrow}

\def\Li{\mathop{\rm Li}\nolimits}

\def\tree{{\rm tree}}
\def\oneloop{{1 \mbox{-} \rm loop}}

\def\cg{c_\Gamma}

\def\Kh{{\hat K}}

\def\sandp#1.#2.#3{%
\left\langle\smash{#1}{\vphantom1}^{+}\right|{#2}%
\left|\smash{#3}{\vphantom1}^{+}\right\rangle}
\def\ksl{\s{k}}
\def\Ksl{\s{K}}

\def\Den#1#2 {\prod\limits_{k=#1}^{#2} \spa{k}.{(k+1)}}

\def\Fact{{\cal F}}

\def\Wsix#1{W_6^{(#1)}}

\def\Res{\mathop{\rm Res}}
\def\tlambda{{\tilde\lambda}}

\def\Ll{\mathop{\rm L{}}\nolimits}

\def\Kz{\mathop{\hbox{\rm K}}\nolimits_0}
\def\Lz{\mathop{\hbox{\rm L}}\nolimits_0}

\def\Cuth{{\widehat {C}}}
\def\CuthRat{{\widehat {CR}}}
\def\Remaining{{\widehat {R}}}

\def\Ph{{\hat P}}
\def\Pb{{\overline P}}
\def\Vertex{R}
\def\Rational{R}
\def\DiagrammaticRational{R^D}
\def\DiagrammaticRationalS#1{R^{D,#1}}
\def\PureCut{C}
\def\Res{\mathop{\rm Res}}

\def\infpole{{{\rm lrg\ }z}}
\def\Ainf{A^{\infpole}}
\def\poleterms{{\rm poles}}
\def\conv{A^\poleterms}

\def\Overlap{O}
\def\Inf{\mathop{\rm Inf}}
\def\InfPart#1#2{\mathop{\rm Inf}_{#1}{#2}}
\def\Disc{\mathop{\rm Disc}\nolimits}

\newbox\ourfigbox
\def\SizedFigureWithCaption#1#2#3{%
\setbox\ourfigbox
  \hbox{\hss\epsfxsize #1 \epsfbox{#2}\hss}
\hbox to \wd\ourfigbox{\vbox{\noindent\copy\ourfigbox\break
\vskip -6mm      \hbox to \wd\ourfigbox{\hss#3\hss}}}
}

\def\llongrightarrow{%
\relbar\mskip-0.5mu\joinrel\mskip-0.5mu\relbar
     \mskip-0.5mu\joinrel\longrightarrow}
\def\inlimit^#1{\buildrel#1\over\llongrightarrow}
\def\dash{\hbox{-\kern-.02em}}

\begin{document}
\hfuzz 25 pt


\ifpreprint \noindent UCLA/06/TEP/08 \hfill SLAC--PUB--11830 \hfill
Saclay/SPhT--T06/036 \hfill hep-ph/0604195 \fi

\title{Bootstrapping One-Loop QCD Amplitudes\\
         with General Helicities%
\footnote{Research supported in part by the US Department of
 Energy under contracts DE--FG03--91ER40662 and DE--AC02--76SF00515}}

\author{Carola F. Berger}
\affiliation{Stanford Linear Accelerator Center \\
             Stanford University\\
             Stanford, CA 94309, USA
}

\author{Zvi Bern}
\affiliation{ Department of Physics and Astronomy, UCLA\\
\hbox{Los Angeles, CA 90095--1547, USA}
}

\author{Lance J. Dixon}
\affiliation{ Stanford Linear Accelerator Center \\
              Stanford University\\
             Stanford, CA 94309, USA
}

\author{Darren Forde and David A. Kosower}
\affiliation{Service de Physique Th\'eorique\footnote{Laboratory
   of the {\it Direction des Sciences de la Mati\`ere\/}
   of the {\it Commissariat \`a l'Energie Atomique\/} of France},
   CEA--Saclay\\
          F--91191 Gif-sur-Yvette cedex, France
}

\date{April 2006}

\begin{abstract}
The recently developed on-shell bootstrap for computing one-loop
amplitudes in non-supersymmetric theories such as QCD combines the
unitarity method with loop-level on-shell recursion.  For generic
helicity configurations, the recursion relations may involve
undetermined contributions from non-standard complex singularities
or from large values of the shift
parameter. Here we develop a strategy for sidestepping 
difficulties through the use of
pairs of recursion relations. To illustrate the strategy, we
present sets of recursion relations needed for obtaining
$n$-gluon amplitudes in QCD.
We give a recursive solution for the one-loop $n$-gluon QCD
amplitudes with three or four color-adjacent gluons of negative
helicity and the remaining ones of positive helicity.  We provide
an explicit analytic formula for the QCD amplitude $A_{6;1}(1^-,
2^-, 3^-, 4^+, 5^+, 6^+)$, as well as numerical results for
$A_{7;1}(1^-, 2^-, 3^-, 4^+, 5^+, 6^+, 7^+)$, $A_{8;1}(1^-, 2^-,
3^-, 4^+, 5^+, 6^+, 7^+,8^+)$, and $A_{8;1}(1^-, 2^-, 3^-, 4^-,
5^+, 6^+, 7^+,8^+)$.  We expect the on-shell bootstrap approach to
have widespread applications to phenomenological studies at
colliders.
\end{abstract}

\pacs{11.15.Bt, 11.25.Db, 11.25.Tq, 11.55.Bq, 12.38.Bx \hspace{1cm}}

\maketitle


\renewcommand{\thefootnote}{\arabic{footnote}}
\setcounter{footnote}{0}


\section{Introduction}
\label{IntroSection}

The success of the forthcoming experimental program at CERN's Large
Hadron Collider will depend in part on theoretical tools.  Our ability
to find and understand new physics at the TeV scale will rely on the
quality of predictions for a variety of known-physics processes.  A
classic example is the $W+4{\rm\ jet}$ background to top-quark
production.  Tools to perform higher-order corrections to a wide
variety of processes in the component gauge theories of the Standard
Model will play an important role.  Tree-level scattering amplitudes
provide the basic predictions for cross sections for Standard Model
processes.  However, next-to-leading order (NLO) QCD corrections
are typically quite large.  One-loop QCD amplitudes, which enter at NLO,
are therefore needed in order to reduce theoretical uncertainties to
the level of 10\% or so.  An important set of Standard Model backgrounds
to new physics dictates the computation of new one-loop
amplitudes for processes containing one or more vector bosons
($W$s, $Z$s, and photons) and multiple jets.

Experience has shown that while methods relying on direct analytical 
evaluation of Feynman diagrams can be used for five-point processes, 
they have not proven powerful enough to compute six-point processes 
or beyond in QCD.  The recent development of
semi-numerical approaches~\cite{GieleGloverNumerical,EGZ,EGZ06}
shows promise for improving traditional capabilities.
All helicity configurations for the six-gluon amplitude have been
evaluated numerically in this way, and numerical results presented
for a single phase-space point~\cite{EGZ06}.
These results are also of utility in confirming analytic expressions.
(For other numerical or semi-numerical approaches, see
ref.~\cite{OtherNumerical}.)

On-shell methods for computing amplitudes can be much more efficient than
Feynman diagrams, because they avoid gauge non-invariant intermediate
states and instead focus on the key analytic properties that any
physical amplitude must satisfy.  The unitarity-based
method~\cite{Neq4Oneloop, Neq1Oneloop,BernMorgan, UnitarityMachinery}
was applied long ago, not only to six-point processes, but also to
all-multiplicity amplitudes, for particular configurations of external
helicities.  Early applications of
the method were generally restricted, for practical
reasons, to supersymmetric theories or to the polylogarithmic part of
QCD amplitudes. This practical restriction arose from the greater
complexity of $D$-dimensional unitarity calculations, required for full
QCD amplitudes in this approach.

A key feature of the unitarity method is that new amplitudes
are constructed with only on-shell tree-level 
amplitudes (which are generally quite simple) as inputs.
 A number of related techniques
have emerged in the past two years, including
the application of maximally-helicity-violating (MHV)
vertices~\cite{CSW,Risager} to loop calculations~\cite{BST,BBSTQCD}
and the use~\cite{CachazoAnomaly,BCF7} of the holomorphic
anomaly~\cite{HolomorphicAnomaly} to evaluate the cuts.

More recent improvements to the unitarity
method~\cite{BCFUnitarity,BMSTUnitarity,BBCFSQCD,BFM}
use {\it complex\/} momenta within generalized
unitarity~\cite{ZFourPartons, TwoLoopSplit,NeqFourSevenPoint},
allowing, for example, a simple and purely algebraic
determination of all box integral coefficients.  (The name
`generalized unitarity', as applied to amplitudes for massive
particles, can be traced back to ref.~\cite{Eden}.)
In ref.~\cite{BBCFSQCD}, Britto, Buchbinder, Cachazo and Feng developed
efficient techniques for evaluating generic one-loop unitarity cuts,
by using spinor variables and performing the cut integration via
residue extraction.  Quite recently, Britto, Feng and
Mastrolia~\cite{BFM} further extended these techniques and completed
the computation of all cut-containing terms for the six-gluon
helicity amplitudes.  The cut-containing terms for other helicity
configurations, and for other components of the amplitudes,
were obtained in
refs.~\cite{Neq4Oneloop,Neq1Oneloop,NeqOneNMHVSixPt,DunbarBoxN1,%
RecurCoeff,BBCFSQCD}.
The only terms now missing in the analytic expressions for the six-gluon
amplitudes are the pure-rational ones.  The computation of the rational
terms, in these and more general amplitudes, is the subject of this paper.

In a previous paper~\cite{Bootstrap}, three of the authors presented a
systematic, recursive bootstrap approach to making high-multiplicity QCD
calculations practical within the framework of the unitarity-based method.
It complements the use of four-dimensional unitarity for logarithmic
and polylogarithmic terms with an on-shell recursion
relation~\cite{BCFRecurrence, BCFW, OnShellRecurrenceI, Qpap}
for the purely-rational terms. This approach systematizes a
unitarity-factorization bootstrap previously applied to
the amplitudes for $e^+e^-\to~4$~partons~\cite{ZFourPartons}.
It has already been used to solve for infinite sequences
of one-loop $n$-gluon helicity amplitudes,
in particular the MHV amplitudes containing two color-adjacent
negative-helicity gluons and $(n-2)$ positive-helicity
ones~\cite{FordeKosower}.  These papers do not explain 
how to attack more general helicity configurations.
That is the purpose of the present paper: to extend the range of applicability
of the recursive bootstrap method to cover as generic a helicity
configuration as possible.

Recursion relations have long been used in
QCD~\cite{BGRecurrence,DAKRecurrence}, and are an elegant and
efficient means for computing tree-level amplitudes.  Other related
approaches~\cite{Alpgen}, as well as computer-driven approaches such
as {\sc MadGraph}~\cite{Madgraph}, have also been employed.
Stimulated by the compact forms of seven- and higher-point tree
amplitudes~\cite{NeqFourSevenPoint,NeqFourNMHV,RSVNewTree} that
emerged from studying infrared consistency
equations~\cite{UniversalIR} for one-loop amplitudes (computed using
the unitarity-based method), Britto, Cachazo and Feng wrote
down~\cite{BCFRecurrence} a new set of tree-level recursion relations.
The new recursion relations differ in that they employ only {\it
on-shell\/} amplitudes (at {\it complex\/} values of the external
momenta).  A simple and very general proof of the relations, using
special continuations (shifts) of the external momenta in terms of a
complex variable $z$, was then given by Britto, Cachazo, Feng and
Witten~\cite{BCFW}. The power of this type of recursion relation
follows from the generality of the proof, which relies only on
factorization and Cauchy's theorem.  (The numerical efficiency of
these recursion relations, with respect to the older, off-shell
recursion relations~\cite{BGRecurrence,DAKRecurrence} and those based
on MHV vertices~\cite{CSW,BBKR}, has been studied
recently~\cite{DTW}.)  On-shell recursive methods have also yielded
compact expressions for tree amplitudes in gravity~\cite{Gravity} as
well as gauge theory~\cite{TreeRecurResults}, and have been extended to
theories with massive scalars and fermions~\cite{GloverMassive,Massive}.
They even provide a derivation~\cite{Risager} of 
the Cachazo--Svr\v{c}ek--Witten representation of amplitudes 
in terms of MHV vertices~\cite{CSW}.  Many of these
developments, as well as the resurgence of interest in unitarity
methods, were inspired by the development of twistor string
theory~\cite{WittenTopologicalString}.

The unitarity-based method~\cite{Neq4Oneloop, Neq1Oneloop} turns a
general property of field theories --- the unitarity of the
(perturbative) $S$-matrix --- into a practical technique for computing
cut-containing terms in amplitudes.  In a similar spirit, on-shell
recursion relations turn another general property --- factorization on
poles in intermediate states --- into a technique for computing
rational terms in amplitudes.  The idea of using factorization as a
computational tool goes back to the computation of the $Z\rightarrow
4$~parton one-loop matrix elements~\cite{ZFourPartons} (or
equivalently, by crossing, the virtual diagrams for $pp \rightarrow
W,Z + {\rm jets}$), wherein all terms consistent with the helicity
assignments were written down, and collinear limits used to isolate
the correct ones and their coefficients.  This approach gets harder to
apply as the number of external legs increases, because of the
difficulty of finding terms with the correct factorization
properties. The one-loop on-shell recursion relations~\cite{Bootstrap}
provide a practical and systematic method for constructing the
rational terms, avoiding this difficulty.  Moreover, in special cases,
when certain criteria on the unitarity cuts are
satisfied~\cite{RecurCoeff}, it is also possible to obtain the
rational coefficients of the cut-containing (poly)logarithmic terms
via on-shell recursion relations.

The factorization properties of one-loop amplitudes in gauge theories, as
a function of real Minkowski four-momenta, have been known for a long
time.  We may distinguish two different cases.  In the first case, dubbed
``multi-particle'' factorization, the momentum going on shell is a sum of
three or more external momenta.  In the second case, called ``collinear''
factorization, it is a sum of two momenta.  The standard derivations
describe how amplitudes factorize in either of these limits, when all
momenta involved are real.  The implementation of on-shell recursion
relations requires a generalization of these factorizations to {\it
complex\/} momenta.  This generalization is straightforward, both at tree
level and at one loop, for multi-particle factorization.  The
generalization is also straightforward at tree level for collinear
factorization.  This is no longer true at one loop.

The heuristic reason why collinear factorization is more intricate with
complex momenta is that one cannot define a nonsingular all-massless
three-point kinematics with real momenta, while one can using complex
momenta.  At the loop level, the complexity of complex collinear
factorization is reflected in the appearance of double poles and `unreal'
poles in scattering amplitudes~\cite{OnShellRecurrenceI,Qpap}.  As yet, we
have no general theorems providing universal factorization formul\ae{} in
these ``non-standard'' cases.  
In previous computations~\cite{Bootstrap,FordeKosower} of
one-loop amplitudes with two color-adjacent negative legs, these problems
could be sidestepped by making special choices in constructing recursion
relations.  Within the framework of ref.~\cite{Bootstrap}, we must choose
momentum shifts under which the amplitude vanishes at large shift
parameter $z$.  Otherwise, the contour integral over $z$ which gives 
rise to the recursion relation would receive an undetermined contribution
from large $z$. In the case of MHV $n$-gluon amplitudes, which contain
only two negative-helicity gluons, it is possible to make such a
choice and yet avoid channels with unknown
factorizations~\cite{Bootstrap,FordeKosower,MHVQCDLoop}.

For general helicity configurations, however, it is no longer possible
to do this.  It might seem that we should therefore study the
`difficult' channels, and attempt to derive a universal form for
their complex-momentum factorization.  It turns out, however, that
it is easier to relax the other requirement, that of a vanishing
amplitude at large shift parameter $z$.  Indeed, in
ref.~\cite{OnShellRecurrenceI} it was shown that,
if we somehow knew the large-$z$ behavior of an amplitude
in a recursion, then a non-vanishing behavior posed no problems; the
recursion relations still reconstructed the remaining terms in an
amplitude correctly.  Our aim here is to show how to determine the
large-$z$ behavior of amplitudes from scratch.  We will
do so by using an auxiliary recursion relation, constructed by
considering pairs of momentum shifts, one in the parameter $z$
and a second involving different external legs and another parameter $w$.
With these additional terms in hand, we can follow the approach
of ref.~\cite{Bootstrap} for the remainder of the calculation,
computing recursive and overlap diagrams
to add to the cut-containing terms.

As an illustration of our method, we will compute one-loop corrections to
a class of next-to-maximally-helicity-violating (NMHV)
$n$-gluon amplitudes in QCD,
those with three adjacent negative helicities in the color ordering,
$A_{n;1}^\oneloop({-}{-}{-}{+}{+}\cdots{+}{+})$.
Under a supersymmetric decomposition~\cite{GGGGG}, 
these amplitudes may be thought of as
composed of $\NeqFour$ and $\NeqOne$ supersymmetric pieces together
with a non-supersymmetric ($\NeqZero$) scalar loop contribution.  The
$\NeqFour$ contributions were computed in ref.~\cite{NeqFourNMHV}, and
the $\NeqOne$ terms in ref.~\cite{BBDPSQCD}.  The logarithmic parts of
the $\NeqZero$ scalar loop amplitudes were determined in
ref.~\cite{RecurCoeff}, by constructing an on-shell recursion relation
for integral-function coefficients appearing in the amplitudes.  
We shall complete the QCD computation in this paper by obtaining
the rational-function contributions.

We also describe a recursive solution for the rational-function parts
of the scalar loop amplitudes with four color-adjacent negative
helicities, using the logarithmic terms computed in ref.~\cite{RecurCoeff}
as a starting point.  We have computed in this way the $\NeqZero$
terms in the eight-gluon amplitude 
$A_{8;1}^\oneloop({-}{-}{-}{-}{+}{+}{+}{+})$.

We present numerical values for the six-, seven- and
eight-gluon amplitudes with ``split'' helicity configurations,
in which all the negative helicities are color-adjacent, as a reference
point for future implementations of these amplitudes in phenomenological
studies.

This paper is organized as follows. In the next section, we review
notation and the organization of color-ordered amplitudes used
in this paper. In \sect{FivePointSection}, we present a known
five-point amplitude, to illustrate and guide our strategy for
obtaining the rational parts of one-loop amplitudes with general
helicity configurations.  In \sect{ThreeMinusSixPtSection}, we then
apply this strategy to determine a sample six-point amplitude. Before
continuing to more general cases in \sect{DerivationSection}, we
review and extend the on-shell bootstrap formalism~\cite{Bootstrap} 
to cases where the shifted amplitudes $A_n(z)$
do not vanish for large shift parameter $z$.  In
\sect{GeneralHelicitySection}, we observe various empirical
properties, which we use to construct a procedure for general
helicities, focusing on $n$-gluon amplitudes.
As a non-trivial confirmation of the general procedure, in
\sect{ProcedureSampleSection} we present examples of applications of our
procedure for determining the behavior of amplitudes for large values
of the shift parameter.  This procedure is then used in
\sect{ThreeMinusAmplitudeSection} to determine a recursive solution
of the rational functions for $n$-point amplitudes with three 
nearest-neighboring negative helicities in the color ordering.  We also
describe a recursive solution to the eight-gluon amplitude with four
color-adjacent negative helicities,
$A_{8;1}^\oneloop({-}{-}{-}{-}{+}{+}{+}{+})$.
In \sect{NumericalSection} we present numerical values of the scattering
amplitudes at select kinematic points.  In \sect{ConclusionSection}
we present our conclusions and outlook for the future.
We include an appendix collecting previously computed amplitudes
that feed into our recursive computations.


\vskip 15pt
\section{Notation}
\label{NotationSection}

In this section we summarize the notation used in the remainder of
the paper.  Following the notation of previous
papers~\cite{OnShellRecurrenceI,Qpap,Bootstrap}, we
use the spinor helicity formalism~\cite{SpinorHelicity,TreeReview},
in which the amplitudes are expressed in terms of spinor inner-products,
\begin{equation}
\spa{j}.{l} = \langle j^- | l^+ \rangle = \bar{u}_-(k_j) u_+(k_l)\,,
\hskip 2 cm
\spb{j}.{l} = \langle j^+ | l^- \rangle = \bar{u}_+(k_j) u_-(k_l)\, ,
\label{spinorproddef}
\end{equation}
where $u_\pm(k)$ is a massless Weyl spinor with momentum $k$ and positive
or negative chirality. The notation used here follows the QCD
literature, with $\spb{i}.{j} = \sign(k_i^0 k_j^0)\spa{j}.{i}^*$ so
that,
\begin{equation}
\spa{i}.{j} \spb{j}.{i} = 2 k_i \cdot k_j = s_{ij}\,.
\end{equation}
Our convention is that all legs are outgoing.
We also define,
\begin{equation}
 \lambda_i \equiv u_+(k_i), \qquad \tlambda_i \equiv u_-(k_i) \,.
\label{lambdadef}
\end{equation}

We denote the sums of cyclicly-consecutive external momenta by
\begin{equation}
K^\mu_{i\ldots j} \equiv
   k_i^\mu + k_{i+1}^\mu + \cdots + k_{j-1}^\mu + k_j^\mu \,,
\label{KDef}
\end{equation}
where all indices are mod $n$ for an $n$-gluon amplitude.
The invariant mass of this vector is
\be
s_{i\ldots j} \equiv K_{i\ldots j}^2\,.
\ee
Special cases include the two- and three-particle invariant masses,
which are denoted by
\begin{equation}
s_{ij} \equiv K_{ij}^2
\equiv (k_i+k_j)^2 = 2k_i\cdot k_j,
\qquad \quad
s_{ijk} \equiv (k_i+k_j+k_k)^2 \,.
\label{TwoThreeMassInvariants}
\end{equation}
We also define spinor strings,
\begin{eqnarray}
\spab{i}.{(a\pm b)}.{j} &=& \spa{i}.{a} \spb{a}.{j} \pm \spa{i}.{b} \spb{b}.{j} \,,
          \nonumber   \\
\spbb{i}.{(a+b)}.{(c+d)}.{j} &=&
     \spb{i}.{a} \spab{a}.{(c+d)}.{j} +
     \spb{i}.{b} \spab{b}.{(c+d)}.{j} \,.
\end{eqnarray}

We use the trace-based color
decomposition of amplitudes~\cite{TreeColor,BGSix,MPX,TreeReview}.
For tree-level amplitudes with $n$ external gluons, this decomposition
is,
\begin{equation}
{\cal A}_n^\tree(\{k_i,h_i,a_i\}) = g^{n-2}
\sum_{\sigma \in S_n/Z_n} \Tr(T^{a_{\sigma(1)}}\cdots T^{a_{\sigma(n)}})\,
A_n^\tree(\sigma(1^{h_1},\ldots,n^{h_n}))\,.
\label{TreeColorDecomposition}
\end{equation}
Here $g$ is the QCD coupling, $S_n/Z_n$ is the group of non-cyclic
permutations on $n$ symbols, and $j^{h_j}$ denotes the $j^{\rm th}$ gluon,
with momentum $k_j$, helicity $h_j$, and adjoint color index $a_j$.
The SU$(N_c)$ color matrices in the fundamental
representation are normalized by $\Tr(T^a T^b) = \delta^{ab}$.  

For spin-$J$ adjoint particles circulating in the loop,
the color decomposition for one-loop $n$-gluon amplitudes
is given by~\cite{BKColor},
\begin{equation}
{\cal A}_n^{\rm adjoint} ( \{k_i,h_i,a_i\} ) = g^n
 \sum_{J} n_J  \sum_{c=1}^{\lfloor{n/2}\rfloor+1}
      \sum_{\sigma \in S_n/S_{n;c}}
     \Gr_{n;c}( \sigma ) \,A_{n;c}^{[J]}(\sigma) \,.
\label{AdjointColorDecomposition}
\end{equation}
The notation in \eqn{AdjointColorDecomposition} has been described
repeatedly elsewhere~\cite{BKColor,OnShellRecurrenceI,Qpap,Bootstrap}.
Here we just note that we need to compute only
the leading-color partial amplitudes
$A_{n;1}^{[J]}(1^{h_1},\ldots,n^{h_n})$, because the subleading-color 
partial amplitudes for a gluon in the loop, $A_{n;c}^{[1]}$ for $c>1$, 
are given by a sum over permutations of the leading-color
ones~\cite{Neq4Oneloop}.  The analog of
\eqn{AdjointColorDecomposition} for fundamental-representation particles
in the loop (such as quarks, with spin $J=1/2$) is also expressed
in terms of $A_{n;1}^{[J]}$,
\begin{equation}
{\cal A}_n^{\rm fund} ( \{k_i,h_i,a_i\} ) = g^n
 \sum_{J = 0,1/2} n_J
      \sum_{\sigma \in S_n/Z_n}
  \Tr(T^{a_{\sigma(1)}}\cdots T^{a_{\sigma(n)}}) \, A_{n;1}^{[J]}(\sigma) \,.
\label{FundamentalColorDecomposition}
\end{equation}

The contributions of different spin states can be
rewritten in terms of supersymmetric and non-supersymmetric
parts~\cite{GGGGG},
\ba
A_{n;1}^{[1/2]} &=& A^{\NeqOne}_{n;1} - A^{\NeqZero}_{n;1} \,, \label{An1/2}\\
A_{n;1}^{[1]} &=& A^{\NeqFour}_{n;1} -4 A^{\NeqOne}_{n;1}+ A^{\NeqZero}_{n;1}
\,.
\label{An1}
\ea
The non-supersymmetric amplitudes, denoted by $\NeqZero$, are
just the contributions of a complex scalar circulating in the loop,
$A^{\NeqZero}_{n;1} \equiv A^{[0]}_{n;1}$.  The supersymmetric and 
non-supersymmetric pieces have different analytic properties. 
The supersymmetric pieces can be constructed completely from 
four-dimensional unitarity cuts~\cite{Neq4Oneloop,Neq1Oneloop} 
and have no additional rational contributions.  
The polylogarithms and logarithms of the $\NeqZero$
non-supersymmetric contributions may also be computed from the
four-dimensional unitarity cuts.  (In certain cases,
the coefficients of integral functions containing the logarithms
and polylogarithms may instead be determined
recursively~\cite{RecurCoeff}.)

The leading-color QCD amplitudes are expressible in terms of
the different supersymmetric components via,
\begin{eqnarray}
A_{n;1}^{\rm QCD} &=&
A^{\NeqFour}_{n;1} -4 A^{\NeqOne}_{n;1}+ (1-\eps\delta_R ) A^{\NeqZero}_{n;1}
+ {\nf\over\Nc} \Bigl( A^{\NeqOne}_{n;1}- A^{\NeqZero}_{n;1}\Bigr)
+ {n_{s}\over\Nc} A^{\NeqZero}_{n;1} \,,
\label{AnQCD}
\end{eqnarray}
where $\nf$ is the number of active quark flavors in QCD.  We also
allow for a term proportional to the number of active fundamental
representation scalars $\ns$, which vanishes in QCD.
We regulate the infrared and ultraviolet divergences of one-loop
amplitudes dimensionally.  (In this paper, we will not treat divergent 
and finite parts separately.)
The regularization-scheme-dependent parameter $\delta_R$
specifies the number of helicity states of internal gluons
to be $(2-\epsilon\delta_R)$. For the 't~Hooft-Veltman scheme~\cite{HV}
$\delta_R = 1$, while in the four-dimensional helicity (FDH)
scheme~\cite{BKStringBased,Neq1Oneloop,OtherFDH}
$\delta_R = 0$.

The amplitudes in this paper have not been renormalized.
To perform an \MSbar\ renormalization, subtract from the leading-color 
partial amplitudes $A_{n;1}^{\rm QCD}$ the quantity,
\begin{equation}
  c_\Gamma \left[{n-2\over2}{1\over\e}\left({11\over3}
  - {2\over3} {\nf\over N_c} - {1\over3}{\ns\over N_c} \right) \right]
  A_n^\tree\,,
\label{MSbarsubtraction}
\end{equation}
with the universal prefactor,
\begin{equation}
\cg = {1\over(4\pi)^{2-\eps}}
  {\Gamma(1+\eps)\Gamma^2(1-\eps)\over\Gamma(1-2\eps)}\, .
\label{cgdefn}
\end{equation}
%

\section{A Five-Point Amplitude the Hard Way}
\label{FivePointSection}

\subsection{Overview of the On-shell Bootstrap}

In this paper we will continue the development of the method of
ref.~\cite{Bootstrap} for obtaining complete one-loop amplitudes in QCD
and other non-supersymmetric theories.  In
refs.~\cite{Bootstrap,FordeKosower}, amplitudes with two color-adjacent
negative-helicity legs were considered.  In that case, it was possible to
choose a shift so that:
\begin{enumerate}
\item the recursion relations do not contain any terms
with non-standard complex factorizations, and
\item the amplitude vanishes for large values of the complex shift
  parameter $z$.
\end{enumerate}
However, for general helicity configurations it
is not possible to satisfy both conditions at once.  Here we will provide
a simple procedure for sidestepping this apparent difficulty.

Before we attempt to calculate some six- and higher-point helicity 
amplitudes with three negative helicities, for which both conditions 
cannot be satisfied, let us examine a five-point amplitude, 
$A_{5;1}^{\rm QCD}(1^-,2^-,3^-,4^+,5^+)$.  
We shall focus on the scalar-loop contribution, $A_{5;1}^{\NeqZero}$,
which is the only component not computable from the $D=4$ unitarity cuts. 
We know the answer for
this amplitude~\cite{GGGGG}, which makes it a good test case. Of course,
it is maximally-helicity-violating in the conjugate spinors,
and therefore could be computed as in ref.~\cite{Bootstrap}.  Here, we
will instead compute it in a way that foreshadows our computation of
higher-point amplitudes.

We now briefly review the construction of ref.~\cite{Bootstrap};
in \sect{DerivationSection} we will present a more systematic
review and extension of the on-shell bootstrap.  We first choose
a complex-valued shift~\cite{BCFW} of the momenta of a pair of
external particles, $k_j \to \hat{k}_j(z)$, $k_l \to \hat{k}_l(z)$.
We describe a $\Shift{j}{l}$ shift in terms of the spinor
variables $\lambda$ and $\tlambda$ defined in \eqn{lambdadef},
\begin{equation}
\Shift{j}{l}:\hskip 2 cm
\tlambda_j \rightarrow \tlambda_j - z\tlambda_l \,,
\hskip 2 cm
\lambda_l \rightarrow \lambda_l + z\lambda_j \,.
\label{SpinorShiftI}
\end{equation}

To compute the amplitude, we must first examine the cut terms,
obtained from the unitarity method and/or recursion on the
coefficients of integral functions, for the presence
of spurious singularities.  Spurious singularities refer to
kinematic regions where the full amplitude is nonsingular,
but different components of it can contain (cancelling) divergent
behavior.  If we use the unitarity-based method to obtain coefficients
of complete loop integrals (including associated rational pieces),
then the cut terms should be free of spurious singularities.
If we extract pure (poly)logarithmic expressions, then we must
generally add rational terms to cancel such singularities.
The result of this procedure is referred to as the
completed-cut term $\Cuth_n$.

\begin{figure}[t]
\centerline{\epsfxsize 5 truein\epsfbox{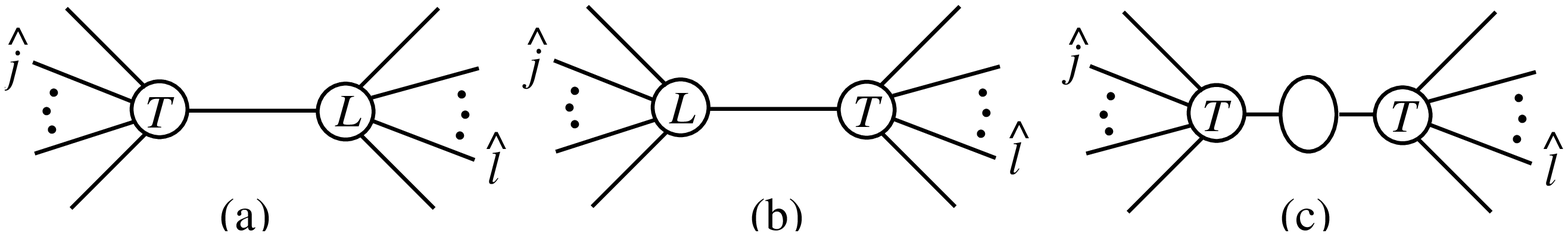}}
\caption{
Schematic representation of recursive contributions.  The labels `$T$' and
`$L$' refer to tree and loop vertices.  The multi-particle
factorization-function
contribution (c) does not appear for MHV amplitudes.
}
\label{LoopGenericFigure}
\end{figure}

Next, we have to compute a set of recursive rational terms
$\DiagrammaticRational_n$, corresponding to all diagrams in which
the shifted legs are attached to different amplitudes.  For each
arrangement of legs, we must sum over the different (complex)
factorizations in that channel, schematically shown in
\fig{LoopGenericFigure}:
tree times loop, loop times tree,
or tree times tree times a factorization function~\cite{BernChalmers}.
The factorization-function contribution --- which is equivalent to a
propagator or vacuum-polarization correction in the $\NeqZero$ case ---
does not appear for MHV amplitudes, and was therefore unnecessary
in ref.~\cite{Bootstrap}.

Finally, we compute the residues of the rational part of the
completed-cut terms, denoted by $\CuthRat_n$, in the
channels affected by the shift. The computation of
the residues of $\CuthRat_n(z)/z$ on the physical poles gives us
the overlap terms $\Overlap_n$, which correct for double-counting of
terms between the recursive diagrams and the completed cut.

The amplitude is the sum of these three terms,
\be A_n = \cg\Bigl[ \Cuth_n +\DiagrammaticRational_n + \Overlap_n \Bigr]\,.
\label{BasicEquation0}
\ee
(See section V of ref.~\cite{Bootstrap} for a relatively simple example
of overlap contributions for $n=5$.)  In the
present paper we will modify this construction somewhat to allow also
for non-trivial contributions from $z\rightarrow \infty$.  Note that
in the present paper, as in the derivation in section~3 of
ref.~\cite{Bootstrap},
these individual contributions are defined with respect to
$A$, whereas in the explicit calculations in ref.~\cite{Bootstrap}, these
quantities were defined with respect to pure-finite terms ($F^x$ parts
of amplitudes). This means the explicitly computed quantities
in ref.~\cite{Bootstrap} differ from the quantities in the
present paper by a factor of $i$.

\subsection{Choice of Shifts}
\label{ChoiceShiftSubSection}

What shift should we choose?  The computation in ref.~\cite{Bootstrap}
corresponds to choosing a $\Shift45$ shift ($j=4, l=5$, in
\eqn{SpinorShiftI}) here.  However, several properties of the
amplitude under this shift do not generalize to higher-point
amplitudes.  In particular, the amplitude may not vanish as the shift
parameter $z$ is taken to infinity.  A $\Shift{-}{+}$ shift, such as
$\Shift34$ or $\Shift15$ in the present case, appears quite generally
to have good behavior at infinity.  For the five-point case, we can verify
this explicitly using the known answer~\cite{GGGGG}, given in
\eqn{A5Neq0mmmpp} of the appendix.  For reasons we shall comment on in
\sect{ThreeMinusSixPtSection}, for our purposes, here it is convenient
to introduce a modified $\Ll_2$ function,
\be
\Ll_{2a}(r) = {\Ll_1(r)+{1\over2}\over 1-r} = \Ll_{2}(r) - {1\over 2r} \,,
\label{ModifiedL2}
\ee
rather than the more standard $\Ll_2(r)$ defined in \eqn{Lsdef}. These
functions differ only in terms that are nonsingular as $r \rightarrow 1$.
Using $\Ll_{2a}(r)$, we can give an alternate
expression for $A^{\NeqZero}_{5;1}$, instead of the form in
\eqn{A5Neq0mmmpp}.
\ba
A^{\NeqZero}_{5;1}(1^-,2^-,3^-,4^+,5^+) &=& \nn\\
&&\outdent
  {1\over3} A^{\NeqOne}_{5;1}(1^-,2^-,3^-,4^+,5^+)
  +{2\over9} \cg A^\tree_5(1^-,2^-,3^-,4^+,5^+)
    + \cg\Remaining_{5a}
\nn \\
&&\outdent \null
      +i {\cg\over 3}
   {\spa1.2\spb2.4\spb5.2\spa2.3
 \biggl(\spb5.1\spa1.2\spb2.4+\spb5.2\spa2.3\spb3.4 \biggr)
        \over\spb1.2\spb2.3}
     {\Ll_{2a}\biggl( {-s_{34}\over -s_{51}} \biggr)\over s_{51}^3}
\,, \hskip 1 cm \label{A5Neq0a}
\ea
where
\be
\Remaining_{5a} =
 i \Biggl[
 {1\over3} {\spa1.3^2 (s_{12}+s_{23})
             \over \spb1.2 \spb2.3 \spa3.4 \spa4.5 \spa5.1 }
 + {1\over6} {\spa1.2 \spa2.3 \spb2.4 \spb2.5 \sand3.{(2-1)}.5
             \over \spb1.2 \spb2.3 \spa3.4 s_{51}^2 } \Biggr]
\,.
\ee

\begin{figure}[t]
\centerline{\epsfxsize 6.5 truein\epsfbox{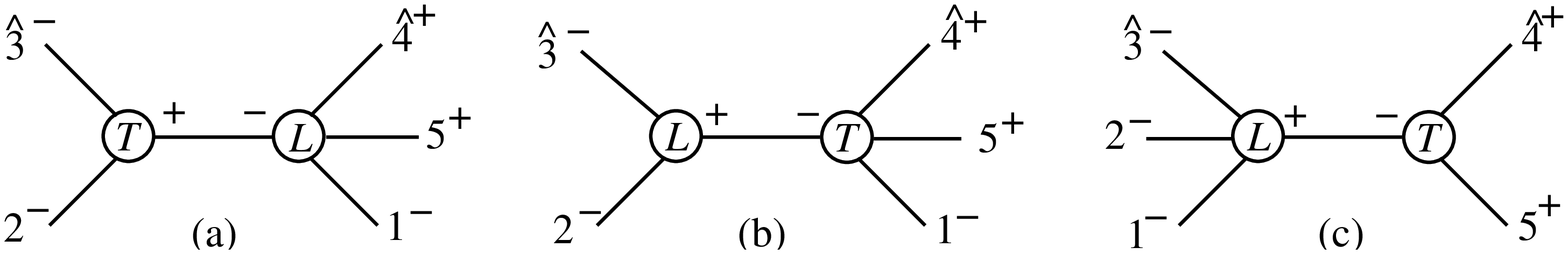}}
\caption{
The recursive diagrams arising from a $\Shift34$ shift
in $A^{\NeqZero}_{5;1}(1^-,2^-,3^-,4^+,5^+)$.
Diagram (b) has a non-standard complex singularity.
}
\label{Loop3mFivePt34Figure}
\end{figure}

Consider now the $\Shift34$ shift,
\ba
&& \tlambda_3 \rightarrow \tlambda_3  - z \tlambda_4\,, \hskip 2 cm
 \lambda_4 \rightarrow \lambda_4 + z \lambda_3\,.
\label{Shift34}
\ea
The three non-vanishing recursive diagrams are shown in
\fig{Loop3mFivePt34Figure}.  These diagrams correspond to
residues of the shifted amplitude at poles in $z$ where
intermediate states go on shell.
Diagrams~\ref{Loop3mFivePt34Figure}(a)
and~\ref{Loop3mFivePt34Figure}(c) are straightforward to evaluate,
because the three-point vertex is one which appears at tree level,
and which can have only a single pole.
Diagram~\ref{Loop3mFivePt34Figure}(b), however, involves a one-loop
``vertex'' $A_3^{(1)}(2^-,\hat 3^-;\hat K^+)$.
From refs.~\cite{OnShellRecurrenceI,Qpap}, we know that
the related ``vertex'', with opposite intermediate helicity,
$A_3^{(1)}(2^-,\hat 3^-;\hat K^-)$, does not factorize in
complex momenta as a naive generalization of the factorization
in real momenta.  This property is related to the appearance 
of double poles at the loop level. 
In that case it was possible to deduce the relatively
simple nonfactorizing structure, at least for the finite one-loop
helicity amplitudes studied in refs.~\cite{OnShellRecurrenceI,Qpap}.
For the case of $A_3^{(1)}(2^-,\hat 3^-;\hat K^+)$, however, we do not
know the general structure.  Analysis of the behavior under shifting of
$A^{\NeqZero}_{5;1}(1^-,2^-,3^-,4^+,5^+)$
(see \eqn{DiagrambShift34mmmppSIMPLE} below),
and of other known amplitudes, reveals that it is more subtle
than the case of $A_3^{(1)}(2^-,\hat 3^-;\hat K^-)$.
(It may even be that, in situations where double poles can appear,
additional contributions arise which cannot be 
interpreted as factorized diagrams at all.  However, an analysis 
of the diagrams such as those in \fig{Loop3mFivePt34Figure}, 
which incorporates some empirical information about the non-standard terms,
appears to cover any such additional 
contributions as well.)

\begin{figure}[t]
\centerline{\epsfxsize 6 truein\epsfbox{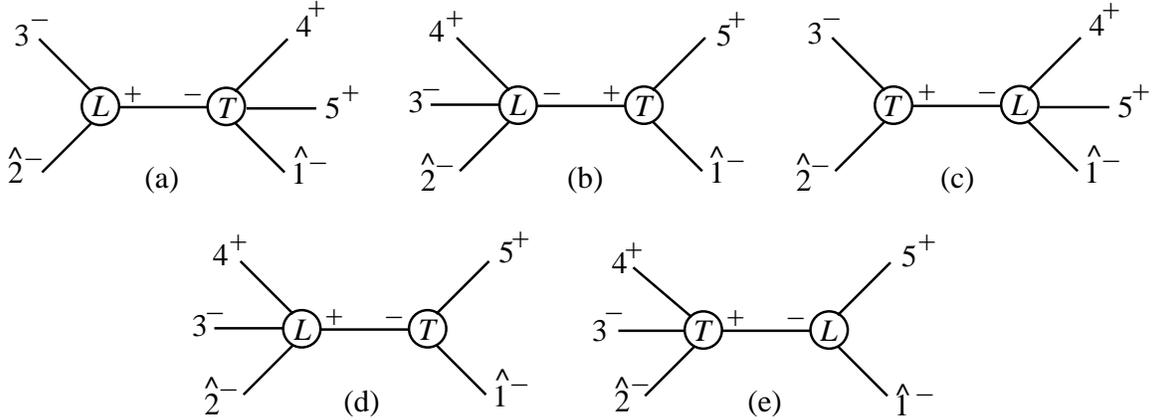}}
\caption{
The recursive diagrams arising from a $\Shift12$ shift
in $A^{\NeqZero}_{5;1}(1^-,2^-,3^-,4^+,5^+)$.
As discussed in the text, only diagram (d) is nonvanishing.
}
\label{Loop3mFivePt12Figure}
\end{figure}

Can we avoid diagrams like~\ref{Loop3mFivePt34Figure}(b)?
To study this, let us consider a $\Shift12$ shift,
\ba
&& \tlambda_1 \rightarrow \tlambda_1  - z \tlambda_2\,, \hskip 2 cm
\lambda_2 \rightarrow   \lambda_2 + z \lambda_1\,,
\label{Shift12}
\ea
which does in fact avoid generating diagrams whose
complex factorization is as-yet unknown.  The recursive diagrams for
this shift are shown in \fig{Loop3mFivePt12Figure}.
(In the five-point case, choosing a $\Shift45$ or
$\Shift54$ shift would avoid non-standard complex singularities,
but as noted above its properties do not generalize simply
to higher-point amplitudes.)

Before proceeding to inspect the specific diagrams 
in \fig{Loop3mFivePt12Figure}, we make a few general remarks about
the properties of three-point vertices, at one loop and beyond,
which will be relevant for diagrams \fig{Loop3mFivePt12Figure}(a) and
\fig{Loop3mFivePt12Figure}(e).  Prior to assigning definite
helicities to a three-point vertex with external legs $a$ and $b$,
it can be written as $A_3^\mu(\pol_a,\pol_b)$, where
$\pol_a,\pol_b$ are the external polarization vectors, and 
$\mu$ is the Lorentz index for the intermediate gluon.  This
gluon is going on shell in a particular way; either $\spa{a}.b$ or
$\spb{a}.b$ is vanishing, depending on the choice of shift.
Due to Bose symmetry, and the antisymmetry of the extracted color factor,
$A_3^\mu(\pol_a,\pol_b)$ is antisymmetric under the exchange $k_a \lr k_b$,
$\pol_a \lr \pol_b$.  Using Bose symmetry and gauge invariance,
there are only two possible terms in the
tensor decomposition of 
$A_3^\mu(\pol_a,\pol_b)$~\cite{BDSSplit,OneLoopSplitUnitarity},
\ba
A_3^\mu(\pol_a,\pol_b) &=& 
g_1(s_{ab},{\textstyle{k_a\cdot\eta\over (k_a+k_b)\cdot\eta}})
\, {1 \over s_{ab}}
 \, ( \pol_a^\mu \, \pol_b\cdot k_a - \pol_b^\mu \, \pol_a\cdot k_b
    + k_b^\mu \,\pol_a\cdot\pol_b )
\nn\\
&&\hskip0.0cm
+ \, g_2(s_{ab},{\textstyle{k_a\cdot\eta\over (k_a+k_b)\cdot\eta}})
  \, k_a^\mu \, {1 \over s_{ab}}
  \,
 \biggl( \pol_a\cdot \pol_b 
       - { \pol_a\cdot k_b \, \pol_b\cdot k_a \over k_a\cdot k_b } \biggr)
       \,,
\label{splitabstract}
\ea
where the form factors $g_1$ and $g_2$ are symmetric under $k_a \lr k_b$.
(The required antisymmetry in $k_a \leftrightarrow k_b$ follows from
the subleading nature of terms proportional to $K^\mu = -(k_a+k_b)^\mu$.) 
We have introduced a fixed external vector $\eta$ 
to indicate that $g_1$ and $g_2$ may depend on how the intermediate
gluon is going on shell.  For example, in a real collinear limit,
${k_a\cdot\eta\over (k_a+k_b)\cdot\eta}$ is the longitudinal momentum
fraction carried by gluon $a$.  The form factors can also depend on
the vanishing quantity $s_{ab}$.  However, the leading dependence 
can only be logarithmic, and so it is subdominant to the power-law 
behavior of the tensor structures.

The first tensor structure in \eqn{splitabstract} is the one that 
appears at tree level,
\be
A_3^{\rm tree, \, \mu}(\pol_a,\pol_b) = 
{1 \over s_{ab}}
 \, ( \pol_a^\mu \, \pol_b\cdot k_a - \pol_b^\mu \, \pol_a\cdot k_b
    + k_b^\mu \,\pol_a\cdot\pol_b ) \,,
\label{splitabstracttree}
\ee
so we know a lot about its behavior in complex on-shell kinematics.
The second tensor structure vanishes for opposite-helicity
gluons; with reference vectors $q_a$ and $q_b$,
\be
\pol_a^+ \cdot \pol_b^- 
       - { \pol_a^+\cdot k_b \, \pol_b^-\cdot k_a \over k_a\cdot k_b }
= - { \spb{a}.{q_b} \spa{b}.{q_a}
      \over \spa{a}.{q_a} \spb{b}.{q_b} }
      + { \spb{a}.{b} \spa{b}.{q_a} \spa{b}.{a} \spb{a}.{q_b}
        \over \spa{a}.{q_a} \spb{b}.{q_b} \spa{a}.{b} \spb{b}.{a} }
= 0 \,.
\label{oppositevanish}
\ee
Therefore, in the case that the two external gluons have opposite
helicity, if the tree-level vertex vanishes, the loop-level vertex
(at any number of loops) should also vanish, since the same
tensor structure is all that enters.

If the gluons have the same helicity, say both positive, the second 
tensor structure is nonvanishing off shell,  
\ba
\pol_a^+ \cdot \pol_b^+ 
       - { \pol_a^+\cdot k_b \, \pol_b^+\cdot k_a \over k_a\cdot k_b }
&=& { \spb{a}.{b} \spa{q_b}.{q_a}
      \over \spa{a}.{q_a} \spa{b}.{q_b} }
      - { \spb{a}.{b} \spa{b}.{q_a} \spb{b}.{a} \spa{a}.{q_b}
        \over \spa{a}.{q_a} \spa{b}.{q_b} \spa{a}.{b} \spb{b}.{a} }
\nn\\
&=&  - { \spb{a}.{b} \over \spa{a}.{b} \spa{a}.{q_a} \spa{b}.{q_b} }
   ( \spa{a}.{b} \spa{q_a}.{q_b} + \spa{b}.{q_a} \spa{a}.{q_b} )
\nn\\
&=& - { \spb{a}.{b} \over \spa{a}.{b} } \,.
\label{samenonvanish}
\ea
However, it vanishes if we approach the complex on-shell kinematics 
such that $\spb{a}.{b} \to 0$.   Similarly, the structure relevant
when both gluons have negative helicity vanishes as $\spa{a}.{b} \to 0$.
These configurations are those for which the corresponding tree-level
vertices are also known to vanish.  In summary, whenever a tree-level
three-point vertex vanishes, the corresponding loop-level vertex
should vanish as well.  

Note that the identical-helicity tensor structure~(\ref{samenonvanish}), which 
has a vanishing form factor $g_2$ at tree level, but not at one loop and beyond, 
has the form of an ``unreal pole''~\cite{Qpap}; that is, it is nonsingular
for real collinear limits, but blows up or vanishes in complex on-shell
kinematics.  In the case that it blows up, the additional factor of
$1/s_{ab}$ can produce a double pole in $\spa{a}.{b}$,
for the appropriate helicity of the intermediate gluon $P$.
(The contraction $\pol_P \cdot k_a$ is proportional to either
$\spb{a}.{b}$ or $\spa{a}.{b}$, depending on the helicity of $P$; the
former case leads to the double pole.)  The subleading terms in
the expansion around such a double pole are non-standard, and are
not yet understood. But again, this problem can occur only for the 
identical-helicity case, and only for the complex kinematical
configuration for which the tree vertex is nonvanishing.

Now we return to the diagrams of \fig{Loop3mFivePt12Figure}.
First note that diagram~\ref{Loop3mFivePt12Figure}(c) contains 
a tree-level vertex which vanishes in the complex on-shell kinematics,
because
\be
A_3^\tree(\hat 2^-, 3^-, -\Kh^+_{23}) \propto \spash{\hat 2}.3^3 = 0 \,.
\ee
As a result, diagram~\ref{Loop3mFivePt12Figure}(c) vanishes.
Note that $A_3^\tree(\hat 2^-, 3^-, -\Kh^+_{23})$ would not have
vanished with a shift of $\tlambda_2$ instead of $\lambda_2$.
From the above discussion, the loop vertex for the same
helicity configuration and type of shift should also vanish.
This vertex appears in diagram~\ref{Loop3mFivePt12Figure}(a).
Thus diagram~\ref{Loop3mFivePt12Figure}(a) also vanishes.
In the particular case at hand, we can easily verify that
diagram~\ref{Loop3mFivePt12Figure}(a) must vanish: If we examine the
amplitude~(\ref{A5Neq0a}), we see that there is no pole in the $(23)$ channel.
The reason is that only $\spb2.3$ products, not $\spa2.3$ products,
are present in the denominators.  The $\spb2.3$ products are left
untouched by the $\Shift12$ shift.  Hence they cannot give rise to
a pole in $z$ corresponding to diagram~\ref{Loop3mFivePt12Figure}(a),
which would be located at $\spash{\hat 2}.3 = 0$, or
$z = -\spa2.3/\spa1.3$.

Diagram~\ref{Loop3mFivePt12Figure}(b) vanishes for the same reason
as diagram~\ref{Loop3mFivePt12Figure}(c). 
Diagram~\ref{Loop3mFivePt12Figure}(e) contains a loop vertex
with a configuration for which the corresponding tree vertex
(which appears in diagram~\ref{Loop3mFivePt12Figure}(d)) is nonvanishing.
However, because it has opposite-helicity external gluons,
we know that only the ``standard'' tree-type tensor structure 
in \eqn{splitabstract} contributes.  The one-loop form factor 
$g_1$ for the $\NeqZero$ case of a scalar in the loop can then
be extracted from the one-loop splitting amplitude, and it vanishes.
(In the notation of ref.~\cite{Neq4Oneloop}, 
$r_S^{[0]}(\lambda_P,a^\pm,b^\mp)=0$, for either sign of $\lambda_P$.)
In summary, all the recursive diagrams in \fig{Loop3mFivePt12Figure} are 
under control, and only diagram~\ref{Loop3mFivePt12Figure}(d) is nonvanishing.

The $\Shift12$ shift, however, fails to satisfy the second requirement that
$A_n(z)\rightarrow0$ as $z\rightarrow \infty$.  For large $z$, we find that
\be
A^{\NeqZero}_{5;1}(1^-,2^-,3^-,4^+,5^+; z) \longrightarrow
i\, {\cg\over3}{\spa1.3^3 \bigl( z \spa1.5 + \spa2.5)
   \over \spa1.5^2 \spa3.4\spa4.5\spb1.2} + {\cal O}\Bigl({1\over z}\Bigr)\,.
\ee
It is easy to see that the following rational function has the same
large-$z$ behavior as the full amplitude,
\be
\InfPart{\Shift12}{A^{\NeqZero}_{5;1}(1^-,2^-,3^-,4^+,5^+)} =
i \, {\cg\over3}{\spa1.3^3 \spa2.5 \over \spa1.5^2 \spa3.4\spa4.5\spb1.2}\,,
\label{LargeZFunction}
\ee
where $\InfPart{\Shift12}{A}$ denotes those terms in $A$ that would
give rise to a non-vanishing contribution upon performing a $\Shift12$
shift and taking $z \rightarrow \infty$.  (See \sect{AnalyticSubsection}
for further discussion of such terms.)

We note in contrast that $A^{\NeqOne}_{5;1}$ {\it does\/} vanish in
the limit.  More generally, supersymmetric loop amplitudes appear to
vanish at large $z$ whenever the corresponding tree-level amplitudes do.

Based on our discussion, it is not surprising that with more
 positive-helicity gluons, higher-point $\NeqZero$ amplitudes in
general suffer from one of these two problems (or perhaps both) under
any shift: either a `bad' channel with non-standard complex
singularities, or `bad' large-$z$ behavior.

\begin{figure}[t]
\centerline{\epsfxsize 4 truein\epsfbox{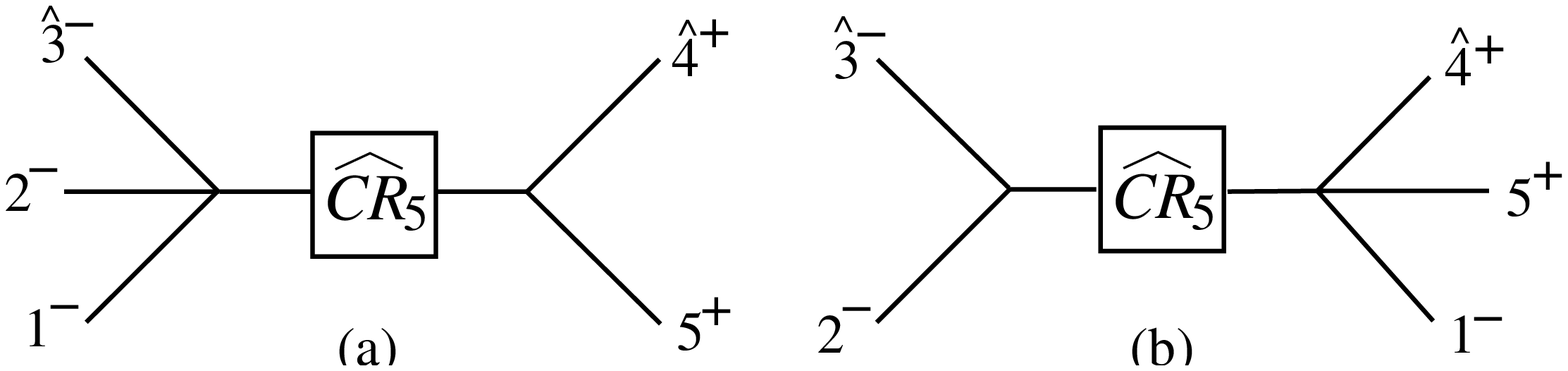}}
\caption{
The overlap diagrams arising from an auxiliary
$\Shift{3}{4}$ shift in $A^{\NeqZero}_{5;1}(1^-,2^-,3^-,4^+,5^+)$.
}
\label{Overlap3mFivePt34Figure}
\end{figure}

\subsection{Pairs of Shifts to the Rescue}

To get around these problems, we use a combination of shifts,
one particular shift to obtain one set of
contributions, and a different shift to obtain the remaining
contributions. In our five-point
example, we take the $\Shift12$ shift as the primary one for
determining all terms except for the large-$z$ contributions given in
\eqn{LargeZFunction}.  We then use the $\Shift34$ shift as an auxiliary
shift for determining the large-$z$ contributions of the $\Shift12$
shift.  To see how this might be possible, imagine constructing
the amplitude $A^{\NeqZero}_{5;1}(1^-,2^-,3^-,4^+,5^+)$ via the
{\it auxiliary} $\Shift34$ shift
\ba
&& \tlambda_3 \rightarrow \tlambda_3  - w \tlambda_4\,, \hskip 2 cm
\lambda_4 \rightarrow   \lambda_4 + w \lambda_3\,,
\label{Shift34Aux}
\ea
where we have taken the shift parameter to be $w$, to distinguish it
from the primary shift parameter in \eqn{Shift12}.
As above, we decompose the result into three pieces:
the completed-cut terms, the overlap diagrams, and the recursive diagrams.
Let us examine how each of these $\Shift34$-decomposed pieces behaves
under the action of the primary $\Shift12$ shift, in particular as
the parameter $z$ of that shift goes to infinity.
To the recursive diagrams for the $\Shift34$ shift 
shown in fig.~\ref{Loop3mFivePt34Figure},
we add the overlap diagrams in fig.~\ref{Overlap3mFivePt34Figure}.

In computing the overlap diagrams in \fig{Overlap3mFivePt34Figure}, we
choose a completed-cut expression based on \eqn{A5Neq0a},
\ba
\Cuth_{5}(1^-,2^-,3^-,4^+,5^+) &=& \nn\\
&&\outdent
  {1\over3\cg} A^{\NeqOne}_{5;1}(1^-,2^-,3^-,4^+,5^+)
 + {2 \over 9} A^{\tree}_{5}(1^-,2^-,3^-,4^+,5^+)
\nn \\
&&\outdent \null
      +{i\over 3}
   {\spa1.2\spb2.4\spb5.2\spa2.3
 \biggl(\spb5.1\spa1.2\spb2.4+\spb5.2\spa2.3\spb3.4 \biggr)
        \over\spb1.2\spb2.3}
     {\Ll_{2a}\biggl( {-s_{34}\over -s_{51}} \biggr)\over s_{51}^3}
\,, \hskip .5 cm \label{Cuth5}
\ea
where we have kept the $2/9 A_5^\tree$ term in the completed-cut term
for consistency with our later computations.  With this choice, it is
a simple matter to check that under the $\Shift12$ shift $\Cuth_{5}(z)$
vanishes for $z\rightarrow \infty$, using
the $\NeqOne$ amplitude given in
\eqn{A5Neq1mmmpp}, as well as \eqn{ModifiedL2} for $\Ll_{2a}(r)$.

Now consider the overlap contributions.  These are obtained by extracting
the residues of $\CuthRat_{5}(w)/w$, on the physical poles,
\be
w^{\rm (a)} = - {\spa4.5 \over \spa3.5} \,, \hskip 2cm
w^{\rm (b)} =   {\spb2.3 \over \spb2.4} \,,
\ee
after applying the $\Shift34$ shift.  (The rational part of the 
completed-cut terms, $\CuthRat_{5}$, is obtained by setting 
all logarithms to zero in \eqn{Cuth5}.)  The absence of a $\spa4.5$
product in the denominator of \eqn{Cuth5} tells us that the first
overlap contribution, fig.~\ref{Overlap3mFivePt34Figure}(a), vanishes.
The other overlap contribution is straightforward to evaluate,
\ba
\Overlap^{\rm (b)} &=&
-  \biggl({1\over 3\eps} + {8 \over 9} \biggr)
                 A^{\tree}_{5}(1^-,2^-,3^-,4^+,5^+) \\
&& \null
     - {i\over 3} {\spa1.2\spb2.4\spb5.2\spa2.3
       \biggl(\spb5.1\spa1.2\spb2.4+\spb5.2\spa2.3\spb3.4 \biggr)
        \over\spb1.2\spb2.3}
       \, {3 s_{51} - s_{34} \over 2 (s_{51} - s_{34})^2 s_{51}^2}
\nn \,.
\ea
Applying a $\Shift12$ shift to this expression and taking
$z\rightarrow \infty$, we
see that this expression also vanishes.  This is, of course, not
surprising, given that it is extracted from the completed-cut piece
$\Cuth_5$, which vanishes at large $z$.  Thus, there are no
$\Shift12$ large-$z$ contributions arising from cut or overlap terms.

This leaves us with the recursive diagrams to inspect.  The first
recursive diagram in \fig{Loop3mFivePt34Figure}(a) is simple to
evaluate, yielding,
\be
D^{\rm (a)} =
i \, \biggl({1\over 3\e} + {8\over 9} \biggr)
      {\spb4.5^3 \over \spb1.2 \spb1.5 \spb2.3 \spb3.4}
 \,. \label{DiagramaShift34mmmpp}
\ee
The $\Shift12$ shift of $D^{\rm (a)}$ clearly vanishes as $z \to\infty$.

While we do not know how to evaluate the second recursive
diagram, we can nonetheless extract its value by `reverse engineering'
from the final answer.  That is, we start from the final answer
in~\eqn{A5Neq0a} (or equivalently, the form in \eqn{A5Neq0mmmpp}),
shift it, and extract the residue of $-A^{\NeqZero}_{5;1}(w)/w$
at $w = \spb2.3/\spb2.4$, with the logarithms set to zero.  We find
the contributions corresponding to the sum of diagrams
\ref{Loop3mFivePt34Figure}(a) and \ref{Loop3mFivePt34Figure}(b).
Subtracting off the value of diagram
\ref{Loop3mFivePt34Figure}(a), \eqn{DiagramaShift34mmmpp},
gives us the reverse-engineered value for the non-standard residue
encoded in diagram \ref{Loop3mFivePt34Figure}(b),
\ba
D^{\rm (b)} &=&
 {i \over 6} \Biggl\{
- {\spa1.2 \spa2.3 \spb2.4 \spb2.5
 (s_{15} + s_{34})   (\sandpm5.{12}.4 + \sandpm5.{23}.4) \over
  \spa1.5 \spa3.4 \spb1.2 \spb1.5 \spb2.3 \spb3.4 (s_{15} - s_{34})^2} \nn\\
&& \null \hskip .6 cm
 + {\spa1.2 \spa2.3 \spb2.4 \spb2.5 \spb4.5\over
    \spb1.2 \spb2.3 s_{34} s_{15} }
+  2 {\spa1.3^2 \spa1.2 \spb2.4 \over
      \spb1.2 \spb2.3 \spa3.4 \spa5.1^2  }  \Biggr\}
\,.
\label{DiagrambShift34mmmpp}
\ea
Using some spinor-product identities, this result can be
simplified to
\ba
D^{\rm (b)} &=&
 {i \over 3}
{{\spa1.2}^2 {\spb2.4}^2 \spa2.3
\over \spb1.2 \spb2.3 {\spa5.1}^2\ (s_{34} - s_{15})^2}
\Bigl( 2 \spa1.5 \spb5.2 + { \spa1.3 \over \spa2.3 } (s_{34} - s_{15})
\Bigr)
\,.
\label{DiagrambShift34mmmppSIMPLE}
\ea
The appearance of the unreal pole $\spa2.3/\spb2.3$ in this
diagrammatic contribution is the hallmark of a non-standard
factorization.  If we now shift this expression under the primary
$\Shift12$ shift, and take the large-parameter limit, we see that
both diagrams~\ref{Loop3mFivePt34Figure}(a) and (b)
vanish.  In summary, the only piece of
$A^{\NeqZero}_{5;1}(1^-,2^-,3^-,4^+,5^+)$, decomposed according to
the auxiliary $\Shift34$ shift, which survives in the large-$z$ limit of the
$\Shift12$ shift, is that given by the $\Shift34$ recursive diagram
in \fig{Loop3mFivePt34Figure}(c).

Therefore, while the $\Shift34$ shift is not useful for
evaluating the entire amplitude, because of the non-standard
factorization in diagram \ref{Loop3mFivePt34Figure}(b), it {\it is\/}
useful for deriving a recursion for those terms with bad large-$z$ behavior
under a different shift.  And once we have those terms, we can make
use of the primary $\Shift12$ shift to compute the entire amplitude.

Retaining only those terms from diagram
\ref{Loop3mFivePt34Figure}(c) that contribute in the large-$z$ limit of
the $\Shift12$ shift, we find a remarkably simple recursion formula
for these terms,
\be
\InfPart{\Shift12}{A^{\NeqZero}_{5;1}(1^-,2^-,3^-,4^+,5^+)}
= \InfPart{\Shift12}{ A^{\NeqZero}_{4;1}(1^-,2^-,\hat{3}^-,\hat{K}_{45}^+)}
        \times {i\over s_{45}} \times
    A_3^\tree(-\hat{K}_{45}^-,\hat{4}^+,5^+)\,.
\label{LargeZRecursion5Pt}
\ee
Here, we denote the operation of extracting the large-$z$ behavior
of the $\Shift12$ shift by $\InfPart{\Shift12}{\mbox{}}$. We will give a
formal definition of this operation in \sect{AnalyticSubsection} below.
In this recursion relation, $\hat{a}$ denotes a shifted momentum
with the shift parameter frozen to the value,
\be
w^{\rm (c)} = - {\spa4.5 \over \spa3.5} \,,
\ee
according to the auxiliary $\Shift34$ shift.  The right-hand
side of the recursion relation (\ref{LargeZRecursion5Pt}) simplifies,
because both legs $1$ and $2$ that are shifted under the primary shift
lie on one side of the pole, so only the first factor is affected by the
extraction of the large-$z$ behavior.

To evaluate the recursion relation (\ref{LargeZRecursion5Pt}), we
need the source of the large-$z$ behavior in
the four-point amplitude,
$A^{\NeqZero}_{4;1}(1^-,2^-,\hat{3}^-,\hat{K}_{45}^+)$, which may be
obtained by relabeling the parity conjugate of
\eqn{A4Neq0amppp} in the appendix.  We find,
\ba
\InfPart{\Shift12}{A^{\NeqZero}_{5;1}(1^-,2^-,3^-,4^+,5^+)} &=&
-i\, {\cg \over 3} \, {\spbsh1.{\hat3} \spash1.{\hat3}^3
      \over \spash{\hat{K}_{45}}.1 \spb1.2 \spb2.{\hat3}
               \spash{\hat 3}.{\hat{K}_{45}}}
\, {1\over s_{45}} \,
{\spbsh{\hat4}.5^3 \over
     \spbsh{\hat{K}_{45}}.{\hat 4} \spbsh{5}.{\hat K_{45}} }  \nn \\
&=&
i \, {\cg\over3}{\spa1.3^3\spa2.5\over\spa1.5^2\spa3.4\spa4.5\spb1.2}
\,,
\label{FivePoint45Channel}
\ea
an expression identical to \eqn{LargeZFunction}.  We see that the
recursion relation (\ref{LargeZRecursion5Pt}), containing only the
$(45)$ channel, reproduces the entire large-$z$ behavior of the
five-point amplitude.

With this term in hand, we can compute the five-point amplitude using
the $\Shift12$ shift, even though the amplitude does not vanish as
$z\rightarrow \infty$, because the difference $A^{\NeqZero}_{5;1}-
\InfPart{\Shift12}{A^{\NeqZero}_{5;1}}$ does have good behavior at
infinity.  The difference is given by the sum~(\ref{BasicEquation0}) of the
completed-cut expression~(\ref{Cuth5}), the recursive diagrams (only
diagram~\ref{Loop3mFivePt12Figure}(d) is nonvanishing),
and the overlap diagrams (only one is nonvanishing).
This gives us a modified formula for the amplitude,
\be
A_n = \InfPart{\null}{A_n} +
 \cg\Bigl[\Cuth_n + \DiagrammaticRational_n + \Overlap_n\Bigr]
\,,
\label{BasicEquation}
\ee
where the new term, $\InfPart{\null}{A_n}$, accounts for a
non-vanishing contribution at $z \rightarrow \infty$, compared to
\eqn{BasicEquation0}.  (We will need to modify this
formula in subsequent sections to account for the possibility
that the completed-cut terms also have nonvanishing large-$z$ behavior.)

\subsection{Summary of Strategy}

In summary, the above example suggests a simple strategy for
constructing loop amplitudes, avoiding difficulties
from either non-vanishing large-$z$ behavior or from non-standard
complex factorizations:

\begin{itemize}

\item Use a primary shift whose recursion relation does not
contain any non-standard complex factorizations, but under which the
amplitude might not vanish at large $z$.

\item Use an auxiliary shift and recursion relation to determine
the large-$z$ behavior of the amplitude under the primary shift.  This
auxiliary recursion relation may contain non-standard complex
factorizations, but these will be harmless if the contributions from these
channels vanish in the large-$z$ limit of the primary shift.

\end{itemize}

In general, of course, we may not know ahead of time whether an
amplitude has non-trivial large-parameter behavior for a chosen shift.
We should therefore first derive an auxiliary recursion for this
behavior (which might of course reveal that it is absent).  The
derivation of an auxiliary recursion relation may require assumptions
about the behavior of contributions in `bad' channels, such as that of
the vanishing of diagram~\ref{Loop3mFivePt34Figure}(b) in the
large-parameter limit of the $\Shift12$ shift.  These assumptions can
be verified once we have obtained a (candidate) final answer for the
full amplitude, through a check of all (real-momentum) factorization
limits.  If the assumptions are not valid, and the auxiliary recursion
relation yields an incomplete expression for the large-parameter
behavior, the final answer will not have the correct collinear or
multi-particle factorization limits.  In \sect{GeneralHelicitySection},
we present shifts that we expect will satisfy all criteria necessary for
constructing the rational parts of any $n$-gluon amplitude.  We
also expect this strategy to be widely applicable to other
phenomenologically interesting amplitudes.

\section{A Non-MHV Six-Point Example: $A_6(1^-,2^-,3^-, 4^+, 5^+, 6^+)$ }
\label{ThreeMinusSixPtSection}

In the previous section we have developed
a strategy for computing amplitudes even
with shifts under which $A_n(z)\not\rightarrow 0$ as the shift parameter
$z\rightarrow\infty$.  We now apply these ideas to the calculation
of a previously unknown amplitude with three negative helicities,
$A^{\NeqZero}_{6;1}(1^-,2^-,3^-, 4^+, 5^+, 6^+)$.

As in the five-point case, we shall use a $\Shift12$ shift as the
primary shift for computing the amplitude.  For the same reasons as
in the five-point case, the six-point amplitude is also free of
contributions in `bad' channels with non-standard complex
singularities.  So we can evaluate all of the recursive diagrams.  It
is clear from a consideration of the factorization properties as the
momenta of gluons 5 and 6 become collinear that the
amplitude must have non-trivial behavior at large $z$, because the
corresponding five-point amplitude has such behavior under the shift.
Our first task is to obtain a rational function which reproduces that
behavior.

The full amplitude is obtained by combining the large-$z$ terms with
the other standard terms --- completed-cut terms, recursive diagrams, and
overlap diagrams --- via \eqn{BasicEquation}.  It turns out that we need
to modify \eqn{BasicEquation} a bit more to subtract out
nonvanishing large-$z$ contributions from the completed-cut terms,
$\Cuth_n$.  As we shall discuss more fully in \sect{DerivationSection},
we are really performing a contour-integral
analysis on the rational function of $z$,
$(A_n(z) - [\InfPart{\null}{A_n}](z)) - \Cuth_n(z)$.
It is this expression that must vanish as $z\to\infty$ to allow closing
the contour at infinity.
In the five-point example of the previous section, we avoided the need for
subtracting the large-$z$ behavior of $\Cuth_n$, by use
of the modified $\Ll_2$ function in \eqn{ModifiedL2}; we simply
ensured that ${\Cuth_5}(z) \to 0$ as $z\to\infty$.
For more general amplitudes it is simpler to subtract out any
nonvanishing large-$z$ contributions arising from $\Cuth_n$,
before adding back the complete large-$z$ behavior.  This amounts
to performing a contour-integral analysis on
$(A_n(z) - [\InfPart{\null}{A_n}](z))
- (\Cuth_n(z)-[\InfPart{\null}{\Cuth_n}](z))$,
which automatically vanishes at infinity.  The full result is then,
\be
A_n = \InfPart{\null}{A_n} +
 \cg\Bigl[\Cuth_n - \InfPart{\null}{\Cuth_n} + \DiagrammaticRational_n
   + \Overlap_n\Bigr]
\,,
\label{BasicEquation2}
\ee
where $\InfPart{\null}{A_n}$ contains the large-$z$ behavior of
the entire amplitude,
$\DiagrammaticRational_n$ are the contributions from the recursive
diagrams, and $\Overlap_n$ are the overlap pieces.  In
\sect{DerivationSection}, we will provide a more systematic derivation
of this formula.

\subsection{A Recursion Relation for Large-Parameter Behavior}

\begin{figure}[t]
\centerline{\epsfxsize 6 truein\epsfbox{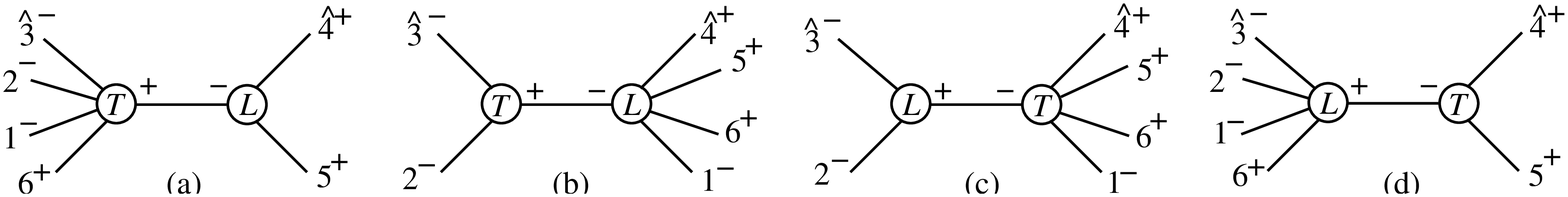}}
\caption{
The recursive diagrams arising from an auxiliary $\Shift34$ shift in
$A^{\NeqZero}_{6;1}(1^-,2^-,3^-,4^+,5^+,6^+)$.
Diagram (c) has non-standard complex singularities.  }
\label{Loop3mSixPt34Figure}
\end{figure}

We begin by conjecturing a recursion relation for the large-$z$
behavior under the $\Shift12$ shift that is the simplest possible
generalization of the one we found for the five-point amplitude,
namely,
\ba
\InfPart{\Shift12}{A^{\NeqZero}_{6;1}(1^-,2^-,3^-,4^+,5^+,6^+)}
&=&  \InfPart{\Shift12}{A^{\NeqZero}_{5;1}(1^-,2^-,\hat{3}^-,\hat
       K_{45}^+,6^+)} \times {i\over s_{45}} \nn \\
&& \hskip 2 cm \null
 \times
    A_3^\tree(-\hat{K}_{45}^-,\hat{4}^+,5^+)\,.
\label{LargeZRecursion6Pt34}
\ea
As in \eqn{LargeZRecursion5Pt} for the five-point case, $\hat{a}$
indicates a momentum shifted according to the auxiliary $\Shift34$ shift in
\eqn{Shift34Aux}, with $w$ frozen to the value
\be
w^{\rm (d)} = - {\spa4.5 \over \spa3.5} \,.
\label{w34frozen}
\ee

We will return to a more systematic construction of the large-$z$
contributions via on-shell recursion in
sections~\ref{DerivationSection}-\ref{ProcedureSampleSection}.  Here
we only wish to provide some motivation for the large-$z$ recursion
relation (\ref{LargeZRecursion6Pt34}), following the same procedure
as at five points.

The recursive diagrams for computing the amplitude
using a $\Shift34$ shift are displayed in \fig{Loop3mSixPt34Figure}.
However, we are only interested in the values of the diagrams in the
large-$z$ limit of the $\Shift12$ shift.
Diagram~\ref{Loop3mSixPt34Figure}(a) has legs 1 and 2 attached to a tree
which vanishes in the large-$z$ limit~\cite{BCFW}.
As we shall discuss in \sect{InfinityPoleSubsection}, 
we assume, based on empirical evidence from known amplitudes,
that the full diagram is also suppressed at large $z$, even though
the loop vertex is non-standard.
Diagram~\ref{Loop3mSixPt34Figure}(b) is more complicated, 
in that the shifted legs straddle the pole, but its
value may be determined easily since its components are known; it is
then not difficult to verify that it is suppressed at large $z$.
Diagram~\ref{Loop3mSixPt34Figure}(c) is problematic, since it also contains
a non-standard complex pole.  For this case, we assume that the
addition of the extra leg on the tree side, compared to the
corresponding five-point diagram shown in
\fig{Loop3mFivePt34Figure}(b), does not upset the suppression at large
$z$; we take it to have a vanishing contribution in this limit.
One must also check that there are no additional
large-$z$ contributions from the cut terms (after subtracting
$\InfPart{\null}{\Cuth_n}$) or from the overlap terms.  Following a similar
analysis as for the five-point case, it is not difficult to verify
that there are none in this case.  We are
left with the single diagram~\ref{Loop3mSixPt34Figure}(d), as the
sole surviving contribution in the large-$z$ limit of the primary
$\Shift12$ shift, motivating \eqn{LargeZRecursion6Pt34}.

To evaluate the large-$z$ recursion relation in \eqn{LargeZRecursion6Pt34},
we use the five-point large-$z$ result in \eqn{FivePoint45Channel} as input.
We obtain,
\ba
\InfPart{\Shift12}{A^{\NeqZero}_{6;1}(1^-,2^-,3^-,4^+,5^+,6^+)}
&=& - i \, {\cg\over3}{\spash1.{\hat 3}^3
\spa2.6\over\spa1.6^2\spash{\hat 3}.{\hat K_{45}}
\spash{\hat K_{45}}.6\spb1.2}\,
{1\over s_{45}} \, {\spbsh{\hat 4}.5^3 \over \spbsh{\hat K_{45}}.{\hat
4} \spbsh{5}.{\hat K_{45}} } \nn \\ &=&
 i\,{\cg\over3}{\spa1.3^3\spa2.6\over\spa1.6^2\spa3.4\spa4.5\spa5.6\spb1.2}\,,
\label{Boundarymmmppp12}
\ea
where, as above, $\hat{a}$
indicates a momentum shifted according to the $\Shift34$ shift in
\eqn{Shift34Aux}, with $w$ frozen to the value~(\ref{w34frozen}).

\subsection{The Completed-Cut Terms}

The cut-containing terms of our target amplitude were computed in
ref.~\cite{RecurCoeff} using a recursion relation on {\it
coefficients\/} of integral functions,
along with known lower-point results~\cite{GGGGG,Neq1Oneloop}.  
In the six-point case, this procedure yields,
\def\indentA{\hskip 0mm} 
\ba
   \Cuth_6(1^-,2^-,3^-,4^+,5^+,6^+)  &=&
 \frac{1}{3 \cg}\,A_{6;1}^{\,\NeqOne}(1^-,2^-,3^-,4^+,5^+, 6^+)
\nn\\
&& \hskip0.0cm 
 + {2\over 9} A_6^\tree(1^-,2^-,3^-,4^+,5^+, 6^+)
 + \hat{C}_6^a + \hat{C}_6^a \Bigr|_{\rm flip\; 1}
\,,~~ \label{Cuth6} 
\ea
where
\ba 
  \hat{C}_6^a &=&
{i\over3} \Biggl[ { \spa1.2\! \spa2.3\!\spb2.4 \!\spab1.{(3\!+\!4)}.2
\bigl[ \spaa3.{4}.{2}.1 s_{234}
      - \spaa3.{2}.{(3+4)}.1 s_{34} \bigr]
  \over \spa3.4\spa5.6 \spa6.1\spb2.3 \spab5.{(3\!+\!4)}.2 }
{ \Lzz ( {-s_{234}\over -s_{34}} ) \over s_{34}^3  }
\nonumber\\
&& \hskip0.0cm
+ { \!\spa3.5\!\spb4.5\!\spb5.6\! \spab5.{(1\!+\!2)}.6
    \bigl[ \sand3.{(5-4)}.6 \!s_{345}
              +\sand3.{(4+5)}.6\! s_{34}\bigr] \,
  \over \spa4.5\spb1.2 \spb1.6 \spab5.{(3+4)}.2 }
{ \Lzz ( {-s_{345}\over -s_{34}} ) \over s_{34}^3  } \Biggr]
%
\,, \nonumber\\
&& \hskip0.0cm{~} \label{Cuth6a}
\ea
and where we have introduced the flip symmetry operation,
\be
X(1,2,3,4,5,6)\Bigr|_{\rm flip\; 1} \equiv X(3,2,1,6,5,4) \,.
\label{mmmpppflip1def}
\ee
The first term in \eqn{Cuth6} is proportional to the contribution 
of an $\NeqOne$ chiral multiplet in the loop.  
This contribution is fully constructible from the four-dimensional 
cuts~\cite{Neq1Oneloop}.  The result is~\cite{NeqOneNMHVSixPt},
\def\indentA{\hskip 10mm} 
\be
A_{6;1}^{\,\NeqOne}(1^-,2^-,3^-,4^+,5^+,6^+) = 
S_6^a + S_6^a \Bigr|_{\rm flip\; 1}
\,, \label{AsusyN1mmmppp}
\ee
where
\ba 
S_6^a &=&
{i \cg \over 2} \Biggl[ 
{1\over i} A_6^\tree(1^-,2^-,3^-,4^+,5^+,6^+) \, \Kz( s_{34} ) 
\label{S6a} \\
&&\hskip0.7cm
- {\sand1.{(2+3)}.4^2\,
\bigl[ \spaa3.{4}.{2}.1 s_{234}
      - \spaa3.{2}.{(3+4)}.1 s_{34}\bigr]
\over \spa{5}.{6} \spa{6}.1 \spb2.3\,s_{234}\,s_{34}
   \sandmm5.{(3+4)}.{2}}
{\Lz ( {-s_{234}\over -s_{34}} ) \over s_{34}  } \nonumber \\
&&\hskip0.7cm
- {\sand3.{(1+2)}.6^2
\bigl[\sand3.{(5-4)}.6 s_{345} + \sand3.{(4+5)}.6 s_{34}\bigr]
\over \spa3.4\spa4.5\spb1.2\spb1.6\,s_{345}\,\sandmm5.{(3+4)}.2 }
{ \Lz ({-s_{345}\over -s_{34}} ) \over s_{34}  } \Biggr]
\,.  \nonumber
\ea

After performing the $\Shift12$ shift,
the completed-cut expression given in \eqn{Cuth6}
does not vanish as $z\rightarrow \infty$, but tends
to a purely-rational constant,
\ba
&& \InfPart{\Shift12}{\Cuth_6(1^-,2^-,3^-,4^+,5^+, 6^+)} =
       \lim_{z\rightarrow \infty}
      \Cuth_6(1^-,2^-,3^-,4^+,5^+, 6^+; z) \nn\\
&& \hskip 2 cm \null
=
 {i\over6} { \spa1.2 \spa1.3 \spab3.{(4+5)}.2
  \Bigl[ - \spaa1.2.{(4+5)}.3 + \spa1.3 s_{345} \Bigr]
  \over \spb1.2 \spa3.4 \spa4.5 {\spa6.1}^2
   \, s_{345} \, \spab5.{(3+4)}.2 }
\nonumber\\
&&  \hskip 2.5 cm \null
 + {i\over6} { \spa1.2 \spb2.4 \spa1.3
   ( \spa1.2 \spb2.4  - \spa1.3 \spb3.4 )
  \over \spb2.3 \spa5.6 \spa6.1 \, s_{34} \, \spab5.{(3+4)}.2 }
\,.
\label{Cuth6infpole}
\ea
Since we have already determined the complete large-$z$ behavior
of the full amplitude in \eqn{Boundarymmmppp12}, we must subtract
from $\Cuth_6$ this rational constant, $\InfPart{\Shift12}{\Cuth_6}$,
so that the difference $\Cuth_6 - \InfPart{\Shift12}{\Cuth_6}$ vanishes as
$z \rightarrow \infty$.  When computing the overlap contribution we
may use either this difference or the original $\Cuth_6$.  They are
equivalent because $\InfPart{\Shift12}{\Cuth_6}$ has no poles in $z$
under a $\Shift12$ shift; therefore it does not generate an overlap
contribution.

\subsection{Recursive Contributions}

\begin{figure}[t]
\centerline{\epsfxsize 5.5 truein\epsfbox{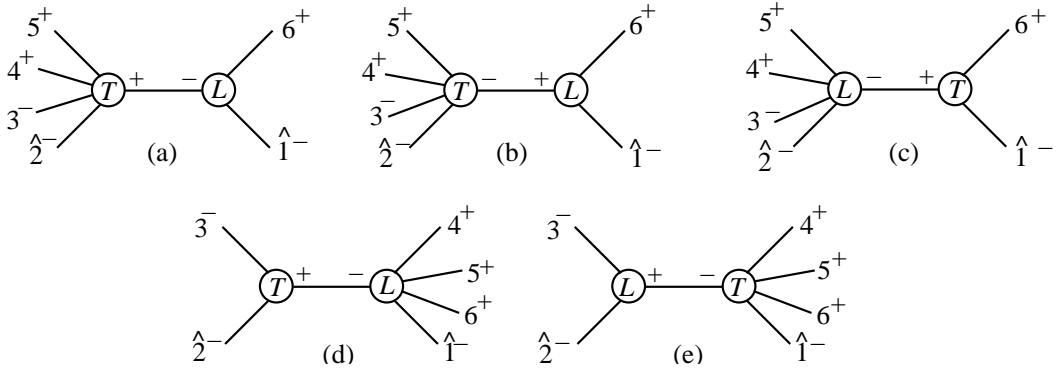}}
\caption{
Some vanishing recursive diagrams for the $\Shift{1}{2}$ shift
of $A^{\NeqZero}_{6;1}(1^-,2^-,3^-,4^+,5^+,6^+)$.
}
\label{Vanish3mSixPt12Figure}
\end{figure}

\begin{figure}[t]
\centerline{\epsfxsize 3.6 truein\epsfbox{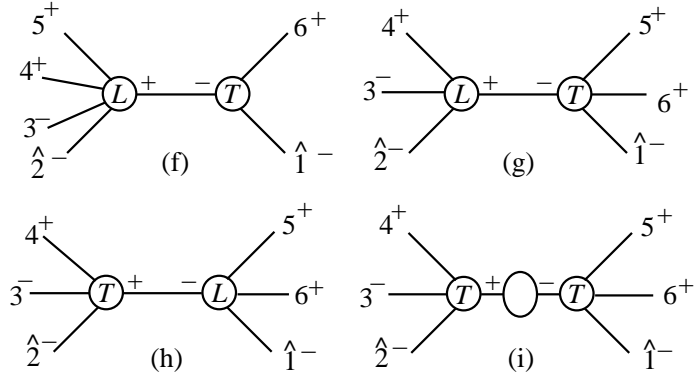}}
\caption{
Non-vanishing recursive diagrams.
Diagram (i) is the factorization-function contribution.
}
\label{Loop3mSixPt12Figure}
\end{figure}

Next we evaluate the recursive diagrams for the $\Shift12$ shift
of $A^{\NeqZero}_{6;1}(1^-,2^-,3^-,4^+,5^+,6^+)$.
Most of them vanish.  Some of the vanishing diagrams are shown in
\fig{Vanish3mSixPt12Figure}.  (We omitted two diagrams in the
$(23)$ channel, where $(-\Kh_{23})$ carries negative helicity, which
vanish even more trivially.)  The first two diagrams,
\ref{Vanish3mSixPt12Figure}(a) and \ref{Vanish3mSixPt12Figure}(b),
vanish because the loop vertices
$R_3(6^+, \hat 1^-, -\hat K_{61}^\pm)$ vanish, as discussed 
in \sect{ChoiceShiftSubSection} and 
in the appendix.
Diagrams \ref{Vanish3mSixPt12Figure}(c) and \ref{Vanish3mSixPt12Figure}(d)
vanish because
\be
A_3^\tree(6^+,\hat 1^-,-\Kh_{61}^+) \propto \spbsh{6}.{\Kh_{61}}^3 = 0\,,
\hskip 1.5 cm
A_3^\tree(\hat 2^-,3^-, -\Kh_{23}^+) \propto \spash{\hat 2}.{3}^3 = 0\,.
\ee
As discussed for the five-point case,
diagram~\ref{Vanish3mSixPt12Figure}(e) vanishes because its
loop vertex is of the same type as the vanishing
tree vertex in diagram~\ref{Vanish3mSixPt12Figure}(d).
Summarizing, we have
\be
D_6^{\rm (a)} = D_6^{\rm (b)} = D_6^{\rm (c)}
              = D_6^{\rm (d)} = D_6^{\rm (e)} =  0\,.
\label{Diagram6abcdemmmppp}
\ee

The four non-vanishing recursive diagrams,
\ba
\DiagrammaticRational_6  = D^{\rm (f)} + D^{\rm (g)}
                         + D^{\rm (h)} + D^{\rm (i)} \,,
\ea
are shown in \fig{Loop3mSixPt12Figure}. These diagrams are
straightforward to evaluate because all channels involve standard
factorizations.  Diagram \ref{Loop3mSixPt12Figure}(f) yields
\ba
D_6^{\rm (f)} &=&
A_3^\tree(\hat{1}^-,-\hat{K}_{61}^-,6^+) \, {i\over s_{61}}
\, R_5(\hat{2}^-,3^-,4^+,5^+,\hat{K}_{61}^+)
\nonumber\\
&=&
 - i \biggl( {1 \over 3\e} + {8\over9} \biggr)
  { {\spab3.{(1+2)}.6}^3
   \over \spb1.2 \spa3.4 \spa4.5 \spb6.1
   \, s_{345} \, \spab5.{(3+4)}.2 }
\nonumber\\
&& \null
- {i\over6} \, { s_{345} + s_{34} \over s_{34} s_{345} (s_{345} - s_{34})^2 }
  { \spa3.5 \spb4.5 \spb5.6  \spab5.{(1+2)}.6
   \over \spb1.2 \spa4.5 \spb6.1 \, \spab5.{(3+4)}.2 }
\nonumber\\
&&\hskip1cm \times
   ( \spab3.4.5  \spab5.{(1+2)}.6 + \spab3.5.6 s_{345} )
\nonumber\\
&& \null
+ {i\over3}\, { {\spb4.6}^3 \spab4.{(3+5)}.2
    \over \spb1.2 \spb2.3 \spb3.4 \spa4.5 \spb6.1 \, \spab5.{(3+4)}.2 }
+ {i\over3}\, { {\spb4.6}^2 \spab3.{(1+2)}.6
    \over \spb1.2 \spb3.4 \spa4.5 \spb6.1 \, \spab5.{(3+4)}.2 }
\nonumber\\
&& \null
- {i\over6} \, {\spa3.5  \spb4.5  \spb5.6 \spab3.{(1+2)}.6 \spab5.{(1+2)}.6
      \over \spb1.2\spa4.5 \spb6.1
    \, s_{34} s_{345} \, \spab5.{(3+4)}.2 }
 \,,
\label{Diagram6fmmmppp}
\ea
where  the
five-point scalar loop vertex is given in \eqn{R5mmmpp}.
Diagram~\ref{Loop3mSixPt12Figure}(g) is
\ba
D^{\rm (g)} & = &
\, A_4^\tree(\hat{2}^-,3^-,4^+,\hat{K}_{561}^+)
\, {i\over s_{561}}
R_4(\hat{1}^-,-\hat{K}_{561}^-,5^+,6^+)
\nonumber\\
&=&
 \biggl( {1 \over 3\e} + {8\over9} \biggr)
 A_4^\tree(\hat{2}^-,3^-,4^+,\hat{K}_{561}^+)  \, {i\over s_{561}} \,
 A_4^\tree(\hat{1}^-,-\hat{K}_{561}^-,5^+,6^+)
\nonumber\\
&=&
i \biggl( {1 \over 3\e} + {8\over9} \biggr)
 { {\spab1.{(2+3)}.4}^3
    \over \spb2.3 \spb3.4 \spa5.6 \spa6.1
    \, s_{234} \, \spab5.{(3+4)}.2  }
\,,
\label{Diagram6gmmmppp}
\ea
where we used the four-point tree amplitude, \eqn{A4treemmpp},
and the loop four-vertex given in \eqn{R4mmpp}.
It is easy to see that diagram \ref{Loop3mSixPt12Figure}(h)
gives the same value,
\be
D^{\rm (h)}  =
  D^{\rm (g)} \,.
\label{Diagram6hmmmppp}
\ee

Diagram~\ref{Loop3mSixPt12Figure}(i) contains the factorization function
contribution~\cite{BernChalmers}, which for the scalar loop
case amounts to a vacuum polarization insertion.  The value of the
factorization function vertex appearing in this
diagram is given in~\eqn{FactFunctionVertex}.  Using
this value of the factorization function,
diagram~\ref{Loop3mSixPt12Figure}(i) is given by
\ba
D^{\rm (i)} & = &
- \biggl( {1 \over 3\e} + {8\over9} \biggr)
 A_4^\tree(\hat{2}^-,3^-,4^+,\hat{K}_{561}^+) \, {i\over s_{561}}
\, A_4^\tree(\hat{1}^-,-\hat{K}_{561}^-,5^+,6^+) \nn \\
& = &
 - D^{\rm (g)} \,.
\label{Diagram6immmppp}
\ea
The fact that diagrams \ref{Loop3mSixPt12Figure}(g), (h) and (i) are
equal, up to signs, is rather special to this amplitude.

\subsection{The Overlap Contributions}

\begin{figure}[t]
\centerline{\epsfxsize 6 truein\epsfbox{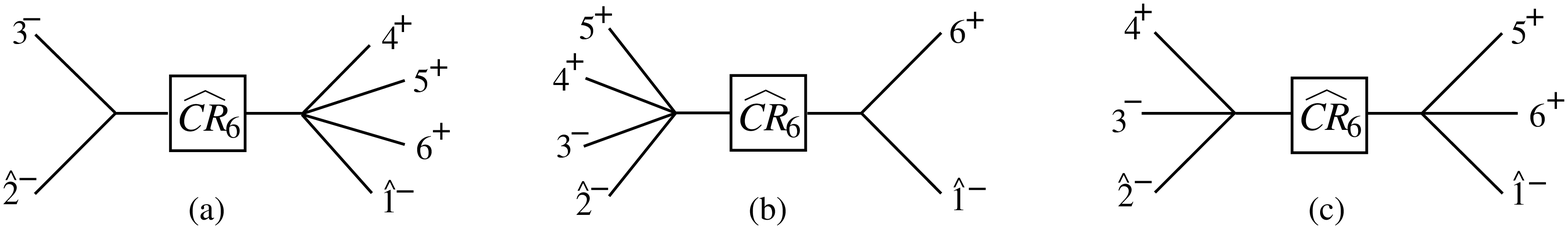}}
\caption{
The overlap diagrams for the $\Shift{1}{2}$ shift.
Diagram (a) vanishes.
}
\label{Overlap3mSixPt12Figure}
\end{figure}

To evaluate the overlap contributions we start from the rational
parts of the completed-cut terms, obtained by setting all logarithms
to zero in \eqn{Cuth6}.  That is, we replace the $\Kz$ and $\Ll_i$
functions defined in \eqn{Lsdef} with their rational parts,
\ba
\Kz(r) &\rightarrow & {1\over \e} + 2 \,, \nn \\
\Ll_0(r) &\rightarrow& 0\,, \nn \\
\Ll_2(r) &\rightarrow&  {1\over 2} {(r+1)\over r(1-r)^2} \,.
\ea
Making these replacements in \eqn{Cuth6} for $\Cuth_6$ gives us
$\CuthRat_6$.  Applying the $\Shift12$ shift, \eqn{Shift12},
yields $\CuthRat_6(z)$, which we use to evaluate the overlap
contributions.

As depicted in \fig{Overlap3mSixPt12Figure}, for the $\Shift12$ shift
there are three channels which can potentially contribute to the
overlap,
\be
\Overlap = \Overlap^{\rm (a)} + \Overlap^{\rm (b)} + \Overlap^{\rm (c)} \,,
\ee
corresponding to the three
residues of $\CuthRat_6(z)/z$ located
at the values of $z$,
\be
z^{\rm (a)} = -{\spa2.3 \over \spa1.3}\,, \hskip 2 cm
z^{\rm (b)} = {\spb1.6 \over \spb2.6} \,, \hskip 2 cm
z^{\rm (c)} = -{s_{234} \over \sand1.{(3+4)}.2} \,.
\ee
Evaluating these residues gives us the overlap contributions.
After simplification, they are given by,
\ba
\Overlap^{\rm (a)} &=&
 0
\,, \nn \\
\Overlap^{\rm (b)}
&=&
i \biggl( {1 \over 3\e} + {8\over9} \biggr)
  { {\spab3.{(1+2)}.6}^3
   \over \spb1.2 \spa3.4 \spa4.5 \spb6.1
   \, s_{345} \, \spab5.{(3+4)}.2 }
\nonumber\\
&&
- {i\over6}\, { \spa1.2 \spab3.{(4+5)}.2 {\spab3.{(1+2)}.6}^2
   \over \spb1.2 \spa3.4 \spa4.5 \, s_{61} s_{345}
   \, \spab5.{(3+4)}.2 }
   \nonumber \\
&& + {i \over 6} \, { \spb4.5 \spb5.6 \spa3.5 \spab5.{(1+2)}.6
\left( \spa3.5 \spb5.6 s_{345} - \spa3.4 \spb4.5 \spab5.{(3+4)}.6
\right) \over \spb1.2 \spa4.5 \spb6.1 \spab5.{(3+4)}.2 } \nn \\
&& \quad \times
{ s_{34} + s_{345} \over s_{34} s_{345} \left( s_{345}-s_{34} \right)^2 }
\nonumber \\
&& - {i \over 6} {\spa1.5 \spb4.5 \spb4.6 \spab1.{(5-6)}.4
\over \spb2.3 \spb3.4 \spa5.6 s_{61} \spab5.{(3+4)}.2 }
 \,, \nn \\
\Overlap^{\rm (c)}
&=&
- i\biggl( {1 \over 3\e} + {8\over9} \biggr)
 { {\spab1.{(2+3)}.4}^3
    \over \spb2.3 \spb3.4 \spa5.6 \spa6.1
    \, s_{234} \, \spab5.{(3+4)}.2 }
\nonumber\\
&&
+ {i\over6}\, { \spa1.2 \spb2.4  {\spab1.{(2+3)}.4}^2
    \over \spb2.3 \spb3.4 \spa5.6 \spa6.1 \, s_{234} \, \spab5.{(3+4)}.2 }
\nonumber\\
&&
+ {i\over6} \, { \spa1.5 \spa3.4  \spb4.5 \spa1.6  {\spab1.{(2+3)}.4}^2
    \over \spb2.3 \spa5.6 {\spa6.1}^2 \, s_{34} s_{234}
    \, \spab5.{(3+4)}.2 }
 \,.
\label{mmmpppoverlap}
\ea
Although there is no need to keep the $1/\e$ terms (their values are known
{\it a priori}), we have carried them along.


\subsection{The Full Amplitude and Consistency Checks}

We may now combine all the pieces to obtain the full amplitude,
\be
A_{6;1}^{\NeqZero}(1^-, 2^-, 3^-, 4^+, 5^+, 6^+) =
     \cg \Bigl[  \Cuth_6 + \Remaining_6 \Bigr] \,,
\label{FullA6mmmppp}
\ee
where $\Cuth_6$ is given in \eqn{Cuth6}.  The rational remainder
$\Remaining_6$, consisting of recursive diagrams, overlap diagrams and
large-$z$ contributions, is,
\ba
\Remaining_6 &\equiv& {1 \over \cg}
\InfPart{\Shift12}{A_{6;1}^{\NeqZero}}
-  \InfPart{\Shift12}{\Cuth_6} +
D^{\rm (f)} + D^{\rm (g)} + D^{\rm (h)} + D^{\rm (i)}
+ \Overlap^{\rm (b)} + \Overlap^{\rm (c)} \, ,
\ea
where $\InfPart{\Shift12}{A_{6;1}^{\NeqZero}}$ is given in
\eqn{Boundarymmmppp12}  and
$\InfPart{\Shift12}{\Cuth_6}$ in \eqn{Cuth6infpole}.
The values of the recursive and overlap contributions are given in
eqs.~(\ref{Diagram6fmmmppp})--(\ref{Diagram6immmppp}), and
\eqn{mmmpppoverlap}.  Simplifying the result for the rational
remainder and making use of the flip symmetry~(\ref{mmmpppflip1def}),
we can write the result for the remaining rational part in
a rather compact form,
\be
\Remaining_6 = \Remaining_6^a  + \Remaining^a_6 \Bigr|_{\rm flip\; 1} \, ,
\label{mmmpppRsimple}
\ee
where
\ba
\Remaining_6^a &=&
 {i\over6} { 1 \over \spb2.3 \spa5.6 \, \spab5.{(3+4)}.2 }
   \Biggl\{   - { {\spb4.6}^3 \spb2.5 \spa5.6 \over \spb1.2 \spb3.4 \spb6.1 }
  - { {\spa1.3}^3 \spa2.5 \spb2.3 \over \spa3.4 \spa4.5 \spa6.1 }
\nonumber\\
&&\hskip1cm \null
 + { {\spab1.{(2+3)}.4}^2 \over \spb3.4 \spa6.1 }
          \biggl( { \spab1.{2}.4 - \spab1.{5}.4 \over s_{234} }
          + { \spa1.3 \over \spa3.4 }
          - { \spb4.6 \over \spb6.1 } \biggl)
\label{mmmpppRasimple}\\
&&\hskip1cm \null
  - { {\spa1.3}^2  ( 3  \spab1.{2}.4 + \spab1.{3}.4 )
     \over \spa3.4 \spa6.1 }
  + { {\spb4.6}^2 ( 3  \spab1.{5}.4 + \spab1.{6}.4 )
     \over \spb3.4 \spb6.1 }
    \Biggr\}
\,. \hskip 1 cm
\nn
\ea
The result~(\ref{mmmpppRsimple}) is manifestly symmetric,
not only under the flip~(\ref{mmmpppflip1def}), but also under
the second flip symmetry, involving spinor conjugation,
\be
X(1,2,3,4,5,6)\Bigr|_{\rm flip\ 2} \equiv X^*(6,5,4,3,2,1).
\label{mmmpppflip2def}
\ee
The remarkable simplicity of the rational remainder is rather
striking.

We have performed a number of checks on our result for the amplitude,
\eqn{FullA6mmmppp}.  We have confirmed that it has the proper
factorization properties in real momenta in all two- and three-particle
channels and that all spurious singularities indeed cancel as they
should.  Since the large-$z$ contribution in
\eqn{Boundarymmmppp12} contains kinematic poles in a variety
of channels, an omitted piece would necessarily be detected in
some of the collinear limits.  Finally, the numerical value
at one phase-space point agrees with that in ref.~\cite{EGZ06}.
The consistency of the amplitude demonstrates the validity of our
procedure for determining the large-$z$ terms, for a new and
rather non-trivial analytic amplitude.

\section{On-Shell Recursion Relations for Loop Amplitudes}
\label{DerivationSection}

Before continuing to the case of general helicities, we present
in this section a more systematic discussion of loop-level 
on-shell recursion relations,
emphasizing the extensions beyond ref.~\cite{Bootstrap}.
The derivation of such loop recursion relations is similar in spirit
to the tree-level case, but it does require the treatment of
factorizations which differ from the `ordinary' factorization in {\it
real\/} momenta.  It also differs because of the appearance of branch
cuts and spurious singularities associated with logarithms or
polylogarithms.  The cut-containing parts of the amplitude are an
input to the loop recursion relations.  We assume that they
have been computed by other means, such as the unitarity-based method.

\subsection{Analytic Behavior of Shifted Loop Amplitudes}
\label{AnalyticSubsection}

The starting point for the loop recursion relations is a
complex-valued shift of the momenta of a pair of external particles in
an $n$-point amplitude, $k_j \to \hat{k}_j(z)$, $k_l \to
\hat{k}_l(z)$.  We describe a $\Shift{j}{l}$ shift in terms of the
spinor variables $\lambda$ and $\tlambda$ defined in
\eqn{lambdadef} ,
\begin{equation}
\Shift{j}{l}:\hskip 2 cm
\tlambda_j \rightarrow \tlambda_j - z\tlambda_l \,,
\hskip 2 cm
\lambda_l \rightarrow \lambda_l + z\lambda_j \,,
\label{SpinorShift}
\end{equation}
following the tree-level construction~\cite{BCFW}.
The shift maintains overall momentum conservation as well as the
masslessness of the external momenta, $\hat{k}_j^2 = \hat{k}_l^2 =
0$.  Let us denote the original $n$-point amplitude by
$A_n \equiv A_n(0)$, and the shifted one by $A_n(z)$.  We seek an
equation for $A_n(0)$ relying on the analytic properties of $A_n(z)$.
(We will also denote by $f(z)$ other functions $f$ of the momenta, such as
the cut-containing terms, after the shift~(\ref{SpinorShift}).)

At tree level, $A_n(z)$ is a meromorphic function of $z$.  Its
poles are determined by the factorization properties of $A_n$ as
multi-particle invariants or spinor products vanish.  The former are
identical for real and complex momenta, and at tree level, the
singularities for spinor products of complex momenta are also
completely determined by the corresponding singularities (collinear
factorizations) for real momenta.  At tree level, one can
show~\cite{BCFW,GloverMassive,VamanYao}, that there are always choices
of shift momenta for which $A_n(z)\rightarrow 0$ as
$z\rightarrow\infty$.  This allows the derivation of a recursion
relation through consideration of a contour integral on a circle at
infinity.

At one loop, we must consider several new aspects.
The most obvious of these is the presence of branch cuts, which arise
from logarithms or polylogarithms in the amplitudes.  But there are a
number of other important features.  While factorization in real
momenta is
understood~\cite{Neq4Oneloop,BernChalmers,OneLoopSplitUnitarity}, this
does not completely determine the singularity structure in complex
momenta.  In particular, we must in general handle double poles as
well as `unreal' poles, present for complex momenta but absent for
real ones.  We obtained some of the required factorizations
heuristically~\cite{OnShellRecurrenceI,Qpap}, confirming them by
explicit calculation.  In certain two-particle channels, however, the
structure of complex factorization is not yet completely clear;
accordingly, we will design our calculations to avoid relying on them.

While it does appear in general possible to find momentum shifts under
which $A_n(z)\rightarrow 0$ as $z\rightarrow\infty$, it is not always
possible to do so while avoiding poles in channels with obscure
complex-factorization behavior.  Accordingly, we must
deal with amplitudes that have either a pole at infinity, so that
$A_n(z)\rightarrow\infty$ as $z\rightarrow\infty$, or that behave as
a finite constant.
We will, however, assume that this
behavior is given by a {\it rational\/} function of $z$.  In all cases
we deal with here, where the corresponding tree amplitudes vanish in
the large-$z$ limit, the large-$z$ terms are purely rational.  For the moment,
let us assume we know these terms, along with the cut-containing
terms; below we return to the issue of their computation.

More precisely, we wish to define an operator $\InfPart{\null}{\null}$
that yields a pole-free rational function reproducing the large-$z$
behavior of an amplitude, when both are shifted,
\be
\lim_{z\rightarrow\infty} 
\Bigl(\Bigl[\InfPart{\null}{A_n}\Bigr](z) -A_n(z)\Bigr) = 0\,.
\ee
We implement this operator via a series expansion of the amplitude
around the point $z=\infty$, or $u=1/z = 0$.  We take $A_n(z)$ to be
an analytic function with a series expansion of the form,
\be
A_n(z) = \sum_{i=0}^{i_{\rm max}} {1\over u^i}\, a^{(-i)} +
               \sum_{i=1}^\infty u^i\,
                      b^{(i)} + \sum_{i=1}^\infty \ln(u) u^i \, c^{(i)} +
                      \sum_{i=1}^\infty \ln^2(u) u^i\, d^{(i)} \,,
\ee
where the coefficients, $a^{(-i)}$, $b^{(i)}$, $c^{(i)}$ and $d^{(i)}$
depend on spinor products. The leading
degree of large-$z$ behavior, $i_{\rm max}$,
depends on the helicity configuration. In writing this expansion,
we have assumed that no $\ln(z)$ factors appear in terms which survive
as $z\rightarrow \infty$.  (We can always confirm the validity of this
assumption for a given amplitude, since all logs and
polylogarithms will have already been computed via the unitarity
method or other means.)  The large-$z$ terms are those which do not
vanish as $z\rightarrow \infty$, 
\be
\Ainf_n(z) \equiv \sum_{i=0}^{i_{\rm max}} z^i\, a^{(-i)} 
= \Big[\Inf A_n \Big](z)\,,
\ee
given by shifting their progenitor $\Inf A_n$.

We may then write $A_n(z)$ as a sum of the large-$z$ terms
and the remaining terms,
\begin{equation}
A_n(z) = \Ainf_n(z) + \conv_n(z)\,,
\label{InfinitySplit}
\end{equation}
so that all finite-$z$ poles, logarithms, and polylogarithms in $z$
are put in the second term.  We will call
the latter object the `finite-pole' terms.

Since we will be interested in the unshifted physical amplitude obtained
by setting $z=0$, we will need those terms in $A_n(0)$ that
generate $\Ainf_n(z)$ under the shift.  We find in practice
that these are given by setting $z=0$ in  $\Ainf_n(z)$,
that is keeping only the $z^0=u^0$ terms in the expansion of $A_n(z)$,
\be
\InfPart{\null}{A_n} = a^{(0)} \,.
\ee
A key aim of this paper is to provide a means for
computing this quantity.

\begin{figure}[t]
\centerline{\epsfxsize 2.0 truein\epsfbox{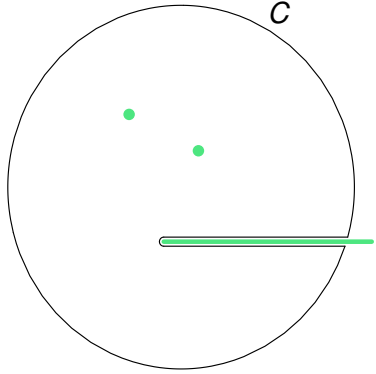}}
\caption{
A configuration of poles and branch cuts for a term in a one-loop
amplitude, with a branch-cut-hugging contour.}
\label{BranchCutIntegralFigure}
\end{figure}

Since $\conv_n(z)$ vanishes for $z\rightarrow \infty$, we may apply
Cauchy's residue theorem. Consider the contour going along the circle
at infinity, but avoiding the branch cuts by integrating inwards along
one side, and then outwards along the other, as shown in
\fig{BranchCutIntegralFigure}.  We route the branch cuts so that no
two overlap.  (As discussed in ref.~\cite{Bootstrap} the case of poles
touching the ends of branch cuts works in the same way as the basic
case, and we shall not distinguish it further in our discussion.)
The integral along this contour is given by the sum
of residues.  This contour, however, includes
a branch-cut-hugging integral,
\begin{equation}
{1\over2\pi i}\int_{B^\uparrow+i\epsilon} {dz\over z}\; \conv_n(z)
+ {1\over2\pi i}\int_{B^\downarrow-i\epsilon} {dz\over z}\; \conv_n(z) \,,
\end{equation}
where $B^\uparrow$ is directed from an endpoint $B_0$ to infinity,
and $B^\downarrow$ is directed in the opposite way.  Now, $\conv_n(z)$ has
a branch cut along $B$, which means that it has a non-vanishing discontinuity,
\begin{equation}
2\pi i\Disc_B \conv_n(z) = \conv_n(z+i\epsilon)-\conv_n(z-i\epsilon),
\qquad z{\rm\ on\ }B \,.
\end{equation}
(Because $\Ainf_n$ is taken to be purely rational, $\Disc_B \conv_n(z)
= \Disc_B A_n(z)$.)
Thus, using the vanishing of  $\conv_n(z)$ for $z\rightarrow \infty$,
we have,
\begin{equation}
0 = \conv_n(0) +
  \sum_{{\rm poles}\ \alpha} \Res_{z=z_\alpha} {\conv_n(z)\over z}
+ \int_{B_0}^\infty {dz\over z}\;\Disc_B \conv_n(z) \,.
\label{BasicAmplitudeEquation}
\end{equation}
%

\subsection{Cuts and Cut Completion}
\label{CutContainingSubsection}

To proceed further, let us assume that we have already computed all
terms having branch cuts, along with certain closely related terms
that can generally be obtained from the same computation.  That is, we
have computed all polylog terms, all log terms, and all $\pi^2$ terms.
As discussed in ref.~\cite{Bootstrap}, there are also certain classes
of rational terms that are natural to include with the cut-containing
terms.

In particular, there are rational terms whose presence is required to
cancel spurious singularities in the (poly)logarithmic terms.
Such spurious singularities arise in the course of integral reductions.
They are not singularities of the final amplitude, because they
are unphysical, and not singularities of any Feynman diagram.
A simple example comes from a `two-mass' triangle integral for
which two of the three external legs are off shell (massive),
with momentum invariants $s_1$ and $s_2$, say.
When there are sufficiently many loop momenta
inserted in the numerator of this integral, it gives rise to
functions such as,
\begin{equation}
{\ln(r)\over (1-r)^2} \,,
\label{sampleL1pure}
\end{equation}
where $r$ is a ratio of momentum invariants (here $r=s_1/s_2$).
The limit $r\rightarrow 1$ (that is, $s_1\rightarrow s_2$)
is a spurious singularity; it does not correspond to any physical
factorization. Indeed, this function always appears in the amplitude
together with appropriate rational pieces,
\begin{equation}
{\ln(r)+1-r\over (1-r)^2} \,,
\label{sampleL1}
\end{equation}
in a combination which is finite as $r\rightarrow 1$.
From a practical point of view, it is most convenient to `complete'
the unitarity-derived answer for the cuts by replacing functions
like~\eqn{sampleL1pure} with non-singular combinations
like~\eqn{sampleL1}.
Such completions are of course not unique; one could add
additional rational terms free of spurious singularities.

The reason we want to eliminate spurious singularities from sub-expressions
has to do with the sum over pole residues in \eqn{BasicAmplitudeEquation}.
The sum runs over all poles, whether they arise from a shift in a
physical singularity variable, or in a spurious one.  In practice, it
is sufficient to eliminate all spurious singularities that acquire a
$z$ dependence under the momentum shift.  By construction, $\Ainf_n(z)$
is free of such `dangerous' spurious singularities, since it cannot contain
poles in $z$.

Singularities that look like collinear ones, but involve
non-adjacent momenta, are also spurious.
For example, in the scalar contributions to
the five-gluon $({-}{+}{-}{+}{+})$ amplitude~\cite{GGGGG}, there are factors of
$\spa2.4$ and $\spa2.5$ appearing in the denominators of certain
coefficients.  These might appear to give rise to non-adjacent collinear
singularities in complex momenta; but by expanding the polylogarithms
and logarithms in that limit, one can show that these singularities
are in fact absent in the full amplitude.

Let us define two decompositions of the 
amplitude.  The first
is into `pure-cut' and `rational' pieces.
The rational parts are defined by setting all logarithms,
polylogarithms, and $\pi^2$ terms to zero,
\begin{equation}
\Vertex_n(z) \equiv {1\over \cg} A_n\Bigr|_{\rm rat} =
{1\over \cg} A_n\biggr|_{\ln, \Li, \pi^2 \rightarrow 0} \,.
\label{RationalDefinition}
\end{equation}
(The normalization constant $\cg$, defined in \eqn{cgdefn},
plays no essential role in the following arguments, we carry it
along for completeness.)
The `pure-cut' terms are the remaining terms, all of which must
contain logarithms, polylogarithms, or $\pi^2$ terms,
\begin{equation}
\PureCut_n(z) \equiv {1\over  \cg} A_n\Bigr|_{\rm pure-cut} =
{1\over  \cg} A_n\biggr|_{\ln, \Li, \pi^2} \,.
\label{PureCutDefinition}
\end{equation}
In other words,
\begin{equation}
A_n(z) =  \cg \Bigl[ \PureCut_n(z) + \Vertex_n(z) \Bigr] \,,
\label{ACREq}
\end{equation}
where we have explicitly taken $\cg$ outside of $\PureCut_n(z)$
and $\Vertex_n(z)$.

The second decomposition uses the
`completed-cut' terms, obtained from $\PureCut_n(z)$
by replacing logarithms and polylogarithms by corresponding
functions free of spurious singularities (at least those
that suffer a shift).  We call this completion $\Cuth_n$.
The decomposition defines the remaining
rational pieces $\Remaining_n$,
\begin{equation}
A_n(z) =  \cg \Bigl[ \Cuth_n(z) + \Remaining_n(z) \Bigr] \,.
\label{CompletedCutDecomposition}
\end{equation}
This reorganization has effectively moved some of the rational
terms into the completed-cut terms.
In general, $\PureCut_n(z)$ and $\Cuth_n(z)$ will not vanish as
$z\rightarrow\infty$.  We will assume that the large-$z$
behavior is given by a rational function of the spinor
products, as in the previous examples.
The rational function may include contributions from series expansions
of the logarithms and polylogarithms in $\Cuth_n(z)$.  
Taking this into account, we can define a 
useful decomposition of the finite-pole terms,
\be
\conv_n(z) = \cg\Bigl[\Cuth_n(z) + \Remaining_n(z) 
                      -[\Inf \Cuth_n](z)
                      -[\Inf \Remaining_n](z)\Bigr] \,.
\ee
In the cases relevant to the present paper, $[\Inf\Cuth_n](z)$
is in fact at worst a constant in $z$.
Note that the additional terms will not contain any `dangerous'
spurious singularities, as the act of taking the large-$z$ limit will eliminate
them.  

We also need to define the rational part of the completed-cut
terms, $\widehat{CR}_n(z)$.  We write,
\begin{equation}
\Cuth_n(z) = C_n(z) + \CuthRat_n(z) \,,
\label{CuthCCR}
\end{equation}
where
\begin{equation}
\CuthRat_n(z) \equiv \Cuth_n(z)\Bigr|_{\rm rat} \,.
\label{CuthRatDefinition}
\end{equation}
Combining eqs.~(\ref{ACREq}), (\ref{CompletedCutDecomposition}),
and (\ref{CuthCCR}), we see that the full rational part
is the sum of the rational part of the completed-cut terms,
and the remaining rational pieces,
\begin{equation}
\Vertex_n(z) = \CuthRat_n(z) + \Remaining_n(z)\,.
\label{RCRRhat}
\end{equation}

Now, because we know all the terms containing branch
cuts, we could compute the branch-cut-hugging integral in
\eqn{BasicAmplitudeEquation},
\ba
\int_{B_0}^\infty {dz\over z}\;\Disc_B \conv_n(z)
&=& \int_{B_0}^\infty {dz\over z}\;
    \Disc_B \bigl[\Cuth_n(z) -[\Inf\Cuth_n](z)\bigr]\nonumber\\
&=& \int_{B_0}^\infty {dz\over z}\;
    \Disc_B \Cuth_n(z)\,,
\ea
where the second line follows from the rational nature of $[\Inf\Cuth_n](z)$.
However, there is no need to do the integral
explicitly, because we already know the answer for the integral, plus the
associated residues.  Up to a contribution coming from $\Inf \Cuth_n$,
it is just $\Cuth_n(0)$, part of the final
answer.  That is, applying the same logic to $\Cuth_n(z)-[\Inf \Cuth_n](z)$
as was applied to $\conv_n(z)$ in \eqn{BasicAmplitudeEquation}, we have,
\begin{equation}
 \Cuth_n(0) = \Inf \Cuth_n
- \sum_{{\rm poles}\ \alpha} \Res_{z=z_\alpha} {\Cuth_n(z)\over z}
- \int_{B_0}^\infty {dz\over z}\;\Disc_B \Cuth_n(z) \,,
\label{BasicCuthEquation}
\end{equation}
where by construction $\Inf \Cuth_n$ does not contribute to the sum
over residues either.

Using \eqn{BasicAmplitudeEquation},
the decomposition~(\ref{CompletedCutDecomposition}),
and \eqn{BasicCuthEquation} to evaluate the terms involving
$\Cuth_n(z)$, we can write our desired answer as follows,
\begin{eqnarray}
\conv_n(0) &=&
- \cg \Biggl[ \int_{B_0}^\infty {dz\over z}\;\Disc_B \Cuth_n(z)
       +\sum_{{\rm poles}\,\alpha} \Res_{z=z_\alpha} {\Cuth_n(z)\over z}
       +\sum_{{\rm poles}\,\alpha} \Res_{z=z_\alpha}
                 {\Remaining_n(z)\over z} \Biggr]
\nonumber\\
&=&  \cg \Biggl[ \Cuth_n(0)- \Inf \Cuth_n
       -\sum_{{\rm poles}\,\alpha} \Res_{z=z_\alpha}
                 {\Remaining_n(z)\over z}  \Biggr]
\,.
\label{FormII}
\end{eqnarray}
By construction, the completed-cut terms $\Cuth_n(z)$ contain
no spurious singularities, and so the sums over the poles in \eqn{FormII}
are only over the genuine, `physical' poles in the amplitude.
As we are working with {\it complex\/} momenta, these are the poles that arise
for such momenta, and not merely those that arise for {\it real\/}
momenta.  In particular, this means that double poles and unreal poles
may appear, as discussed in detail in refs.~\cite{OnShellRecurrenceI,Qpap}.

\subsection{Residues of the Remaining Rational Pieces $\Remaining_n(z)$}
\label{RemainingRatSubsection}

Our next task is to evaluate the residues of the remaining rational
terms, $\Remaining_n$.  We will do this by setting up an on-shell
recursion relation.  As discussed in ref.~\cite{Bootstrap}, pure-cut
and rational terms factorize independently, and so one can use factorization
to construct an on-shell recursion relation for the
full rational terms $\Vertex_n$.
However, there is no fundamental factorization distinction between the
rational terms in $\Cuth_n$ and those in $\Remaining_n$, so we cannot
set up a direct recursion relation for them.  We will instead compute
them indirectly, by first computing the full rational terms, and then
subtracting terms which are present in both the full rational terms and
in the completed-cut terms.  These are exactly terms coming from
$\CuthRat_n$, which we therefore call `overlap' terms,
\begin{equation}
-\sum_{{\rm poles}\,\alpha} \Res_{z=z_\alpha}
                 {\Remaining_n(z)\over z} =
-\sum_{{\rm poles}\,\alpha} \Res_{z=z_\alpha}
                 {\Vertex_n(z)\over z}
+\sum_{{\rm poles}\,\alpha} \Res_{z=z_\alpha}
                 {\CuthRat_n(z)\over z} \,.
\label{RemainingExpression}
\end{equation}
Since we know $\CuthRat_n$ explicitly, we can compute the last sum
by shifting and extracting poles,
\begin{equation}
\Overlap_n \equiv \sum_{{\rm poles}\,\alpha} \Res_{z=z_\alpha}
                 {\CuthRat_n(z)\over z} \,.
\label{OverlapDefn}
\end{equation}

To obtain the first term on the right-hand side of \eqn{RemainingExpression},
we must analyze the poles in $\Vertex_n(z)$, that is to say its
properties at appropriate null complex momenta.  The behavior of
$\Vertex_n(z)$ can be extracted from the factorization properties
of the amplitude as a whole, by following the analysis of
ref.~\cite{Bootstrap}, and separating
two classes of terms --- pure-cut and rational ---
in the factorized amplitudes.  Only rational terms in the factorization
can contribute to the required sum of residues.

Given the $\Shift{j}{l}$ shift~(\ref{SpinorShift}), we
define a partition $P$ to be a set of two or more cyclicly-consecutive
momentum labels containing $j$, such that the complementary set $\Pb$
consists of two or more cyclicly-consecutive labels containing $l$:
\begin{eqnarray}
 P &\equiv& \{ P_1, P_2, \ldots, j, \ldots, P_{-1} \} \,,
\label{PartitionDef} \\
 \Pb &\equiv& \{ \Pb_1, \Pb_2,
  \ldots, l, \ldots, \Pb_{-1} \} \,, \nn \\
 P &\cup& \Pb = \{ 1,2,\ldots,n \} \,. \nn
\end{eqnarray}
This definition ensures that the sum of momenta in each partition is
$z$-dependent, so that it can go on shell for a suitable value of $z$.
The complex on-shell momenta $\hat k_j$,
$\hat k_l$ and $\Ph$ are determined by solving the on-shell condition,
$\Ph^2 = 0$, for $z$.

At one loop, there are in general three contributions to factorization
in any given channel,
\begin{equation}
A^\oneloop\Bigl|_{s_{i\ldots j}} =
A_L^\oneloop \times {i\over s_{i\ldots j}} \times A_R^\tree
+A_L^\tree \times {i\over s_{i\ldots j}} \times A_R^\oneloop
+A_L^\tree \times {i\,\Fact^\oneloop\over s_{i\ldots j}} \times A_R^\tree
\,.
\end{equation}
In the first two terms, one of the factorized amplitudes is a one-loop
amplitude and the other is a tree amplitude.  The last term
contributes only in multi-particle channels, and contains
a one-loop `factorization function'.  For the case
of a scalar in the loop ($\NeqZero$), this function
is equal to the scalar contribution to the gluon vacuum
polarization~\cite{BernChalmers}.  Accordingly, in addition to the sum over
channels, we will have a sum over these different factorization
contributions.  Taking the rational parts, we obtain,
\def\indentA{\hskip 7mm}
\begin{eqnarray}
 - \sum_{{\rm poles}\ \alpha} \Res_{z=z_\alpha} {\Vertex_n(z)\over z}
&\equiv&
\DiagrammaticRational_n(k_1,\ldots,k_n) \nonumber\\
 &=& \hskip -.1cm
 \sum_{{\rm partitions}\, P}\, \sum_{h = \pm} \Biggl\{
\Vertex(k_{P_1},\ldots,\hat k_j,\ldots,k_{P_{-1}},-\Ph^h)  \nonumber\\
&& \null \hskip 2.5 cm \indentA\times
{i\over P^2} \times
A^\tree(k_{\Pb_1},\ldots,\hat k_l,\ldots,k_{\Pb_{-1}},\Ph^{-h}) \nn\\
&& \null \hskip 2.5 cm
+ A^\tree(k_{P_1},\ldots,\hat k_j,\ldots,k_{P_{-1}},-\Ph^h)
      \label{RationalRecursion} \\
&& \null \hskip 2.5 cm \indentA \times
{i\over P^2} \times
\Vertex(k_{\Pb_1},\ldots,\hat k_l,\ldots,k_{\Pb_{-1}},\Ph^{-h}) \nn\\
&& \null \hskip 2.5 cm
+ A^\tree(k_{P_1},\ldots,\hat k_j,\ldots,k_{P_{-1}},-\Ph^h)  \nn\\
&& \null \hskip 2.5 cm \indentA\times
{i R_\Fact(P)\over P^2} \times
A^\tree(k_{\Pb_1},\ldots,\hat k_l,\ldots,k_{\Pb_{-1}},\Ph^{-h})
\Biggr\}
 \,,  \nn
\end{eqnarray}
where $P^2$ is the squared momentum associated with the partition $P$,
and $R_\Fact$ is defined in the appendix.
The hatted variables are given as usual by shifting momenta according
to \eqn{SpinorShift}, and freezing $z$ to the value that puts $\hat{P}$
on shell,
\be
z = { P^2 \over \langle j^- | P | l^-\rangle } \,.
\ee
The result~(\ref{RationalRecursion}) follows directly from the 
general factorization behavior
of one-loop amplitudes, plus the separate factorization of pure-cut
and rational terms~\cite{Bootstrap}.
Although the $R$ functions are not complete amplitudes, they can be
thought of as vertices. \Eqn{RationalRecursion} then gives rise
to a set of `recursive diagrams'.

Inserting the result into \eqn{FormII}, and thence into
\eqn{InfinitySplit},
gives us the basic on-shell recursion relation for complete one-loop
amplitudes,
\begin{equation}
A_n(0) =  \InfPart{\null} {A_n} +\cg \biggl[\Cuth_n(0)-
  \InfPart{\null}{\Cuth_n}
+ \DiagrammaticRational_n
 +\Overlap_n \biggr]\,.
\label{BasicBootstrapEquation}
\end{equation}
To compute with this equation, we
construct $\DiagrammaticRational_n$ via recursive diagrams; that is,
via \eqn{RationalRecursion}.
The `overlap' terms $\Overlap_n$ can also be given a
diagrammatic interpretation, associating each pole in \eqn{OverlapDefn}
with a specific diagram, as we have done in
\figs{Overlap3mFivePt34Figure}{Overlap3mSixPt12Figure}.
Although the definition of the completed-cut terms $\Cuth_n$ is not
unique, the ambiguity cancels between $\Cuth_n(0)$,
$\InfPart{\null}{\Cuth_n}$, and the sum over $\CuthRat_n$ residues
in $O_n$.

In general, it is useful to combine all the rational functions
not included in the cut completion into a single function,
\be
\Remaining_{n} \equiv  {1\over \cg} \, \InfPart{\null} {A_n} -
  \InfPart{\null}{\Cuth_n} + \DiagrammaticRational_n
 +\Overlap_n \,.
\label{Remaining}
\ee
so that
\be
A_n(0) =  \cg \biggl[\Cuth_n(0) + \Remaining_{n} \biggl] \,.
\label{BasicBootstrapEquationAlt}
\ee
If $\Cuth_n(0)$ is chosen to preserve a symmetry of the amplitude
({\it e.g.}, under a particular permutation of legs), then
$\Remaining_{n}$ will also have this symmetry, even if the individual
components of $\Remaining_{n}$ do not.

\subsection{Determining Terms Arising from Large Shifts}
\label{InfinityPoleSubsection}

Our remaining task is to determine the large-$z$ contributions set
aside in \eqn{InfinitySplit}, and put back unevaluated in
\eqn{BasicBootstrapEquation}.  As already discussed in
sections~\ref{FivePointSection} and~\ref{ThreeMinusSixPtSection}, to
do so we will use a second, auxiliary shift,
\begin{equation}
\Shift{a}{b}:\hskip 2 cm
\tlambda_a \rightarrow \tlambda_a - w\tlambda_b \,,
\hskip 2 cm
\lambda_b \rightarrow \lambda_b + w\lambda_a \,,
\label{AuxiliarySpinorShift}
\end{equation}
distinct from the primary shift in \eqn{SpinorShift}.
It is useful to choose the auxiliary shift so that $A_n(w)$ vanishes
in the large $w$ limit.  (For the gluon amplitudes we consider in this
paper, and likely in general, such choices can be found, as we discuss
in \sect{GeneralHelicitySection}.)  The price that we must pay is the
presence of contributions to the amplitude from channels with
non-standard complex factorizations.  However, we can arrange matters
so that these channels do not contribute to the terms we are seeking
to compute, those that survive in the limit when the original shift's
parameter $z$ becomes large.

To do that, it suffices to ensure that channels with non-standard
complex factorizations vanish at large $z$ when the primary
shift~(\ref{SpinorShift}) is applied to the auxiliary recursion
relation.  The details of the factorization behavior in those channels
are then unimportant.

\begin{figure}[t]
\centerline{\epsfxsize 3.5 truein\epsfbox{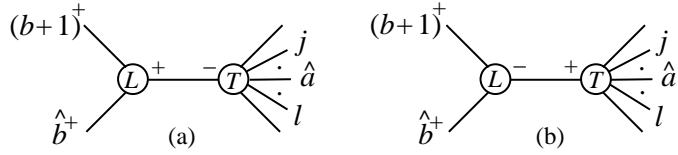}}
\caption{A non-standard factorization using an auxiliary $\Shift{a}{b}$ shift
is assumed to be suppressed in the large-$z$ limit of the
primary $\Shift{j}{l}$ shift, if the tree amplitude naively appearing in the
factorization contains both legs $j$ and $l$ and is suppressed.}
\label{SuppressWeirdFigure}
\end{figure}

For multi-particle channels, factorization in complex
momenta is the same as in real momenta.  For two-particle channels
with opposite-helicity gluons, the discussion in
\sect{ChoiceShiftSubSection} showed that only the standard
tree-level tensor structure contributes, and that for scalars in the
loop the relevant form factor actually vanishes.  For like-helicity
gluons, in the complex kinematics for which the tree vertex is
non-vanishing, the situation is more complicated.  The structure of
the factorization is known empirically for one of the two possible
helicities of the intermediate gluon, for
the case where the other possible intermediate helicity vanishes.
This case occurs in studying the finite series of $n$-gluon amplitudes
$A_{n}^\oneloop(1^-,2^+,\ldots,n^+)$.  The factorization has the
form~\cite{OnShellRecurrenceI},
\begin{eqnarray}
&&A_n^\oneloop(1,\ldots,b^+,(b+1)^+,\ldots,n) \rightarrow\nn\\
&&\hskip 10mm{i\over (s_{b(b+1)})^2} V_3^\oneloop(-K^+,b^+,(b+1)^+)
   A_{n-1}^\tree(1,\ldots,K^-,\ldots,n)  \label{WeirdContribution}  \\
&&\hskip 20mm \null \times (1 \ +\ \mbox{\rm correction factors}) \,.\nn
\nn\end{eqnarray}
This configuration is shown schematically in
\fig{SuppressWeirdFigure}(a).   So
long as the legs shifted under the original, primary, shift are both
contained on the $A^\tree$ side of the factorization, the form of the
correction factors in \eqn{WeirdContribution} is such that the term
vanishes as the primary shift variable $z\rightarrow\infty$, because
the tree amplitude vanishes in that limit.  We will defer a general
study of the structure for the other intermediate helicity shown in
\fig{SuppressWeirdFigure}(b); the only
property that we will need is the vanishing of the primary shift
$z\rightarrow\infty$ limit for contributions where both shifted legs
are on the tree side of the factorization, as in
\fig{SuppressWeirdFigure}.  This property holds for
previously-computed five- and six-point
amplitudes~\cite{GGGGG,Bootstrap}, and we will assume it holds more
generally for both intermediate helicity configurations.  The
validity of this assumption can be tested at the end of a calculation,
by checking all the symmetry and factorization properties of the
fully assembled amplitude.

The auxiliary shift gives us the following expression for the
complete amplitude,
\begin{equation}
A_n(0) = \cg \Bigl[ \Cuth_n(0) - \InfPart{\Shift{a}{b}}{\Cuth_n}
                    +\DiagrammaticRationalS{\Shift{a}{b}}_n
                    +\Overlap_n^{\Shift{a}{b}}\Bigr]\,,
\label{BasicEquationAuxFinal}
\end{equation}
where the superscript $\Shift{a}{b}$ denotes the legs shifted under
the auxiliary shift, and where the recursive diagrams are built
and the overlap contributions determined, with respect to this shift.
 The large-parameter (large-$w$) terms of the auxiliary
shift~(\ref{AuxiliarySpinorShift}) are absent by design.  We
can now extract the large-parameter (large-$z$)
behavior with respect to the primary
$\Shift{j}{l}$ shift,
\begin{equation}
\InfPart{\Shift{j}{l}}{A_n}
=\cg \InfPart{\Shift{j}{l}}{\Bigl[ \Cuth_n(0) - \InfPart{\Shift{a}{b}}{\Cuth_n}
           + \DiagrammaticRationalS{\Shift{a}{b}}_n
           + \Overlap_n^{\Shift{a}{b}}\Bigr]} \,.
\label{BasicEquationAuxInf}
\end{equation}
Following the discussion above, we arrange the shifts so
only channels with standard factorizations
will survive in the large-$z$ limit of the $\Shift{j}{l}$ shift.
In extracting the large-$z$ behavior, we must in general keep {\it all\/}
non-vanishing contributions to the amplitude, which may arise not only
from the recursive diagrams with respect to the auxiliary shift, but also
from completed-cut or overlap terms.  In many practical cases, however,
the only surviving contributions are from a limited set of recursive
diagrams.  Indeed, the typical surviving term will have the form,
\begin{equation}
\InfPart{\Shift{j}{l}}R(k_{P_1},\ldots,
\hat k_a,\ldots,k_{P_{-1}},-\Ph^h)
\times
{i\over P^2} \times
A^\tree(k_{\Pb_1},\ldots,\hat k_b,\ldots,k_{\Pb_{-1}},\Ph^{-h})\,,
\end{equation}
where legs $j$ and $l$ are both on the loop side.

In the next section, we shall present shift choices for
general helicity configurations that implement the approach outlined
in this section: a primary shift free of non-standard channels, but
having non-trivial large shift-parameter behavior, and an auxiliary shift
free of non-trivial large shift-parameter behavior, but containing non-standard
channels that in turn vanish at large values of the primary parameter.


\section{General Helicities}
\label{GeneralHelicitySection}

As an illustration of our strategy, in this section we now present
specific shift choices for determining the rational-function parts of
generic one-loop $n$-gluon amplitudes.  As discussed in previous sections we
must choose a primary shift so that non-standard complex singularities do
not occur in the recursion.  If there are contributions from large
values of the shift parameter $z$, we determine these using
an auxiliary shift and recursion relation.

\subsection{Empirical Structure of the Amplitudes}
\label{EmpiricalStructureSubsection}

\begin{figure}[t]
\centerline{\epsfxsize 6. truein\epsfbox{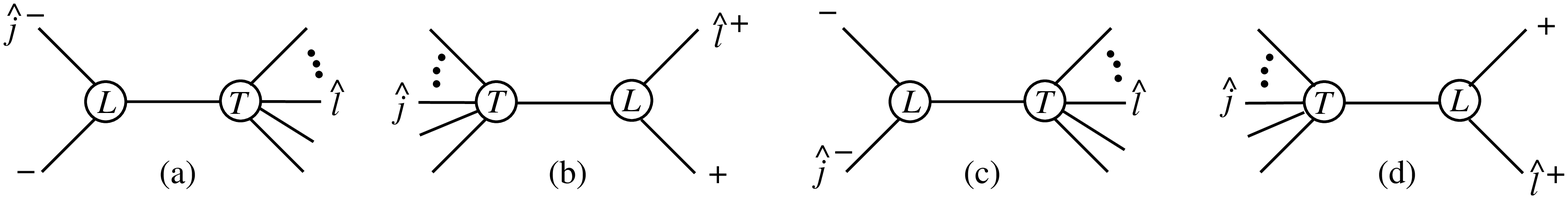}}
\caption{
For a $\Shift{j}{l}$ shift, the four potential channels with
non-standard complex singularities.
Whether these are actually present in a given amplitude
depends on the helicities of the legs 
nearest to $j$ and $l$ in the color ordering.}
\label{WeirdFigure}
\end{figure}

In order to proceed, we will need a few analytic properties of
amplitudes.  Unfortunately, as yet there are no theorems to guide
us on the properties of complex factorizations of amplitudes or on
the behavior of loop amplitudes under large complex shifts $z$. We
therefore follow the empirical approach of
refs.~\cite{OnShellRecurrenceI,Qpap,Bootstrap}. We observe certain
useful properties of known amplitudes and then use these
properties to aid in the computation of new amplitudes.  
We shall not prove these properties, noting that such proofs
would be valuable to help guide future developments.  This
empirical approach has been effective for obtaining a variety of new
one-loop amplitudes~\cite{OnShellRecurrenceI, Qpap,
Bootstrap,FordeKosower}.  By now, a large number of
QCD amplitudes are known 
analytically~\cite{BKStringBased, KunsztEtAl, GGGGG, QQQQG,
TwoQuarkThreeGluon, AllPlus,Mahlon, OnShellRecurrenceI, Qpap,
FordeKosower, MHVQCDLoop}, making it straightforward to develop a
heuristic understanding of their analytic properties.

Our confidence in this pragmatic approach stems from the rather non-trivial
checks that may be performed on any loop amplitude.
In particular, the factorization properties of one-loop
amplitudes in real momenta are well understood~\cite{Neq4Oneloop,
Neq1Oneloop, BernChalmers, BDSSplit, OneLoopSplitUnitarity} and
provide rather non-trivial constraints on the amplitudes by demanding
that {\it every} pole in the amplitude corresponds to a physical
factorization.  The non-trivial consistency requirement stems
from the fact that only a limited number of factorization channels
enter into the recursive construction.

An investigation of the analytic properties of the known one-loop amplitudes
reveals some striking properties:
\begin{enumerate}
\item For {\it any} $\Shift{-}{+}$ shift, {\it all} $n$-gluon amplitudes
vanish for a large shift parameter, $z\rightarrow \infty$.

\item For $n$-gluon amplitudes with the split helicity configuration,
$A_n(1^-, 2^-, \ldots, (m-1)^-, m^+, (m+1)^+, \ldots, n^+)$, 
an alternative set of
shifts where the amplitudes vanish at large $z$ are $\Shift1{m-1}$,
$\Shift{m-1}{1}$, $\Shift{m}{n}$ and $\Shift{n}{m}$.

\item For a given $\Shift{j}{l}$ shift, there are no more than four
channels with non-standard complex singularities, depending on the
helicities of the legs nearest to $j$ and $l$, 
as depicted in \fig{WeirdFigure}.

\end{enumerate}

The above properties are by no means exhaustive. In particular,
there are other shifts where one-loop $n$-gluon amplitudes vanish at
large $z$, although the above observations will be sufficient for
our purposes here.

As already discussed in \sect{InfinityPoleSubsection}, an empirical
rule for suppressing diagrams with non-standard complex singularities
in an auxiliary recursion relation is to ensure that the primary shift
legs are both on the tree side of the naive factorization and that
this tree amplitude is suppressed in the large-$z$ limit of the
primary shift.  Such configurations are displayed in \fig{SuppressWeirdFigure}.

It is worth mentioning that based on our empirical studies of MHV
supersymmetric amplitudes~\cite{Neq4Oneloop,Neq1Oneloop,BST}, it
appears that in the supersymmetric case, the complete set of shifts
where $A_n(z)$ vanishes for $z \rightarrow \infty$ is identical to the
set of shifts where this is true at tree level, {\it i.e.} any
$\Shift{-}{+}, \Shift{+}{+}$ and $\Shift{-}{-}$ shift.


\subsection{Systematics for General Helicities}
\label{ProcedureSection}

Using the above empirical observations, we now present a systematic
procedure for finding pairs of shifts which will allow us to compute
the rational terms in {\it any} $n$-gluon amplitude while avoiding
non-standard complex singularities.

Depending on the helicity configuration, we will use the three independent
shift choices:
\begin{itemize}
\item If the amplitude contains four color-neighboring legs having the
helicity structure $i^+, (i+1)^-, (i+2)^+, (i+3)^-$, then choose the
shift $\Shift{i+1}{i+2}$. With this shift, the amplitude vanishes as
$z\rightarrow \infty$ and no non-standard complex singularities appear
in the recursion relation.  Only a single shift,
and hence only a single recursion relation,
is required in this case.

\item If the amplitude has three nearest neighboring legs $i^-,
(i+1)^-,(i+2)^+$, choose $\Shift{i+1}{i}$ as the primary shift.  For
determining the behavior of the amplitudes for large values of the
primary shift parameter, choose an auxiliary shift $\Shift{a}{b}$ such
that $a$ is a negative-helicity leg, $b$ is a positive-helicity leg and
$a \not = i-1, i, i+1$.

\item For the special case of split helicity configurations, $A_n(1^-,
2^-, \ldots, (m-1)^-, m^+, (m+1)^+, \ldots, n^+)$, a rather convenient
choice is a primary $\Shift{1}{2}$ shift and an auxiliary
$\Shift{n}{m}$ shift.
\end{itemize}

The above choices are not the complete set of choices that we need.
However, all the remaining cases are simply related to the above ones
via parity conjugation or a reversal of legs in the color ordering.
For convenience, we also list these shift choices:

\begin{itemize}

\item If four neighboring legs in the color ordering have helicities
$i^-, (i+1)^+, (i+2)^-, (i+3)^+$ then choose the single shift
$\Shift{i+2}{i+1}$.

\item If the amplitude has three nearest-neighboring legs
$i^+,(i+1)^-,(i+2)^-$, choose $\Shift{i+1}{i+2}$ as the primary shift.  As
the auxiliary shift  choose any $\Shift{a}{b}$ such that $a$ is a
negative-helicity leg, $b$ is a positive-helicity leg and $a \not = i+1,
i+2, i+3$.

\item If the amplitude has three nearest-neighboring legs
$i^-,(i+1)^+,(i+2)^+$ choose $\Shift{i+2}{i+1}$ as the primary shift. As the
auxiliary shift choose any $\Shift{a}{b}$ such that $a$ is a
negative-helicity leg, $b$ is a positive-helicity leg and $b \not = i+1,
i+2, i+3$.

\end{itemize}

\begin{table}
\caption{\label{ShiftPairTable} This table lists helicities and pairs
of shifts that may be used to construct any six- or seven-gluon
amplitude with three negative-helicity legs. The primary shift in the
second column generates a recursion relation which does not have
non-standard complex singularities, but the amplitude may not vanish
for a large shift parameter $z$.  Under the auxiliary shift in the third
column, the recursion relation may have non-standard complex
singularities in the channels listed in the fourth column.  However,
these singularities should be suppressed in the large-$z$ limit of the
primary shift.}

\vskip .4 cm

\begin{tabular}{||l||c|c|c||}
\hline
\hline
\hskip 1.5 cm Helicity& \quad Primary shift \quad & \quad Auxiliary shift \quad
                                 & \quad Suppressed channels \quad \\
\hline
\hline
$1^-, 2^-,3^-, 4^+, 5^+$ & $\Shift12$ & $\Shift54$  &  --- \\
\hline
$1^-, 2^-,3^+, 4^-, 5^+$ & $\Shift43$ &  --- &  --- \\
%
\hline
$1^-, 2^-,3^-, 4^+, 5^+, 6^+$ & $\Shift12$ & $\Shift64$ & $(45)$ \\
\hline
$1^-, 2^-,3^+, 4^-, 5^+, 6^+$ & $\Shift43$ & --- &  --- \\
\hline
$1^-, 2^+,3^-, 4^+, 5^-, 6^+$ & $\Shift12$ & --- &  --- \\
%
\hline
$1^-, 2^-,3^-, 4^+, 5^+, 6^+, 7^+$ & $\Shift12$ & $\Shift74$ &
 $(45)$ \\
\hline
$1^-, 2^-,3^+, 4^-, 5^+, 6^+, 7^+$ & $\Shift43$ & --- & --- \\
\hline
$1^-, 2^-,3^+, 4^+, 5^-, 6^+, 7^+$ & $\Shift12$ & $\Shift56$ &
    $(67)$ \\
\hline
$1^-, 2^+,3^-, 4^+, 5^-, 6^+, 7^+$ & $\Shift34$ & --- & --- \\
\hline
\hline
\end{tabular}
\end{table}

With these choices we should then be able to construct the
rational-function contributions of all unknown $n$-gluon amplitudes, once the
cut-containing pieces are known.  (The above choices are not useful 
for constructing amplitudes with identical helicities, but those 
are already known~\cite{AllPlus,Mahlon}.)
 If more than one of the above
choices is satisfied in a given amplitude, one may choose whichever is
the most convenient.  In Table~\ref{ShiftPairTable} we have listed
all the helicity configurations with three negative-helicity legs and
up to seven external gluons, along with choices of primary and 
auxiliary shifts which may be used to construct the amplitudes.  
(In the first row, for $1^-,2^-,3^-,4^+,5^+$,
our choice of auxiliary shift actually has no non-standard
factorization channels, so it could be used by itself to 
fully determine the amplitude. We display this particular
shift choice because it is based on the above rules and generalizes to
the case of more adjacent positive-helicity gluons.)

It is important to note that there are many other valid shift pairs
besides those in the above construction.  For example, although we
can use the rules to determine all amplitudes with two negative
helicities, it turns out that a somewhat more convenient choice is to
choose a $\Shift{i}{j}$ shift, where legs $i$ and $j$ are the two
negative-helicity legs, as we shall discuss in a companion
paper~\cite{MHVQCDLoop}.  In many cases, it is also possible to relax
the conditions we impose on the shifts.  For example, we have been
demanding that under the auxiliary shift the amplitude vanish for
large shift parameter.  In fact, this restriction is not
necessary; we need only demand that any such terms do not contribute to
the large shift terms of the primary shift.  For example, 
the identical-helicity amplitudes $A_{n;1}(1^+,2^+,\ldots,n^+)$
can be determined using a primary $\Shift57$ shift and an 
auxiliary $\Shift13$ shift (for $n\geq7$),
even though the amplitudes do not vanish~\cite{OnShellRecurrenceI} 
under the large-$z$ limit of either shift. In
Table~\ref{OtherShiftsTable} we have collected a variety of examples
of shift pairs which may be used to determine the amplitudes, but are
outside the class of shifts described above for determining general helicity
configurations.

\begin{table}
\caption{\label{OtherShiftsTable} This table lists examples of
valid shift pairs besides those selected by the procedure discussed in
the text.}

\vskip .4 cm

\begin{ruledtabular}
\begin{tabular}{||l||c|c|c||}
\hskip 1.9 cm  Helicity & \quad Primary shift \quad &
                     \quad Auxiliary shift \quad
                           & \quad Suppressed channels \quad \\
\hline
\hline
$1^+, 2^+, \ldots, n^+$ & $\Shift57$ &  $\Shift13$ & --- \\
\hline
$1^-, 2^+, \ldots, j^-,  (j+1)^+,\ldots, n^+$ & $\Shift{1}{j}$ & --- & --- \\
\hline
$1^-, 2^-,3^-, 4^+, \ldots, n^+$ & $\Shift12$ & $\Shift45$
   & $(56)$ \\
\hline
$1^-, \ldots, (m-1)^-, m^+, \ldots, n^+ $ & $\Shift12$ & $\Shift{m}{m+1}$ &
   $(m+1, m+2) $  \\
\end{tabular}
\end{ruledtabular}
\end{table}

We expect that a similar strategy will be effective for amplitudes 
with massless quarks, and for amplitudes with external massive 
vector bosons or Higgs particles.  With suitable modifications
it should be possible to use the on-shell bootstrap to construct 
amplitudes with massive particles in the loops as well.


\section{Recursive Determinations of Large-$z$ Behavior}
\label{ProcedureSampleSection}

Following the procedure of the last section we now determine the
large shift-parameter behavior of some sample amplitudes.  
We focus on amplitudes with three or
four color-adjacent negative-helicity legs, as a non-trivial
illustration of the method.  Because the logarithmic terms in 
these amplitudes have already been calculated~\cite{RecurCoeff}, 
we can obtain the complete amplitudes by computing the rational terms
recursively, as we do in the next section.  
We can confirm indirectly that our approach to determining 
the large-$z$ behavior is valid in these cases, by verifying
that the amplitudes have the proper symmetries, and that they 
factorize correctly.

To determine the large-$z$ behavior of an amplitude, with the help
of an auxiliary shift, one must include in general the large-$z$ behavior
of the completed-cut, overlap and recursive contributions
(with respect to the auxiliary shift).
In subsection C, we shall encounter a five-point example where all 
three types of contributions are nonvanishing at large $z$.
However, in subsections A and B, we shall arrange the shifts,
for the respective cases of three and four color-adjacent
negative helicities, so that the entire large-$z$ behavior comes from the 
recursive diagrams of the auxiliary shift.  In this way it is very simple
to obtain compact expressions for the large-$z$ behavior of these
amplitudes under the $\Shift12$ shift.

\subsection{$A_{n;1}^{\NeqZero}(1^-,2^-, 3^-, 4^+, 5^+, \ldots, n^+)$}
\label{ThreeMinusBoundarySection}

Consider now the $n$-gluon amplitudes with three color-adjacent
negative-helicity legs.  We use the special split-helicity shift 
choice described in the last section:
a $\Shift12$ shift as our primary shift, and an
$\Shift{n}4$ shift as the auxiliary shift.  This shift allows for an
especially simple determination of the large-$z$ behavior in the
$n$-point case.

\begin{figure}[t]
\centerline{\epsfxsize 6 truein\epsfbox{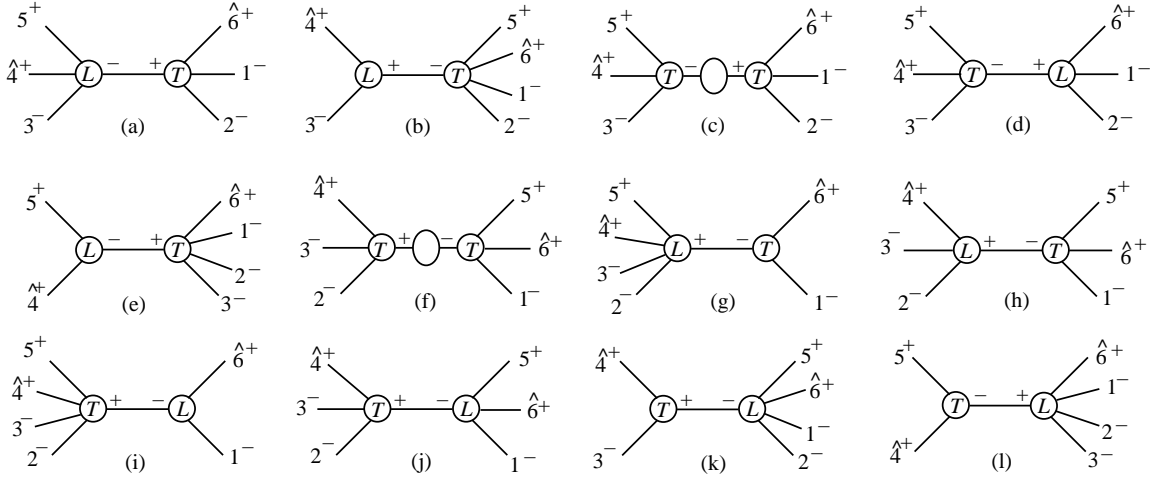}}
\caption{
Recursive diagrams for $A_{6;1}^{\NeqZero}(1^-,2^-, 3^-, 4^+, 5^+, 6^+)$,
for the auxiliary $\Shift{6}{4}$ shift.  We drop the diagrams for which 
the tree vertex vanishes.  Diagram (e) has non-standard
complex singularities.  Only diagrams (k) and (l) contribute to the
large-$z$ limit of the $\Shift12$ shift. Diagrams (c) and (f)
represent the factorization-function contribution.}
\label{Loop3mSixPt64Figure}
\end{figure}

How do the individual contributions behave as the $z$ parameter associated
with a $\Shift12$ shift is taken to infinity?  First consider the completed-cut
contributions, $\Cuth_n$. These contributions were obtained in
ref.~\cite{RecurCoeff} and are given in
\sect{ThreeMinusAmplitudeSection}, \eqn{3minus}, where we describe
them in more detail.  As explained in that section, $\Cuth_n$ is
non-vanishing in the large-$z$ limit of the $\Shift12$ shift
(see \eqn{cuthinfinity}).  However, a computation shows that
the large-$z$ limit of the overlap terms obtained from $\Cuth_n$
cancels this term.  So there are no large-$z$ $\Shift12$ shift 
contributions arising from the sum of the completed-cut
and overlap contributions of our auxiliary $\Shift{n}4$ shift.
All that remains is to inspect the large-$z$ behavior of the recursive
diagrams.

As a warm-up for the all-$n$ case, we consider the six-point case
again, this time with an auxiliary $\Shift64$ shift instead of $\Shift34$.
The recursive diagrams are shown in fig.~\ref{Loop3mSixPt64Figure},
omitting diagrams where the tree vertex vanishes.
The simplest diagrams to analyze are those where both shifted 
legs~$1$ and~$2$ are attached to a tree vertex in a `standard' channel, 
as is true for figs.~\ref{Loop3mSixPt64Figure}(a)--(c).  
The remainder of the diagram (the propagator and other vertex)
is independent of the shift parameter, and so the large-parameter 
behavior is determined by the tree alone.  Because this shift is a 
well-behaved shift at tree level~\cite{BCFW,GloverMassive},
the tree vertex, and hence the diagram, vanishes in the large-$z$ limit.
The same reasoning applies to diagrams where both shifted legs are attached to
a loop vertex, and where we already know from previous computations
that the $\Shift12$ shift is well-behaved.  This is the case for
fig.~\ref{Loop3mSixPt64Figure}(d).  We cannot be certain that
the same logic will apply to diagrams such as
fig.~\ref{Loop3mSixPt64Figure}(e), where both shifted legs are
attached to a tree vertex, when the contribution is in a channel with
non-standard factorization.  Nonetheless, analysis of
specific analogous cases leads us to conclude that these diagrams 
should be suppressed in the large-parameter limit.  
As discussed in the previous section, this
assumption can be tested once the final answer for the amplitude is in
hand.  Diagrams where the shifted legs are attached to different
vertices, such as those in figs.~\ref{Loop3mSixPt64Figure}(f)--(j), 
must be analyzed explicitly.  These diagrams involve only standard
channels, and all do indeed vanish in the large-parameter limit.

This leaves us with two diagrams, those of
figs.~\ref{Loop3mSixPt64Figure}(k) and (l).  Here, the shifted legs are
both attached to the same five-point loop vertex with a standard factorization,
so the large-parameter behavior is determined solely by the loop
vertex.  We know from \eqn{LargeZFunction} that it has non-trivial
behavior in that limit,
which survives to contribute to the auxiliary recursion relation.
Collecting these two contributions, we obtain the desired relation,
\def\indentA{\hskip 20mm}
\ba
&& \InfPart{\Shift12}{A^{\NeqZero}_{6;1} (1^-,2^-, 3^-, 4^+, 5^+,6^+)}
 = \nonumber\\
&& \indentA
\InfPart{\Shift12}{A^{\NeqZero}_{5;1} (1^-,2^-,
        \hat K_{34}^-, 5^+,\hat 6^+)}
   \,  {i \over s_{34}} \, A_3^\tree(3^-, \hat 4^+, -\hat K_{34}^+)
\label{LargeZRecursion6PtB}\\
&& \indentA  \null
+ \InfPart{\Shift12}{A^{\NeqZero}_{5;1} (1^-,2^-, 3^-,
           \hat K_{45}^+, \hat 6^+)}
      \,{i \over s_{45}} \, A_3^\tree(\hat 4^+, 5^+, -\hat K_{45}^-) \, ,
\nonumber
\ea
where $\hat a$ refers to legs shifted and frozen according to the
auxiliary $\Shift64$ shift.  Evaluating this relation (and as we shall
see below, even solving it for all $n$) is straightforward, thanks to
its similarity to a tree-level MHV recursion relation~\cite{BCFW}.  
Plugging in the known values on the right-hand side, we obtain,
\ba
 \InfPart{\Shift12}A^{\NeqZero}_{6;1} (1^-,2^-, 3^-, 4^+, 5^+, 6^+)
&=&
  {i\cg\over 3} {\spash1.{ \hat K_{34}}^3 \spa2.{\hat 6} \over
        \spash1.{\hat 6}^2 \spash{ \hat K_{34}}.5 \spash{5}.{\hat 6} \spb1.2}
      \,  {1 \over s_{34}} \,
            {\spbsh{\hat K_{34}}.{\hat 4}^3  \over \spbsh{3}.{\hat 4}
                                  \spbsh{\hat K_{34}}.3 }  \nonumber \\
&& \null
+    {i\cg\over 3}\, {\spa1.3^3  \spash2.{\hat 6} \over
       \spash1.{\hat 6}^2 \spash3.{\hat K_{45}}
                          \spash{\hat K_{45}}.{\hat 6} \spb1.2 }
      \,{1 \over s_{45}} \,
 {\spbsh5.{\hat 4}^3 \over \spbsh{\hat K_{45}}.5 \spbsh{\hat 4}.{\hat K_{45}}}
      \nonumber \\
& = &
   {i\cg\over 3} { \spa1.3^3\spa2.{6}  \over \spa1.6^2
     \spa3.4 \spa4.5 \spa5.6 \spb1.2 }\,,
\ea
in agreement with \eqn{Boundarymmmppp12}.

It is straightforward to generalize \eqn{LargeZRecursion6PtB} to
obtain the large-$z$ behavior of the $n$-gluon amplitude under a
$\Shift12$ shift.  Following the same logic, we obtain the recursion
relation, using an auxiliary $\Shift{n}4$ shift,
\ba
&&
\InfPart{\Shift12}{A^{\NeqZero}_{n;1}(1^-,2^-, 3^-, 4^+,5^+,\ldots,n^+)}
 =
\nonumber\\
&& \indentA
\InfPart{\Shift12}{A^{\NeqZero}_{n-1;1}
    (1^-,2^-, \hat K_{34}^-, 5^+,\ldots,\hat n^+)}
   \,  {i \over s_{34}} \, A_3^\tree(3^-, \hat 4^+, -\hat K_{34}^+)
\label{LargeZRecursionnPtB}\\
&& \indentA  \null
+ \InfPart{\Shift12}{A^{\NeqZero}_{n-1;1}
              (1^-,2^-, 3^-, \hat K_{45}^+, 6^+, \ldots, \hat n^+)}
      \,{i \over s_{45}} \, A_3^\tree(\hat 4^+, 5^+, -\hat K_{45}^-) \, ,
\nonumber
\ea
where we assumed that diagrams involving non-standard
factorization, and diagrams where legs 1 and 2 appear on different
vertices, are suppressed at large $z$, the same way they were at six points.
Solving this recursion relation by induction, we obtain the large
$z$ behavior at $n$ points,
\be
\InfPart{\Shift12}{A^{\NeqZero}_{n;1} (1^-,2^-, 3^-, 4^+,\ldots,n^+)}
 =
{i\cg\over 3} { \spa1.3^3 \spa2.{n}  \over \spb1.2 \spa{n}.1^2
      \spseq{3}.{n-1}}\, .
\label{Boundary}
\ee
We will use this result in the next section to construct a recursive
solution for the rational functions in the $n$-point amplitude.


\subsection{$A_{n;1}^{\NeqZero}(1^-,2^-, 3^-, 4^-, 5^+, \ldots, n^+)$.}
\label{FourMinusSubsection}

We now consider the case of four color-adjacent negative helicities.
Once again it is useful to illustrate the six-point case, 
$A_{6;1}^{\NeqZero} (1^-,2^-,3^-, 4^-, 5^+, 6^+)$, before
turning to the $n$-point case.  Following the
discussion of the previous section, a convenient pair of shifts is the
$\Shift12$ shift as the primary shift and the $\Shift45$ shift as the
auxiliary shift for determining the large-$z$ behavior of the
amplitude under the primary shift.  (An alternative choice would 
be an auxiliary $\Shift{n}5$ shift.)

As was the case for three color-adjacent negative helicities, an
examination of the cut terms given in ref.~\cite{RecurCoeff} reveals
that in the large-$z$ limit of the $\Shift12$ shift there are no
contributions from the sum of the completed-cut and overlap
contributions.  So here we focus on the non-vanishing rational-recursive
contributions.

\begin{figure}[t]
\centerline{\epsfxsize 5.5 truein \epsfbox{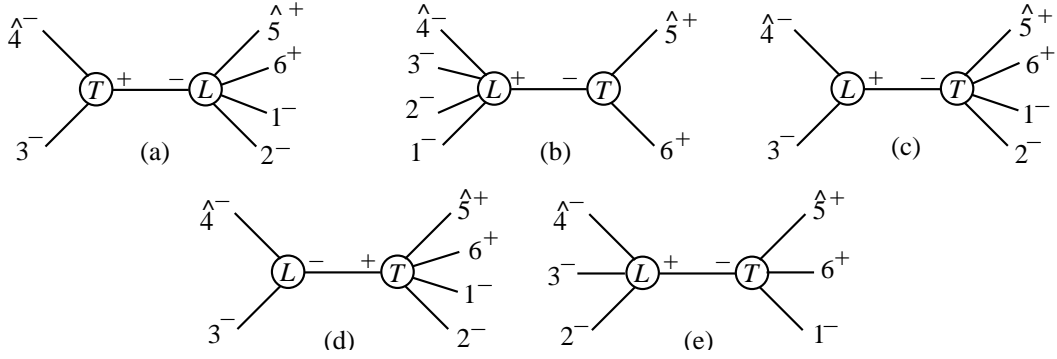}}
\caption[a]{\small Non-vanishing recursive diagrams for
$A_{6;1}^{\NeqZero}(1^-, 2^-, 3^-, 4^-, 5^+, 6^+)$, for
the auxiliary $\Shift{4}{5}$ shift.  We use this recursion 
to determine the large-$z$ behavior under the primary
$\Shift{1}{2}$ shift. Only diagrams (a) and (b) contribute to the
large-$z$ limit.}
\label{Loop4m2p45Figure}
\end{figure}

The auxiliary $\Shift45$ shift diagrams are shown in
\fig{Loop4m2p45Figure} and are given by
\ba
\DiagrammaticRationalS{\Shift{4}{5}}_6(1^-,2^-,3^-,4^-,5^+,6^+)
&=&   D^{\rm (a)} + D^{\rm (b)} +  D^{\rm (c)} + D^{\rm (d)} +  D^{\rm (e)}
\nn \\
&= &
A_3^\tree(3^-, \hat 4^-, -\hat K_{34}^+)
{i \over s_{34}} R_5(1^-,2^-,\hat K_{34}^-,\hat 5^+,6^+) \nonumber  \\
&& \null
+ R_5(1^-,2^-, 3^-, \hat 4^-, \hat K_{56}^+)
{i \over s_{56}} A_3^\tree(\hat 5^+, 6^+, -\hat K_{56}^-) \nonumber  \\
&& \null
+ R_3(3^-, \hat 4^-, -\hat K_{34}^+)
{i \over s_{34}} A_5^\tree(1^-,2^-,\hat K_{34}^-,\hat 5^+,6^+) \nonumber \\
&& \null
+ R_3(3^-, \hat 4^-, -\hat K_{34}^-)
{i \over s_{34}} A_5^\tree(1^-,2^-,\hat K_{34}^+,\hat 5^+,6^+) \nonumber \\
&& \null
+ R_4(2^-, 3^-, \hat 4^-, -\hat K_{234}^+)
{i \over s_{234}} A_4^\tree(1^-,\hat K_{234}^-,\hat 5^+,6^+)
\, , \hskip 1 cm
\ea
where $\hat{a}$ denotes momenta shifted and frozen according to the
auxiliary $\Shift45$ shift.  We apply the $\Shift12$ shift to this
recursion relation and then take the large-$z$ limit of this primary
shift.  We assume that diagrams (c) and (d), which contain
non-standard complex singularities, are suppressed in this limit 
(as discussed in \sect{InfinityPoleSubsection}), 
because the corresponding tree vertices are suppressed.

Diagram (e) is also suppressed, but that requires a closer inspection.
To analyze diagram (e) we need the vertex, obtained by parity 
conjugation from \eqn{A4Neq0amppp},
\be
R_4(1^+, 2^-, 3^-, 4^-) =
{i \over 3} {\spb2.4 \spa2.4^3 \over \spa1.2 \spb2.3 \spb3.4 \spa4.1} \,.
\ee
Then diagram (e) is,
\ba
D^{\rm (e)}  &=&
 R_4(2^-, 3^-, \hat 4^-, -\hat K_{234}^+)
{i \over s_{234}} A_4^\tree(1^-,\hat K_{234}^-,\hat 5^+,6^+) \nonumber \\
& = &
{i \over 3} {\spbsh2.{\hat 4} \spa2.{\hat 4}^3 \over \spash{\hat K_{234}}.2
              \spb2.3 \spbsh3.{\hat 4} \spash{\hat4}.{\hat K_{234}}}
 {1 \over s_{234}} {\spash1.{\hat K_{234}}^3  \over
             \spash{\hat K_{234}}.{\hat 5} \spash{\hat 5}.6 \spa6.1}
 \nonumber \\
& = &
{i \over 3} {\spbsh2.{\hat 4} \spa2.{4}^3 \over \sand2.{\Ksl_{234}}.5
              \spb2.3 \spbsh3.{\hat 4} \sand{4}.{\Ksl_{234}}.5 }
 {1 \over s_{234}} {\sand1.{\Ksl_{234}}.5^3  \over
             \sand{\hat 5}.{\Ksl_{234}}.5 \spash{\hat 5}.6 \spa6.1}
\,. \hskip 1 cm
\ea
We can simplify this term further, by substituting in the
remaining hatted variables, which become independent of $z$ as
$z\to\infty$; but already in this form we can see that it
vanishes in the large-$z$ limit of the $\Shift12$ shift. 
Therefore diagram (e) does not contribute to the recursion relation for
the large-$z$ terms.

Thus we obtain a very simple recursion relation for the
non-vanishing large-$z$ terms, based only on diagrams (a) and (b),
\ba
\InfPart{\Shift12}{A^{\NeqZero}_{6;1}(1^-,2^-,3^-,4^-,5^+,6^+)} &=&
A_3^\tree(3^-, \hat 4^-, -\hat K_{34}^+)
\, {i \over s_{34}}\,
 \InfPart{\Shift12}{ A^{\NeqZero}_{5;1}(1^-,2^-,\hat K_{34}^-,\hat 5^+,6^+)}
 \nonumber  \\
&& \null
+ A_3^\tree(\hat 5^+, 6^+, -\hat K_{56}^-) \,
{i \over s_{56}}\,
\InfPart{\Shift12}
{A^{\NeqZero}_{5;1}(1^-,2^-, 3^-, \hat 4^-, \hat K_{56}^+)}\, .
\nonumber
\\
&& \label{Boundary12Recursion45}
\ea
On the right-hand side there are two large-$z$ contributions.
The first one is familiar from \eqn{FivePoint45Channel},
\be
\InfPart{\Shift12}{A^{\NeqZero}_{5;1}(1^-,2^-,3^-, 4^+,5^+)}  =
i\, {\cg\over 3} {\spa1.3^3 \spa2.5 \over \spb1.2 \spa5.1^2 \spa3.4 \spa4.5}\,.
\label{FirstBoundary}
\ee
To get the second one, we start from formula~(\ref{A5Neq0mpppp}) for
$A^{\NeqZero}_{5;1} (1^-, 2^+, 3^+, 4^+, 5^+)$,
relabel, and take the parity conjugate, to get,
\be
A^{\NeqZero}_{5;1} (1^-, 2^-, 3^-, 4^-, 5^+) =
- i\, {\cg \over 3} \,  {1\over \spb2.3^2}
\Biggl[-{\spa1.4^3 \over \spa5.1 \spa4.5}
       + {\spb5.3^3 \spa3.4 \spb2.4 \over \spb5.1 \spb1.2 \spb3.4^2}
       - {\spb5.2^3 \spa2.1 \spb3.1 \over \spb5.4 \spb4.3 \spb2.1^2}
     \Biggr] 
 \,.
\ee

Applying the $\Shift12$ shift and extracting the $z^0$ terms
gives us the desired contribution, which we wish to feed into
the recursion,
\be
\InfPart{\Shift12}{A^{\NeqZero}_{5;1}(1^-, 2^-, 3^-, 4^-, 5^+)} =
- i \,  {\cg \over 3} \,  {1\over \spb2.3^2}
\Biggl[-{\spa1.4^3 \over \spa5.1 \spa4.5}
    - {\spb5.2^3 \spa2.1 \spb3.1 \over \spb5.4 \spb4.3 \spb2.1^2}
     \Biggr]
\,.
\label{SecondBoundary}
\ee

Relabeling and inserting the large-$z$ terms (\ref{FirstBoundary}) and
(\ref{SecondBoundary}) into the recursion
(\ref{Boundary12Recursion45}) we obtain,
\ba
&& \InfPart{\Shift12}{A^{\NeqZero}_{6;1}(1^-,2^-,3^-,4^-,5^+,6^+)}
 \nonumber \\
&& \null \hskip 1.7 cm
=
A_3^\tree(3^-, \hat 4^-, -\hat K_{34}^+)
\, {i \over s_{34}}\,
\InfPart{\Shift12}{A^{\NeqZero}_{5;1}(1^-,2^-,\hat K_{34}^-,\hat 5^+,6^+)}
 \nonumber  \\
&& \hskip 2.2 cm \null
+ A_3^\tree(\hat 5^+, 6^+, -\hat K_{56}^-) \,
{i \over s_{56}}\,  \InfPart{\Shift12}{A^{\NeqZero}_{5;1}
  (1^-,2^-, 3^-, \hat 4^-, \hat K_{56}^+)} \nonumber
          \\
&& \null \hskip 1.7 cm
 =
i\, {\cg\over 3} {\spash3.{\hat 4}^3 \over \spash{\hat 4}.{\hat K_{34}}
             \spash{\hat K_{34}}.{3}} \, {1 \over s_{34}} \,
    {\spash1.{\hat K_{34}}^3 \spa2.6 \over \spb1.2 \spa6.1^2
              \spash{\hat K_{34}}.{\hat 5} \spa{\hat 5}.6} \nonumber \\
&&\hskip 2.2 cm
\null
 + i \, {\cg \over 3} {\spbsh{\hat 5}.6^3 \over \spbsh{\hat 5}.{\hat K_{56}}
                \spbsh{\hat K_{56}}.{6}} \, {1 \over s_{56}} \,
            {1\over \spb2.3^2}
 \Biggl[-{\spa1.{\hat 4}^3 \over \spash{\hat K_{56}}.1
           \spash{\hat 4}.{\hat K_{56}} }
  - {\spbsh{\hat K_{56}}.2^3 \spa2.1 \spb3.1 \over
         \spbsh{\hat K_{56}}.{\hat 4} \spbsh{\hat 4}.3 \spb2.1^2} \Biggr]
      \nonumber\\
&& \null \hskip 1.7 cm
 =
 i \, {\cg\over 3} \Biggl[
 {\sand1.{(3+4)}.5^3 \spa2.6 \over \spb1.2 \spb3.4 \spb4.5
            \spa6.1^2 s_{345} \sand6.{(4+5)}.3} \label{Result4m2pBoundary} \\
&&\null \hskip 3.0 cm
 + {\spa1.4^3 \over \spb2.3^2 \spa4.5 \spa5.6 \spa6.1 }
 + {\sand4.{(5+6)}.2^3 \spa2.1 \spb1.3 \over \spb1.2^2\spb2.3^2 \spa4.5
                 \spa5.6 s_{456} \sand6.{(4+5)}.3 } \Biggr]
  \,.
\nonumber
\ea
We have checked that this result is in agreement
with the large-$z$ behavior extracted from the parity conjugation of
the six-point amplitude obtained in ref.~\cite{Bootstrap}.

It is straightforward to generalize the discussion to $n$-point amplitudes.
Following the same logic as for six points, we obtain a recursion
relation for the large-$z$ behavior under the $\Shift12$ shift,
\ba
&& \InfPart{\Shift12}{A^{\NeqZero}_{n;1}(1^-,2^-,3^-,4^-,5^+, \ldots,n^+)}
\label{Boundary12RecursionFourMinusNpt}\\
&&\hskip 2 cm  \null
= A_3^\tree(3^-, \hat 4^-, -\hat K_{34}^+)
{i \over s_{34}} \InfPart{\Shift12}{A^{\NeqZero}_{n-1;1}
 (1^-,2^-,\hat K_{34}^-,\hat 5^+,\ldots,n^+)} \nonumber  \\
&& \hskip 2.4 cm \null
+ A_3^\tree(\hat 5^+, 6^+, -\hat K_{56}^-)
\, {i \over s_{56}} \,
\InfPart{\Shift12}{A^{\NeqZero}_{n-1;1}(1^-,2^-, 3^-, \hat 4^-,
                    \hat K_{56}^+, 7^+, \ldots, n^+)} \,.
\nonumber
\ea
We have solved this recursion relation with the result,
\ba
&& \hskip -.3 cm
\InfPart{\Shift12}{A^{\NeqZero}_{n;1}(1^-,2^-,3^-,4^-,5^+, \ldots,n^+)}
\label{FourMinusBoundary} \\
&& \hskip .6 cm  \null =
i\, {\cg\over 3} \Biggl[ 
  {\spa1.4^3 \over \spb2.3^2 \spseq{4}.{n-1}
             \spa{n}.1 }
  + {\sand{4}.{\Ksl_{5\ldots n}}.{2}^3 \spa{2}.1 \spb1.3
                 \over s_{4\ldots n} \sand{n}.{\Ksl_{4\ldots n}}.3 \spb1.2^2
          \spb2.3^2 \spseq{4}.{n-1} } \nonumber\\
&&  \hskip 1.8 cm  \null
 + \sum_{j=5}^{n-1}
   {\spa{j}.{(j+1)}
               \sandmp1.{\Ksl_{3\ldots j} \Ksl_{5\ldots j}}.4^3 \spa2.n \over
             s_{3\ldots j} s_{4 \ldots j} \sand{(j+1)}.{\Ksl_{4 \ldots j}}.3
         \sand{j}.{\Ksl_{4 \ldots j}}.{3} \spb1.2 \spseq{4}.{n-1} \spa{n}.1^2}
 \Biggr]
 \,.
\nonumber
\ea
%

\subsection{Another Look at $A_{5;1}^{\NeqZero}(1^-,2^-,3^-,4^+,5^+)$}

In the two previous subsections, the completed-cut and overlap
contributions did not contribute to the large-$z$ terms of the primary
shift.  However, as we now illustrate, this is not always
true.  We reexamine the large-$z$ limit
of the five-point amplitude $A_{5;1}^{\NeqZero}(1^-,2^-,3^-,4^+,5^+)$
under a $\Shift12$ shift. This example was already considered in
\sect{FivePointSection}. However, here we recursively construct the
large-$z$ behavior by using a different auxiliary shift, a $\Shift45$ shift.
Although there are no non-standard factorization channels here, the
example displays several important features: The large-$z$ recursive
contribution comes entirely from a diagram where legs 1 and 2,
the primary shift legs, are split across the pole of the auxiliary
shift.  Also, as was just mentioned, we need to account for contributions 
from completed-cut and overlap terms under the auxiliary shift.

\begin{figure}[t]
\centerline{\epsfxsize 3 truein\epsfbox{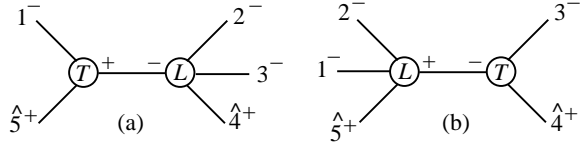}}
\caption{
Nonvanishing diagrams of the $\Shift{4}{5}$ shift recursion relation
for $A_{5;1}^{\NeqZero}(1^-,2^-,3^-,4^+,5^+)$.  Only diagram (a), 
which has legs 1 and 2 on opposite sides of the pole, contributes to 
the large-$z$ limit of the $\Shift12$ shift.}
\label{Loop3mFivePt45Figure}
\end{figure}

The nonvanishing recursive diagrams of the auxiliary $\Shift45$ shift 
are shown in \fig{Loop3mFivePt45Figure}. 
First observe that the behavior of diagram (b) under the $\Shift{1}{2}$ 
shift is determined by the rational part of the four-point one-loop 
MHV amplitude with adjacent negative helicities, which
is proportional to the tree amplitude, according to \eqn{R4mmpp}.
Tree amplitudes vanish at large $z$ for shifts of identical-helicity
pairs of legs~\cite{BCFW,GloverMassive}; hence
diagram (b) does not contribute to the large-$z$ limit.
Therefore, we need only evaluate the contribution of diagram (a).
Here we find a new feature: The shifted legs of the primary
$\Shift12$ shift cross the pole in the auxiliary recursion
relation.  Nevertheless, the large-$z$ contribution can be
determined just as in the previous examples.  This pole-crossing
contribution is given by,
\ba
{\DiagrammaticRationalS{\rm (a)}_5}
 & = &
A_3^{\tree} (1^-,-\hat{K}_{51}^+,\hat{5}^{+})
{i \over s_{51}} R_4(2^-,3^-,\hat{4}^+,\hat{K}_{51}^-)
\nonumber\\
& = &
 {i \over 3}
 { \spbsh{\hat{K}_{51}}.{3} {\spash{\hat{K}_{51}}.{3}}^3
 \over \spb2.3 \spash{3}.{\hat{4}}
 \spash{\hat{4}}.{\hat{K}_{51}} \spbsh{\hat{K}_{51}}.{2}
  }
{1 \over s_{51}} { {\spbsh{\hat{5}}.{\hat{K}_{51}}}^3 \over
{\spbsh{1}.{\hat{K}_{51}}} \spbsh{\hat{5}}.{1} }\, ,
 \label{recurs5}
\ea
where $\hat{a}$ refers to legs shifted according to the auxiliary $\Shift{4}{5}$
shift. We then obtain in the large-$z$ limit of the $\Shift12$ shift
the contribution,
\be
\InfPart{\Shift{1}{2}}{\DiagrammaticRationalS{\Shift{4}{5}}_5
(1^-,2^-,3^-,4^+,5^+)} =
 - {i \over 3} { {\spa1.3}^3  {\spa2.4}
 \over \spb2.3 {\spa3.4}^2 \spa4.5 \spa5.1 }
\, .
\label{fivexptrec}
\ee

\begin{figure}[t]
\centerline{\epsfxsize 4 truein\epsfbox{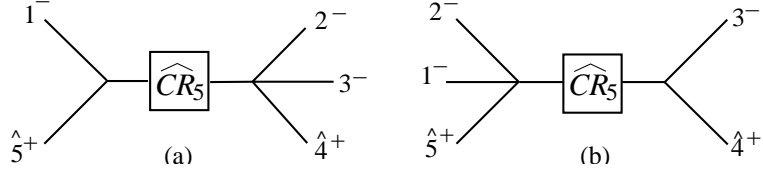}}
\caption{
The overlap diagrams of the $\Shift{4}{5}$ shift.
}
\label{Overlap3mFivePt45Figure}
\end{figure}

Comparing \eqn{fivexptrec} to the known expression in 
\eqn{FivePoint45Channel} we see that the two expressions do not match. 
However, we also have to take into account the contributions to the
large-$z$ limit from the overlap contributions as indicated in
\eqn{BasicBootstrapEquation}.  We apply the $\Shift45$ shift to
$\CuthRat_5$, which can be extracted from
\eqn{Cuth5}. We obtain the overlap contribution from the residues of
$\CuthRat_5(w)/w$ at the following values of $w$, illustrated in
\fig{Overlap3mFivePt45Figure},
\be
w^{\rm (a)} = - {\spa1.5 \over \spa1.4}\,, \hskip 2.5 cm  w^{\rm (b)}
            = {\spb3.4 \over \spb3.5 } \,.
\ee
Evaluating these in the standard way, we obtain the large-$z$ contributions,
\bea
\InfPart{\Shift{1}{2}}{\Overlap_5^{\rm (a),\,\Shift{4}{5}}}
 & = &
- {i \over 3} { {\spa1.3}^2 \spb2.5 \spa2.3 \over
\spb1.2 \spb2.3 {\spa3.4}^2  \spa5.1 }
- {i \over 6} { {\spa1.3}^2 \spb4.5  \over
\spb1.2 \spb2.3 {\spa3.4}  \spa5.1 }
\,, \label{olap1}
\\
\InfPart{\Shift{1}{2}}{\Overlap_5^{\rm (b),\,\Shift{4}{5}}}
& = &
 {i \over 6} { \spa1.3  {\spb4.5}^2 \over {\spb1.2}
\spb2.3 s_{34} }
\, .
\label{olap2}
\eea
Notice that both \eqns{fivexptrec}{olap1} have double poles in
$1/\spa3.4$, which cancel correctly upon adding up all contributions.

Finally, we need to take into account the contribution
from the completed-cut
terms $\Cuth_5$, which are easily found to be,
\be
\InfPart{\Shift12}{\Cuth_5(1^-,2^-,3^-,4^+,5^+)}
=
- {i \over 6} { {\spa1.2} \spb2.4 \spa1.3  \left( \spa1.2
\spb2.4 - \spa1.3 \spb3.4 \right) \over
\spb1.2 \spb2.3  {\spa5.1}^2 s_{34} }\, .
\label{crhat}
\ee

Upon adding all contributions, eqs.~(\ref{fivexptrec}),
(\ref{olap1}), (\ref{olap2}), and (\ref{crhat}), we obtain
\ba
\InfPart{\Shift{1}{2}}{A_{5;1}^{\NeqZero}} &=&
- i \, {\cg } \, \Biggl[ {1 \over 3}
 { {\spa1.3}^3  \spa2.4 \over \spb2.3 \spa3.4^2 \spa4.5 \spa5.1 }
+
\,{1 \over 3} { {\spa1.3}^2 \spb2.5 \spa2.3 \over
\spb1.2 \spb2.3 {\spa3.4}^2  \spa5.1 }
+  {1 \over 6} { {\spa1.3}^2 \spb4.5  \over
\spb1.2 \spb2.3 {\spa3.4}  \spa5.1 } \nn \\
&& \hskip1.2cm
-  {1 \over 6} {\spa1.3  {\spb4.5}^2 \over {\spb1.2} \spb2.3 s_{34} }
+ {1 \over 6} { {\spa1.2} \spb2.4 \spa1.3  \left( \spa1.2
\spb2.4 - \spa1.3 \spb3.4 \right) \over
\spb1.2 \spb2.3  {\spa5.1}^2 s_{34} } \Biggr] \nn \\
&=&
i \, {\cg\over3}{\spa1.3^3\spa2.5\over\spa1.5^2\spa3.4\spa4.5\spb1.2}
\,,
\ea
in agreement with \eqn{FivePoint45Channel}.

This example demonstrates that, in general, if we wish to obtain the
correct large-$z$ behavior using an auxiliary recursion, we must
include the contributions from the completed-cut and overlap terms. 
In practice, however, often only the recursive diagrams contribute.


\section{Complete Split Helicity Amplitudes}
\label{ThreeMinusAmplitudeSection}

In \sect{ThreeMinusBoundarySection} we determined the large-$z$ behavior,
under a $\Shift12$ shift, of the $n$-point amplitudes with three 
nearest-neighboring negative helicities,
$A_{n;1}^{\NeqZero}(1^-,2^-,3^-, 4^+, \ldots,n^+)$.
We now use this result to evaluate the remaining 
rational terms, using the formalism described in
section~\ref{DerivationSection}.  The $n$-point tree amplitudes for these
helicity configurations were determined in
refs.~\cite{DAKRecurrence,CSW,TreeRecurResults,RoibanSplitHelicity},
and are given in \eqn{adjNMHVtree}.

\subsection{Cut Contributions}

The completed-cut contributions for general split-helicity $\NeqZero$
amplitudes have been obtained in ref.~\cite{RecurCoeff}.  
For the case with three negative helicities, the cut part of the amplitude
$A_{n;1}^{\NeqZero}(1^-,2^-,3^-, 4^+, \ldots,n^+)$ is,
\ba
\Cuth_{n}(1^-,2^-,3^-,4^+,\ldots, n^+) &=&
\frac{1}{3 \cg}\,A_{n;1}^{\NeqOne}(1^-,2^-,3^-,4^+,\ldots, n^+)
\nonumber \\
&& \null
 + \frac{2}{9} \, A_n^{\tree}
(1^-,2^-,3^-,4^+,\ldots, n^+)
\nonumber \\
&& \null
 -{i \over 3}
\sum_{r=4}^{n-1}\,
{ \hat d_{n,r}^{\NeqZero}}\,
{   \Lzz ( {-s_{3\ldots r} \over -s_{2 \ldots r}} )\over s_{2\ldots r}^3}
-{i \over 3}
\sum_{r=4}^{n-2}\,
   {\hat g_{n,r}^{\NeqZero}}\,
{   \Lzz ( {-s_{2\ldots r} \over -s_{2\ldots (r+1)}} )\over s_{2\ldots (r+1)}^3}
\nonumber \\
&& \null
- {i \over 3}
\sum_{r=4}^{n-2}\,
   {\hat h_{n,r}^{\NeqZero}}
{  \Lzz ( {-s_{3\ldots r} \over -s_{3\ldots (r+1)}} )\over s_{3\ldots (r+1)}^3}
\,, \hskip 1 cm \label{3minus}
\ea
where
\ba
\hat d_{n,r}^{\NeqZero}  & = &
{ \sandmp3.{{\Ksl}_{3 \ldots r} {\ksl}_{2}}.1 \,
  \sandmp3.{\ksl_{2}\Ksl_{2 \ldots r}}.1 \,
  \sandmp3.{\Ksl_{3\ldots r} \big[\ksl_{2},\Ksl_{2\ldots r}\big]\Ksl_{2\ldots r}}.1
\over
  \sandpp2.{{\Ksl}_{2 \ldots r}}.{r}
  \sandpp2.{{\Ksl}_{2 \ldots r}}.{(r+1)} \spseq{3}.{r-1} \spseq{r+1}.{n} }
 \,, \hskip 1 cm \\
\hat g_{n,r}^{\NeqZero} & = &
\sum_{j=1}^{r-3}
 {\sandmp3.{\Ksl_{3\ldots (j+3)} \Ksl_{2\ldots (j+3)} \big[\ksl_{r+1},\Ksl_{2\ldots r}\big]}.1
            \spa{(j+3)}.{(j+4)}
 \over
  \sandpp2.{\Ksl_{2\ldots (j+3)}}.{(j+3)}
  \sandpp2.{\Ksl_{2\ldots (j+3)}}.{(j+4)} \, \spseq{3}.{n}} 
       \hskip .5 cm  \nonumber \\
&& \hskip 2 cm \null \times
{ \sandmp3.{\Ksl_{3\ldots (j+3)} \Ksl_{2 \ldots (j+3)} \ksl_{r+1} \Ksl_{(r+1) \ldots 1}}.1 
  \over \, s_{3 \ldots (j+3)}
                \,s_{2\ldots (j+3)} }
 \nonumber \\
&& \hskip 2 cm \null \times
   \sandmp3.{\Ksl_{3\ldots (j+3)} \Ksl_{2\ldots (j+3)} \Ksl_{(r+1) \ldots 1}\ksl_{r+1}}.1
\,, \label{gnrscalar} \\
\hat h_{n,r}^{\NeqZero} & = &
(-1)^n \hat g_{n,n-r+2}^{\NeqZero}\Bigl\vert_{(123\ldots n)\to(321n\ldots4)}\,.
\ea
Here $A_{n;1}^{\NeqOne}$ is the contribution of an $\NeqOne$ supersymmetric
chiral multiplet consisting of a scalar and a fermion running in the
loop.  The result for this amplitude is~\cite{BBDPSQCD,RecurCoeff},
\ba
&& \hskip -.5 cm
A_{n;1}^{\,\NeqOne}(1^-,2^-,3^-,4^+, \ldots, n^+)  \nn \\
&& \null \hskip 2 cm  =
\cg \, \frac{A_n^\tree}{ 2} \,\left( \Kz( s_{n1} ) +\Kz( s_{34} )
\right)
-{i \over 2}\, \cg
\sum_{r=4}^{n-1}\,
{\hat d_{n,r}^{\NeqOne}}\,
{ \Lz ( {-s_{3\ldots r} \over -s_{2 \ldots r}} )\over s_{2 \ldots r}}
\nonumber \\
& &  \hskip 2.4 cm \null
-{i \over 2}  \cg
\sum_{r=4}^{n-2}\,
{\hat g_{n,r}^{\NeqOne}}\,
{ \Lz ( {-s_{2\ldots r} \over -s_{2\ldots (r+1)}} )\over s_{2 \ldots (r+1)}}
-{i \over 2}\, \cg
\sum_{r=4}^{n-2}\,
{\hat h_{n,r}^{\NeqOne}}\,
{ \Lz ( {-s_{3 \ldots r} \over -s_{3 \ldots (r+1)}} )\over s_{3 \ldots (r+1)}}
\,,
\hskip 1. cm
\ea
where,
\ba
\hat d_{n,r}^{\NeqOne} & = &
{\sandmp3.{{\Ksl}_{3\ldots r} {\Ksl}_{2 \ldots r}}.1^2\,
 \sandmp3.{\Ksl_{3\ldots r} \big[\ksl_{2},\Ksl_{2\ldots r}\big]\Ksl_{2\ldots r}}.1
\over
  \sandpp2.{{\Ksl}_{2\ldots r}}.r
   \sandpp2.{{\Ksl}_{2\ldots r}}.{(r+1)} \,
   {s}_{2\ldots r}\,s_{3 \ldots r}  \spseq{3}.{r-1} \spseq{r+1}.{n} }
\,,    \hskip .8 cm       \\
\hat g_{n,r}^{\NeqOne} & = &
\sum_{j=1}^{r-3}
{ \sandmp3.{\Ksl_{3 \ldots (j+3)} {\Ksl}_{2 \ldots (j+3)}}.1^2
   \sandmp3.{\Ksl_{3 \ldots (j+3)}\Ksl_{2 \ldots (j+3)}\big[\ksl_{r+1},\Ksl_{2\ldots r}\big]}.1 \over
 \sandpp2.{\Ksl_{2\ldots (j+3)}}.{(j+3)}
 \sandpp2.{\Ksl_{2\ldots (j+3)}}.{(j+4)} \,
s_{3\ldots (j+3)}\,s_{2\ldots (j+3)} }
 \nn \\
&& \hskip 2 cm \null
\times {\spa{(j+3)}.{(j+4)} \over  \spseq{3}.{n} } 
\,, \\
\hat h_{n,r}^{\NeqOne} & = &
(-1)^n
\hat g_{n,n-r+2}^{\NeqOne}\Bigl\vert_{(123\ldots n)\to(321n\ldots 4)}
\,.
\label{ds1}
\ea

From \eqn{3minus} we can extract the rational parts of the cut
completion, $\CuthRat_n$. These terms are given by,
\bea
\CuthRat_n & = &
\left( \frac{1}{3 \epsilon} + \frac{8}{9}\right)
A_n^{\tree} (1^-,2^-,3^-,4^+,\ldots, n^+)
\nonumber \\
& - & \frac{i}{6}
\sum_{r=4}^{n-1}
{ \hat{d}_{n,r}^{\NeqZero}}
{ s_{3 \ldots r} +
s_{2 \ldots r}\over s_{2 \ldots r} s_{3 \ldots r}
 (s_{3 \ldots r}- s_{2 \ldots r})^2}
- \frac{i}{6} \sum_{r=4}^{n-2}
{\hat{g}_{n,r}^{\NeqZero}} \hphantom{!}
{{ s_{2 \ldots r} + s_{2 \ldots (r+1)}}\over
  {s_{2 \ldots r} s_{2 \ldots (r+1)} (s_{2 \ldots (r+1)}- s_{2 \ldots r})^2}}
 \nonumber \\
& - &  \frac{i}{6} \sum_{r=4}^{n-2}
{\hat{h}_{n,r}^{\NeqZero}}
{ s_{3 \ldots r} + s_{3 \ldots (r + 1)}\over
 s_{3 \ldots r} s_{3 \ldots (r+1)} (s_{3 \ldots (r+1)}- s_{3 \ldots r})^2}
\, . \label{CRhat}
 \ea
This cut completion has a (constant) boundary contribution as $z
\rightarrow \infty$ under a $\Shift12$ shift, stemming from the
$\hat{d}^{\NeqZero}$-term. It is given by,
\be
\InfPart{\Shift12}{\Cuth_n} =
  \frac{i}{6} \sum_{r=4}^{n-1}
\frac{  \sandmp 3.{\Ksl_{3 \ldots r} \ksl_2}.1 \,
   \spa1.3 \, \left( s_{3\ldots r}  \spa1.3 +
 \sandmp3.{\Ksl_{3 \ldots r} \ksl_2}.1 \right)
 \spa{r}.{(r+1)} }{ s_{3 \ldots r}
 \sandpp2.{\Ksl_{2 \ldots r}}.r \sandpp2.{\Ksl_{2 \ldots r}}.{(r+1)}
    \,\spseq{3}.{n}}
 \,. \label{cuthinfinity}
\ee
%

\subsection{Recursive Contributions}

\begin{figure}[t]
\centerline{\epsfxsize 5 truein\epsfbox{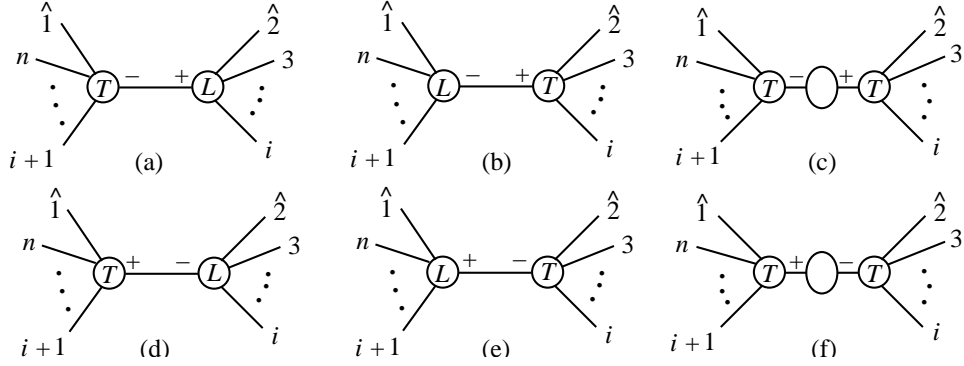}}
\caption{
The recursive diagrams of the $\Shift{1}{2}$ shift for an $n$-point
helicity amplitude.  Some of the diagrams may vanish depending
on the helicity choices.
}
\label{LoopNPt12Figure}
\end{figure}

We now discuss the recursive diagrams of the $\Shift12$ shift,
shown in \fig{LoopNPt12Figure}, for a generic helicity
configuration.  As explained in \sect{DerivationSection}, we must
sum over both helicities of the gluon crossing the pole, and over
loop versus tree vertices appearing on either side of the pole,
as well as factorization-function contributions~\cite{BernChalmers} 
in multi-particle channels.  For specific helicity
configurations some of the diagrams may vanish.

For the case of
$A_{n;1}^{\NeqZero}(1^-,2^-,3^-,4^+,\ldots,n^+)$, the recursive
contributions are,
\def\indentA{\hskip 1.5 cm \null}
\ba
&&\hskip -.3 cm
 \DiagrammaticRational_n(1^-,2^-,3^-,4^+,\ldots,n^+)  = \nonumber \\
&& \hskip 2 cm
\sum_{i = 4}^{n-1}
A^\tree_{n-i+2}(\hat{1}^-,\hat{K}_{2\ldots i}^-,(i+1)^+,\ldots,n^+)
\frac{i}{s_{2\ldots i}}
R_i(\hat{2}^-,3^-,4^+,\ldots,i^+,-\hat{K}_{2 \ldots i}^+)
\nonumber \\
&& \indentA
+ \sum_{i = 4}^{n-2}
R_{n-i+2}(\hat{1}^-,\hat{K}_{2 \ldots i}^-,(i+1)^+,\ldots,n^+)
\frac{i}{s_{2 \ldots i}}
A^\tree_i(\hat{2}^-,3^-,4^+,\ldots,i^+,-\hat{K}_{2 \ldots i}^+)
 \nonumber \\
&& \indentA
+ \sum\limits_{i = 5}^{n-2}
  R_{n-i+2}(\hat{1}^-,\hat{K}_{2 \ldots i}^+,(i+1)^+,\ldots,n^+)
\frac{i}{s_{2 \ldots i}}
A^\tree_i(\hat{2}^-,3^-,4^+,\ldots,i^+,-\hat{K}_{2 \ldots i}^-)
\nonumber
  \hskip .4 cm \\
&& \indentA
 -  \Biggl( \frac{1}{3 \epsilon} + \frac{8}{9}\Biggr)
\sum_{i = 4}^{n-2}
A^\tree_{n-i+2}(\hat{1}^-,\hat{K}_{2 \ldots i}^-,(i+1)^+,\ldots,n^+)
\frac{i}{s_{2 \ldots i}}
\nonumber \\
& & \hskip 4 cm \null
 \times A^\tree_i(\hat{2}^-,3^-,4^+,\ldots,i^+,-\hat{K}_{2 \ldots i}^+)
\,,
\label{rhat}
\ea
where the last line comes from the factorization-function contributions.
We refrain from quoting the explicit expression for $R_n^D$ here.  
It can be obtained straightforwardly by inserting the known tree
amplitudes~\cite{ParkeTaylor,BGRecurrence,TreeRecurResults} 
(quoted in the appendix) and rational parts of loop
amplitudes~\cite{Qpap,FordeKosower} into \eqn{rhat}.

An interesting feature of this recursion relation is that all amplitudes
on the right-hand side have fewer than three negative helicities.  In
contrast, the recursion relation found for two negative-helicity
gluons~\cite{Bootstrap} contains on the right-hand side lower-point loop
amplitudes with the same number of negative helicities as on the left-hand
side, namely two.  For this reason, to solve that recursion relation in
closed form for all $n$ required an `unwinding'
procedure~\cite{FordeKosower}.  Unwinding is not necessary in the present
case because all terms on the right-hand side are known --- given the
two-negative-helicity solution of ref.~\cite{FordeKosower}.


\subsection{Overlap Contribution}

\begin{figure}[t]
\centerline{\epsfxsize 6 truein\epsfbox{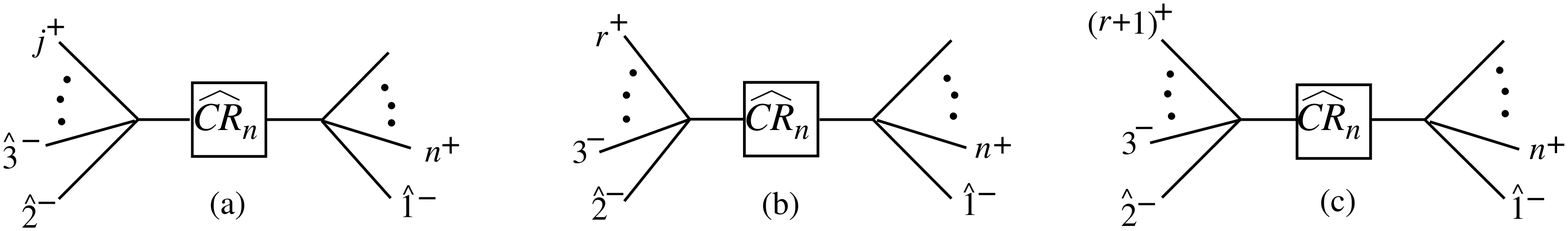}}
\caption{
Overlap diagrams corresponding to the $s_{2 \ldots j}$, $s_{2 \ldots
r}$ and $s_{2 \ldots(r+1)}$ channels.}
\label{Overlap3mNPt12Figure}
\end{figure}

Finally, we need to compute the overlap contribution to avoid double
counting between the recursive diagrams, \eqn{rhat}, and the completed-cut
terms, \eqn{CRhat}.  First we observe that the terms proportional to the
tree amplitude on the first line of \eqn{CRhat}
get shifted only in the denominator, 
namely only in $s_{2 \ldots r}$ in \eqn{adjNMHVtree}.
This means that the overlap contribution from this term
is the negative of the term itself and thus cancels it in the full amplitude,
\eqn{BasicBootstrapEquation}. 
Furthermore, the same happens with
the term involving $\hat{h}^{\NeqZero}$.  Thus, for
these two contributions we have
\ba
O_n^\tree  &=& 
- \left( \frac{1}{3 \e} + \frac{8}{9}\right)
A_n^{\tree} (1^-,2^-,3^-,4^+,\ldots, n^+)
\,, \label{Overlapmmmtree} \\
O_n^{\hat h\mbox{-}{\rm term}} &=& 
  \frac{i}{6} \sum_{r=4}^{n-2}
{\hat{h}_{n,r}^{\NeqZero}}
{ s_{3 \ldots r} + s_{3 \ldots (r + 1)}\over
 s_{3 \ldots r} s_{3 \ldots (r+1)} (s_{3 \ldots (r+1)}- s_{3 \ldots r})^2}
\,. \label{Overlapmmmhterm}
\ea
The $\hat{d}^{\NeqZero}$-term gives a contribution to the overlap,
\be
O_n^{\hat d\mbox{-}{\rm term}} =
- \frac{i}{6}
\sum_{r=4}^{n-1}
\frac{ \sandmp3.{\Ksl_{3 \ldots r} \ksl_2}.1 \,
  \sandmp3.{\Ksl_{3 \ldots r} \Ksl_{2 \ldots r}}.1^2\, \spa{r}.{(r+1)}}{
 \sandpp2.{\Ksl_{2 \ldots r}}.r
 \sandpp2.{\Ksl_{2 \ldots r}}.{(r+1)}\, \spseq{3}.{n} 
     \, s_{2 \ldots r} s_{3 \ldots r}}
\, .
\label{Overlap1}
\ee

In discussing the $\hat{g}^{\NeqZero}$-term, we first
re-index~\eqn{gnrscalar} by letting $j \to j-3$, so the sum
runs from $j=4$ to $j=r$.
The $\hat{g}^{\NeqZero}$-term
has three different kinds of poles, since 
$s_{2 \ldots j}$, $s_{2 \ldots r}$ and $s_{2 \ldots(r+1)}$ are shifted,
\ba
z^{\rm (a)} & = & - \frac{s_{2 \ldots j}}
                           {\sandmm1.{\Ksl_{2 \ldots j}}.2}\,, \nn \\
z^{(\rm b)} & = & - \frac{s_{2 \ldots r}}
                           {\sandmm1.{\Ksl_{2 \ldots r}}.2}\,, \nn \\
z^{\rm (c)} & = & - \frac{s_{2 \ldots(r+1)}}
                           {\sand1.{\Ksl_{2 \ldots(r+1)}}.2}\,,
\ea
corresponding to the overlap diagrams in \fig{Overlap3mNPt12Figure}.
At first glance it may seem that for $j=r$ we obtain a
double pole, since the denominator contains $s_{2\ldots r}^2$. However,
from the numerator of the last factor in $\hat{g}^{\NeqZero}$, we obtain a
factor $s_{2\ldots r}$, which cancels one of the factors in the
denominator, and so we have again only single poles. Nevertheless, we treat
the $j = r$ term separately and obtain for this contribution,
\be
O_n^{{\rm (a),(b)}; j=r}
=
 \frac{i}{6}
\sum_{r=4}^{n-2}
\frac{  \sandmp3.{\Ksl_{3 \ldots r} \ksl_{r+1}}.1 \,
  \sandmp3.{\Ksl_{3 \ldots r} \Ksl_{2 \ldots r}}.1^2\, \spa{r}.{(r+1)}}{
 \sandpp2.{\Ksl_{2 \ldots r}}.r
 \sandpp2.{\Ksl_{2 \ldots r}}.{(r+1)}
\, \spseq{3}.{n} s_{2\ldots r} s_{3\ldots r}}
\label{Overlapmmmtermjeqr} \, .
 \ee
After some straightforward algebra, we find the remaining contributions,
\ba
O_n^{\rm (a)} & = &
\frac{i}{6}
\sum_{r=5}^{n-2} \sum_{j=4}^{r-1}
{ \spa{j}.{(j+1)} \over
 \sandpp2.{\Ksl_{2 \ldots j}}.j
 \sandpp2.{\Ksl_{2 \ldots j}}.{(j+1)} \, \spseq{3}.{n} 
  \, s_{3 \ldots j}\, s_{2\ldots j} }
 \nonumber \\
 & & \null \times
{\sandmp3.{\Ksl_{3 \ldots j} \Ksl_{2 \ldots j}}.1^3
   \sandmm1.{\ksl_{r+1} \Ksl_{(j+1)\ldots r}\Ksl_{2 \ldots j}}.2
    \sandmm1.{\Ksl_{2 \ldots r} \ksl_{r+1}\Ksl_{2 \ldots j}}.2 \over
 { \sandmm1.{(\ksl_{r+1} \Ksl_{(j+1) \ldots r} +\Ksl_{2 \ldots r} \ksl_{r+1}) \Ksl_{2 \ldots j}}.2 }^2 }
   \nonumber \\
 & & \null \times
 \Biggl( \frac{ 1}{\sandmm1.{\Ksl_{2 \ldots j} \Ksl_{(j+1) \ldots r} \Ksl_{2 \ldots r}}.2}
  +  \frac{1}{\sandmm1.{\Ksl_{2 \ldots j} \Ksl_{(j+1) \ldots(r+1)} \Ksl_{2 \ldots (r+1)}}.2}
              \Biggr) \nonumber \\
 & &\null \times
\sandmm1.{( \Ksl_{2 \ldots r} \ksl_{r+1} - \ksl_{r+1} \Ksl_{(j+1) \ldots r}) \Ksl_{2 \ldots j}}.2
\,,
\ea
\ba
O_n^{\rm (b)}  & = &
 - \frac{i}{6}
\sum_{r=5}^{n-2} \sum_{j=4}^{r-1}
\frac{ \spa{j}.{(j+1)}}{
 \sandpp2.{\Ksl_{2 \ldots j}}.j
 \sandpp2.{\Ksl_{2 \ldots j}}.{(j+1)} \,\spseq{3}.{n}  s_{3 \ldots j} \,
                 s_{2\ldots r}}
 \nonumber \\
& & \hskip .1 cm  \null \times
\frac{1}{ \sandmm1.{\Ksl_{2 \ldots r} \Ksl_{(j+1) \ldots r}\, \Ksl_{2 \ldots j}}.2
          \sandmm1.{\Ksl_{2 \ldots r} \ksl_{r+1} \,\Ksl_{2 \ldots r}}.2^2 }
 \nonumber \\
& & \hskip .1 cm \null\times
 \Bigg(
  \sandmp3.{\Ksl_{3 \ldots r} \Ksl_{2 \ldots r}}.1
  \sandmm1.{\Ksl_{2 \ldots r}  \ksl_{r+1} \Ksl_{2 \ldots j}}.2 
\nonumber   \\
& & \hskip 1cm \null
- \sandmp3.{\ksl_{r+1} \Ksl_{2 \ldots r}}.1
  \sandmm1.{\Ksl_{2 \ldots r}  \Ksl_{(j+1) \ldots r} \Ksl_{2 \ldots j}}.2
\Bigg) \nn \\
& & \hskip .1 cm \null \times
\Bigg(
 \sandmp3.{\Ksl_{3 \ldots r} \Ksl_{2 \ldots r}}. 1
  \sandmm1.{\ksl_{r+1} \Ksl_{(j+1) \ldots r} \Ksl_{2 \ldots j}}.2 
 \\
& &  \hskip 1 cm \null
- \sandmp3.{\Ksl_{3 \ldots r}  \ksl_{r+1}}.1
    \sandmm1.{\Ksl_{2 \ldots r} \Ksl_{(j+1) \ldots r} \Ksl_{2 \ldots j}}.2
\Bigg) \nn \\
& & \hskip .1 cm \null \times
 \Bigg(
 \sandmp3.{\Ksl_{3 \ldots r} \Ksl_{2 \ldots r}}.1
 \sandmm1.{(\Ksl_{2 \ldots r}  \ksl_{r+1}-  \ksl_{r+1}  \Ksl_{(j+1) \ldots r}) \Ksl_{2 \ldots j}}.2
  \nonumber \\
 & &  \hskip 1 cm \null
- \sandmp3.{(\ksl_{r+1} \Ksl_{2 \ldots r}- \Ksl_{3 \ldots r} \ksl_{r+1})}.1
  \sandmm1.{\Ksl_{2 \ldots r} \Ksl_{(j+1) \ldots r}  \Ksl_{2 \ldots j}}.2
    \Bigg)
\,,\nonumber
 \ea
\ba
O_n^{\rm (c)}
 & = &
-  \frac{i}{6}
\sum_{r=5}^{n-1} \sum_{j=4}^{r-1}
\frac{   \spa{j}.{(j+1)}}{
  \sandpp2.{\Ksl_{2 \ldots j}}.j
  \sandpp2.{\Ksl_{2 \ldots j}}.{(j+1)} \, \spseq{3}.{n} s_{3 \ldots j}
            \, s_{2\ldots r}}
 \hskip 1 cm
 \nonumber \\
 & & \hskip .1 cm \null \times
\frac{1}{\sandmm1.{\Ksl_{2 \ldots r} \Ksl_{(j+1) \ldots r} \Ksl_{2 \ldots j}}.2
              \sandmm1.{\Ksl_{2 \ldots r} \ksl_{r} \Ksl_{2 \ldots r}}.2^2 }
 \nonumber \\
 & & \hskip .1 cm \null \times
\Bigg(
  \sandmp3.{\Ksl_{3 \ldots r} \Ksl_{2 \ldots r}}.1
 \sandmm1.{\Ksl_{2 \ldots r} \ksl_{r} \Ksl_{2 \ldots j}}.2 
 \nonumber \\
& & \hskip 1 cm \null
 -\sandmp3.{\ksl_{r} \Ksl_{2 \ldots r}}.1
   \sandmm1.{\Ksl_{2 \ldots r} \Ksl_{(j+1) \ldots r} \Ksl_{2 \ldots j}}.2
            \Bigg)   \nonumber \\
& & \hskip .1 cm \null
\times \Bigg(
  \sandmp3.{\Ksl_{3 \ldots r} \Ksl_{2 \ldots r}}.1
  \sandmm1.{\ksl_{r} \Ksl_{(j+1) \ldots r} \Ksl_{2 \ldots j}}.2
  \label{Overlapn}
 \\
& &\hskip 1 cm \null
 -\sandmp3.{\Ksl_{3 \ldots r} \ksl_{r}}.1
  \sand1.{\Ksl_{2 \ldots r} \Ksl_{(j+1) \ldots r} \Ksl_{2 \ldots j}}.2
 \Bigg) \nn \\
& & \hskip .1 cm \null \times
\ \Bigg( 
 \sandmp3.{\Ksl_{3 \ldots r} \Ksl_{2 \ldots r}}.1
 \sandmm1.{( \Ksl_{2 \ldots r} \ksl_{r}-  \ksl_{r} \Ksl_{(j+1) \ldots r}) \Ksl_{2 \ldots j}}.2 
  \nonumber \\
& & \hskip 1 cm \null
 -\sandmp3.{(\ksl_{r} \Ksl_{2 \ldots r}- \Ksl_{3 \ldots r} \ksl_{r})}.1
   \sandmm1.{\Ksl_{2 \ldots r} \Ksl_{(j+1) \ldots r} \Ksl_{2\ldots j}}.2
\Bigg)
\,. \nn
\ea
In term (c) we have relabeled $r + 1 \rightarrow r$ in the sum.

Combining the various overlap contributions from 
eqs.~(\ref{Overlapmmmtree})--(\ref{Overlap1}) 
and~(\ref{Overlapmmmtermjeqr})--(\ref{Overlapn})  gives us the complete 
overlap contribution, 
\be
O_n =  O_n^\tree + O_n^{\hat h\mbox{-}{\rm term}}  +
     O_n^{\hat d\mbox{-}{\rm term}} + O_n^{{\rm (a),(b)}; j=r} + 
      O_n^{\rm (a)} + O_n^{\rm (b)} + O_n^{\rm (c)} \,.
\label{FullOverlapmmmalln}
\ee

\subsection{Assembling the Three-Negative Helicity Amplitude}
\label{ThreeMminusAmplitudeSubSection}

The full amplitude $A_{n;1}^{\NeqZero} (1^-,2^-,3^-,4^+,\ldots,n^+)$ is
obtained by combining the pieces according to
\eqn{BasicBootstrapEquation}. In this equation,
the completed-cut terms and their large-$z$ behavior,
$\Cuth_n(0)$ and $\InfPart{\Shift12}\Cuth_n$, respectively,
are given in \eqn{3minus} and \eqn{cuthinfinity}.
The recursive contribution $\DiagrammaticRational_n$ is given by \eqn{rhat},
and the overlap contribution $O_n$ is given in \eqn{FullOverlapmmmalln}.
Finally, the large-$z$ contribution $\InfPart{\Shift12}{A_{n;1}^{\NeqZero}}$ 
is given in \eqn{Boundary}.


\subsection{Four-Negative Helicity Amplitude}
\label{FourMinusAmplitudeSubSection}

The tree amplitudes for this configuration were first computed
in ref.~\cite{RoibanSplitHelicity},
along with all split helicity configurations, $A_n(1^-, 2^-, \ldots,
m^-, (m+1)^+, \ldots, n^+)$, where legs of like helicity are nearest
neighbors in the color ordering.  At one loop, the
coefficients of all logarithmic terms in the split
helicity amplitudes were then computed in ref.~\cite{RecurCoeff}.
Thus, we have the completed-cut terms of the amplitude with four
negative helicities.  It is straightforward to subtract the
spurious large-$z$ behavior of the completed-cut terms under a
$\Shift12$ shift, and to extract the overlap terms.
The terms with large-$z$ behavior under the $\Shift12$ shift are given
in \eqn{FourMinusBoundary}, leaving only the direct recursive
contributions to be computed, according to \eqn{BasicBootstrapEquation}.

The $\Shift12$ shift recursion relation for four color-adjacent
negative-helicity gluons may be read off from \fig{LoopNPt12Figure},
\ba
&&\hskip -.3 cm
 \DiagrammaticRational_n(1^-,2^-,3^-,4^-, 5^+,\ldots,n^+)  = \nonumber \\
&& \hskip 1.8 cm
\sum_{i = 4}^{n-1}
A^\tree_{n-i+2}(\hat{1}^-,\hat{K}_{2\ldots i}^-,(i+1)^+,\ldots,n^+)
\frac{i}{s_{2\ldots i}}
R_i(\hat{2}^-,3^-,4^-, 5^+,\ldots,i^+,-\hat{K}_{2 \ldots i}^+)
\nonumber \\
&& \hskip 1.5cm \null
+ \sum_{i = 5}^{n-2}
R_{n-i+2}(\hat{1}^-,\hat{K}_{2 \ldots i}^-,(i+1)^+,\ldots,n^+)
\frac{i}{s_{2 \ldots i}}
A^\tree_i(\hat{2}^-,3^-,4^-,5^+,\ldots,i^+,-\hat{K}_{2 \ldots i}^+)
 \nonumber \\
&& \null \hskip 1.5 cm
+ \sum\limits_{i = 6}^{n-2}
  R_{n-i+2}(\hat{1}^-,\hat{K}_{2 \ldots i}^+,(i+1)^+,\ldots,n^+)
\frac{i}{s_{2 \ldots i}}
A^\tree_i(\hat{2}^-,3^-,4^-,5^+,\ldots,i^+,-\hat{K}_{2 \ldots i}^-)
\nonumber
  \hskip .4 cm \\
&& \null \hskip 1.5cm
 - \Biggl( \frac{1}{3 \epsilon} + \frac{8}{9}\Biggr)
\sum_{i = 5}^{n-2}
A^\tree_{n-i+2}(\hat{1}^-,\hat{K}_{2 \ldots i}^-,(i+1)^+,\ldots,n^+)
\frac{i}{s_{2 \ldots i}}
\nonumber \\
& & \hskip 4 cm \null
 \times A^\tree_i(\hat{2}^-,3^-,4^+,5^+,\ldots,i^+,-\hat{K}_{2 \ldots i}^+)
\,,
\label{rhatmmmm}
\ea
where we omitted vanishing diagrams. We have solved this
recursion relation through $n=8$.   Although we will not
present the analytic solution here, in \sect{NumericalSection}
we present the numerical value of 
$A_{8;1}^{\NeqZero}(1^-,2^-,3^-,4^-,5^+,6^+,7^+,8^+)$
at one phase-space point.


\subsection{Consistency of the Computed Amplitudes}

In order to confirm our procedure for constructing the amplitudes, we
have performed a number of checks on the amplitudes.  A simple
check is on the reflection symmetry of the amplitudes under 
a reversal of legs,
\be
A_{n;1}(1^-,2^-,3^-,4^+, \ldots, n^+) = (-1)^{n+1}
      A_{n;1}(3^-,2^-,1^-,n^+, \ldots, 4^+) \,,
\label{FlipSymmetrymmm}
\ee
and
\be
A_{n;1}(1^-,2^-,3^-,4^-, 5^+, \ldots, n^+) = (-1)^{n+1}
      A_{n;1}(4^-,3^-,2^-,1^-,n^+, \ldots, 5^+) \,.
\label{FlipSymmetrymmmm}
\ee
Our choice of shifts makes this symmetry unobvious, so confirming the
symmetry provides a rather non-trivial consistency check.  We have
confirmed the above symmetry numerically through $n=10$ for
the case with three negative helicities and at $n=8$ for the case with
four negative helicities.  We have also confirmed
numerically that the amplitudes have the correct factorization
properties with real momenta through $n=8$.  We checked that all 
unphysical spurious poles are removable, for $n=6$ (as mentioned in
\sect{ThreeMinusSixPtSection}) and for $n=7$.  These checks provide a 
strong confirmation that we found the correct analytic expressions for the
amplitudes.  Moreover, they also validate our recursive
determination of the amplitudes' large-$z$ behavior.


\section{Numerical Points in Phase Space}
\label{NumericalSection}

In order to aid the future implementation of these amplitudes in
numerical codes, we present values of the amplitudes at one point in
phase space.  We present numerics for selected helicity amplitudes
up to eight points, as a demonstration of the utility of our methods 
for high-multiplicity processes.  At six points, where independent 
numerical values are available~\cite{EGZ06}, we can use these as an 
additional check on the amplitudes.

We quote the numerical results in an unrenormalized form.
The renormalization amounts to carrying out the subtraction of
\eqn{MSbarsubtraction}.

Since we have exact analytical expressions for the amplitudes, it is a
simple matter to evaluate the amplitudes to arbitrary precision; we
give ten significant digits.  High precision can be useful for
studying the properties of the amplitudes near (removable) spurious
singularities, in order to investigate numerical instabilities
(roundoff error, {\it etc.}) which might be encountered under 
phase-space integration.  Because our analytic expressions possess 
only a relatively mild set of spurious singularities,
in comparison to more direct evaluations of Feynman diagrams, 
we do not anticipate any significant complications arising from 
roundoff error when constructing an NLO program.

At six points, for ease of comparison, we choose the same numerical
point as given in ref.~\cite{EGZ06}.
\ba
k_1 & = & {\mu \over 2}\, (-1,\, \sin\theta,\, \cos\theta \sin\phi,\,
                           \cos\theta \cos\phi) \,, \nn \\
k_2 & = & {\mu \over 2}\, (-1,\, - \sin \theta,\, -\cos \theta \sin \phi,\,
                            - \cos \theta \cos \phi) \,,\nn \\
k_3 & = & {\mu \over 3}\, (1,\,1,\,0,\,0) \,, \nn \\
k_4 & = &  {\mu \over 7}\, (1,\,  \cos \beta,\, \sin \beta,\,0)\,, \nn \\
k_5 & = & {\mu\over 6} \, (1,\, \cos \alpha \cos \beta,\,
            \cos \alpha \sin \beta,\, \sin \alpha) \,, \nn \\
k_6 & = & -k_1-k_2-k_3-k_4-k_5 \,,
\label{SixPointKinematics}
\ea
where
\ba
&& \theta = {\pi\over 4}\,, \hskip 1 cm
\phi = {\pi\over 6} \,, \hskip 1 cm
\alpha = {\pi \over 3} \,, \hskip 1 cm
\cos \beta = - {7\over 19} \,.
\ea
With this choice, the energies of $k_1$ and $k_2$ are negative,
representing a physical scattering process at a collider.  As in
ref.~\cite{EGZ06}, we choose $\mu = n = 6$ GeV, where $\mu$ is the
scale originating from the dimensionally-regulated integrals.

At seven points, we choose the kinematic point,
\ba
k_1 &=&  {\mu \over 2}\, (-1, \sin \theta,
           \cos \theta \sin \phi, \cos \theta \cos\phi)\,, \nn\\
k_2 &=&  {\mu \over 2}\, (-1, - \sin \theta,
         - \cos \theta \sin \phi, - \cos \theta \cos \phi)\,, \nn \\
k_3 &=&  {\mu\over 3} (1,1,0,0) \,, \nn \\
k_4 &=&  {\mu\over 8} (1, \cos\beta, \sin \beta,0) \,, \nn \\
k_5 &=&  {\mu \over 10} (1, \cos\alpha \cos\beta, \cos \alpha \sin \beta,
                                 \sin \alpha )\,, \nn \\
k_6 &=& {\mu \over 12} (1, \cos\gamma \cos\beta,
            \cos \gamma \sin \beta, \sin \gamma)\,, \nn \\
k_7 &=& -k_1-k_2-k_3-k_4-k_5-k_6\,,
\label{SevenPointKinematics}
\ea
where
\ba
 \theta = {\pi\over 4}\,,\hskip 1 cm
 \phi   =  {\pi\over 6}\,,\hskip 1 cm
 \alpha = {\pi \over 3}\,,\hskip 1 cm
 \gamma = {2 \pi \over 3} \,, \hskip 1 cm
 \cos \beta = - {37\over 128} \,,
\ea
and $\mu = 7$ GeV.

At eight points we choose as our reference kinematic point~\footnote{%
For this kinematic point, the quantities 
$\langle 7^\pm | (1+2) | 5^\pm\rangle$ happen to vanish; 
however, this vanishing causes no problems in evaluating the 
amplitudes considered here.},
\ba
k_1 &=& {\mu \over 2} (-1,\, \sin \theta,\,
           \cos\theta \sin \phi,\, \cos\theta \cos\phi) \,, \nn\\
k_2 &=& {\mu\over 2} (-1,\, - \sin \theta,\,
         - \cos\theta\sin \phi,\, - \cos \theta \cos\phi)\,,\nn\\
k_3 &=& {\mu\over 3} (1,\,1,\,0,\,0) \,, \nn\\
k_4 &=& {\mu\over 3}(1,\, \cos \beta,\, \sin \beta,\,0)\,, \nn\\
k_5 &=&{\mu\over 4} (1,\, \cos \alpha \cos\beta,\,
             \cos \alpha \sin \beta,\, \sin\alpha) \,, \nn\\
k_6 &=&{\mu \over 5} (1,\, \cos \gamma \cos\beta,\,
            \cos \gamma \sin\beta,\, \sin \gamma) \,, \nn\\
k_7 &=& {\mu \over 6}(1,\, \cos\delta \cos\beta,\, \cos \delta \sin \beta,\,
                              \sin\delta) \,, \nn\\
k_8 &=& -k_1-k_2-k_3-k_4-k_5-k_6-k_7\,,
\label{EightPointKinematics}
\ea
where
\be
\theta = {\pi\over 4} \,, \hskip .8 cm
 \phi = {\pi\over 6} \,,\hskip .8 cm
 \alpha = {\pi \over3} \,,\hskip .8 cm
 \gamma = {2 \pi\over 3} \,,\hskip .8 cm
 \delta = - {2 \pi \over 3}\,,  \hskip .8 cm
 \cos\beta = -{10\over 11} \,,  \hskip 1 cm
\ee
and we choose $\mu = 8$ GeV.

\begin{table}
\caption{\label{Neq4Table} Numerical results for the nonvanishing
$\NeqFour$ six- and seven-gluon split helicity amplitudes in the FDH scheme.
The kinematic point is given in eqs.~(\ref{SixPointKinematics}) and
(\ref{SevenPointKinematics}). The analytical expressions used for this
table are from refs.~\cite{Neq4Oneloop,Neq1Oneloop,NeqFourSevenPoint}.
}
\begin{tabular}{||c||c|c||}
\hline
\hline
Helicity& $1/\e$ & $\e^0$ \\
\hline
\hline
${-}{-}{+}{+}{+}{+}$
&  $\;
       448.1350970  + i \, 288.8591589 
$ & $ \;
        231.6837670 + i \, 1219.687214 
$ \\
\hline
${-}{-}{-}{+}{+}{+}$
& $
-\, 12.83149626 + i \, 36.01649473 \hskip .1 cm 
$ & $
-\, 82.74583978 + i \,33.53625588 \hskip.1 cm 
$
\\
\hline
\hline
${-}{-}{+}{+}{+}{+}{+}$
& $
2923.502435 + i \, 683.4723607
$ &  $
 5112.775012 + i \,  6035.881921
$ \\
\hline
${-}{-}{-}{+}{+}{+}{+}$
& $
    -\, 45.77174817 + i \, 14.37948381
%
$ & $
      -\, 118.6756853 - i \, 49.76271406
$ \\
\hline
\hline
\end{tabular}
\end{table}

\begin{table}
\caption{\label{Neq1Table} Numerical results for the nonvanishing
$\NeqOne$ chiral contributions to six-, seven- and eight-gluon split
helicity amplitudes.
The kinematic point is given in eqs.~(\ref{SixPointKinematics}),
(\ref{SevenPointKinematics}) and (\ref{EightPointKinematics}).
The analytical expressions were obtained
from refs.~\cite{Neq1Oneloop,NeqOneNMHVSixPt,BBDPSQCD,RecurCoeff}. }
\begin{tabular}{||c||c|c||}
\hline
\hline
Helicity& $1/\e$ & $\e^0$ \\
\hline
\hline
${-}{-}{+}{+}{+}{+}$
& $
- \,28.11035867 + i \, 4.643367883
$ & $
-\,108.9419206 + i \, 35.02980993
$
\\
\hline
${-}{-}{-}{+}{+}{+}$
& $
-\, 0.7843602985 - i\, 1.886592441
$ & $
 0.1435789412 - i \, 5.332008939
$\\
\hline
\hline
${-}{-}{+}{+}{+}{+}{+}$
& $
-\,104.5611840 + i \, 45.34709475
$ & $
-\,429.6932951 + i \, 209.0560823
$
 \\
\hline
${-}{-}{-}{+}{+}{+}{+}$
& $
 1.069354012 - i \, 1.474267821
$ & $
1.619538641 + i \, 3.617458324
$ \\
\hline
\hline
${-}{-}{+}{+}{+}{+}{+}{+}$
& $
-\,0.1599026654 - i \, 0.1215536994
$ & $
-\,0.4460113565 - i \, 0.2626240110
$ \\
\hline
${-}{-}{-}{+}{+}{+}{+}{+}$
& $
   -\, 0.01086792081 - i \, 0.002373299774
$& $
-\,0.003818661190 - i \, 0.02634017264  \hskip .1 cm 
$ \\
\hline
${-}{-}{-}{-}{+}{+}{+}{+}$
& $
-\,0.006820678758 - i \, 0.003628936145 \hskip .1 cm 
$ & $
0.02086909214 - i \, 0.04355499656
$\\
\hline
\hline
\end{tabular}
\end{table}

\begin{table}
\caption{\label{ScalarTable} Numerical results for the $\NeqZero$
scalar contributions to six-, seven- and eight-gluon split helicity
amplitudes.  The analytical expressions were obtained from
refs.~\cite{AllPlus, Mahlon, Neq1Oneloop,Qpap, Bootstrap, FordeKosower,
RecurCoeff} as well as sections~\ref{ThreeMinusSixPtSection}
and~\ref{ThreeMinusAmplitudeSection} of the present paper.}
\begin{tabular}{||c||c|c||}
\hline
\hline
Helicity& $1/\e$ & $\e^0$ \\
\hline
\hline
${+}{+}{+}{+}{+}{+}$
& $
 0
$ & $
  0.1024706290 + i \, 0.5198025397
%
$ \\
\hline
${-}{+}{+}{+}{+}{+}$
& $
 0
$ & $
 2.749806130 + i\, 1.750985849
$ \\
\hline
${-}{-}{+}{+}{+}{+}$
& $
-\,9.370119558 + i \, 1.547789294
$ & $
-\, 45.80779561 + i \, 13.03695870
$ \\
\hline
${-}{-}{-}{+}{+}{+}$
& $
-\,0.2614534328 - i\, 0.6288641470
$ & $
0.3883482043 -  i \, 5.830791857
$ \\
\hline
\hline
${+}{+}{+}{+}{+}{+}{+}$
& $
0
$ & $
0.1815778027 + i \, 1.941357266
$\\
\hline
${-}{+}{+}{+}{+}{+}{+}$
& $
0
$ & $
22.52927821 + i \, 5.464377788
$ \\
\hline
${-}{-}{+}{+}{+}{+}{+}$
& $
-\,34.85372799 + i \, 15.11569825
$ & $
-\,176.2169235 + i \,87.93931019
$ \\
\hline
${-}{-}{-}{+}{+}{+}{+}$
& $
 0.3564513374 - i \, 0.4914226070
$ & $
0.7087164424 - i \, 11.32916632
$
\\
\hline
\hline
${+}{+}{+}{+}{+}{+}{+}{+}$
& $
 0
$ & $
- \,0.0009856214410 + i \, 0.002143695508 \hskip .1 cm 
$ \\
\hline
${-}{+}{+}{+}{+}{+}{+}{+}$
& $
0
$ & $
 0.001078316199 + i \, 0.03129931739
$ \\
\hline
${-}{-}{+}{+}{+}{+}{+}{+}$
& $
-\,0.05330088846 - i \, 0.04051789981
$ & $
 0.05513350697 + i \, 0.1659518861
$ \\
\hline
${-}{-}{-}{+}{+}{+}{+}{+}$
& $
-0.003622640270 - i \, 0.0007910999246 \hskip .1 cm 
$ & $
 0.02719752089 - i \, 0.02586206549
$ \\
\hline
${-}{-}{-}{-}{+}{+}{+}{+}$
& $
-\,0.002273559586 - i \, 0.001209645382
$& $
 0.01154855076 - i \, 0.0008935357840
$\\
\hline
\hline
\end{tabular}
\end{table}

As explained in \sect{NotationSection}, it is convenient to
decompose QCD amplitudes in terms of $\NeqFour$, $\NeqOne$, and
$\NeqZero$ amplitudes.  QCD amplitudes are recovered from these
components using \eqn{AnQCD}.

We collect the numerical values of the split helicity amplitudes
in Tables~\ref{Neq4Table}-\ref{ScalarTable}.  We have extracted
an overall factor of $i \cg$ from the numerical values presented in 
the tables.
The complete set of analytic expressions for six- and seven-point
$\NeqFour$ amplitudes was obtained in refs.~\cite{Neq4Oneloop,
Neq1Oneloop, BCF7, NeqFourSevenPoint}.  The $\NeqOne$ amplitudes appearing in
Table~\ref{Neq1Table} were computed in refs.~\cite{Neq1Oneloop,
NeqOneNMHVSixPt, BBDPSQCD, BBCFSQCD, RecurCoeff}.  The
finite $\NeqZero$ amplitudes were obtained in refs.~\cite{AllPlus, Mahlon};
a compact representation of the amplitude with a single
negative-helicity was given more recently~\cite{Qpap}.  The
logarithmic parts of the divergent amplitudes were computed in
refs.~\cite{Neq1Oneloop,RecurCoeff}, while the rational-function parts
were determined in refs.~\cite{Bootstrap,FordeKosower} and 
in sections~\ref{ThreeMinusSixPtSection}
and~\ref{ThreeMinusAmplitudeSection} of the present paper.
Multiple analytical calculations exist for
some of the split-helicity amplitudes appearing in the tables.
In particular, the $\NeqFour$, $\NeqOne$, and logarithmic parts
of the $\NeqZero$ MHV amplitudes were also computed in
refs.~\cite{BST,BBSTQCD} using MHV vertices~\cite{CSW}.

Table~\ref{Neq4Table} gives the numerical values for the $1/\e$ and
finite terms of the $\NeqFour$ super-Yang-Mills theory split-helicity
six- and seven-point amplitudes.  We do not include the coefficients of
the leading $1/\e^2$ singularities in the tables, as these are easily
extracted from the values of tree amplitudes, for any helicity
configuration,
\be
A^{\NeqFour}_{n;1} \Bigr|_{1/\e^2} =  -  {n\cg \over \e^2} A^{\tree}_n \, .
\ee
The numerical values of the tree amplitudes may be read off from
the values of the $1/\e$ singularities of either the $\NeqOne$ or the
$\NeqZero$ loop amplitudes,
\be
A^{\NeqOne}_{n;1} \Bigr|_{1/\e} =   {\cg\over \e} A^{\tree}_n \,, \hskip 2cm
A^{\NeqZero}_{n;1} \Bigr|_{1/\e} =   {\cg\over 3\e} A^{\tree}_n \,,
\ee
given in Tables~\ref{Neq1Table} and \ref{ScalarTable}.

We quote numerical values of the $\NeqOne$ and $\NeqZero$ amplitudes
through eight points, but refrain from giving numerical values for
eight-point $\NeqFour$ amplitudes.  (The $\NeqFour$ amplitude with
four negative helicities has not been evaluated explicitly, although
it is straightforward to obtain using the methods of ref.~\cite{BCFUnitarity}. 
The general $\NeqFour$ $n$-point amplitudes
with three negative-helicity legs may be found in ref.~\cite{NeqFourNMHV}.)

We have compared our numerical results for the six-point amplitudes to
those found in ref.~\cite{EGZ06}.  After accounting for differing
overall phase conventions, they agree to within the number of digits
quoted in ref.~\cite{EGZ06}.


\section{Conclusions and Outlook}
\label{ConclusionSection}

In this paper we have continued the development of the on-shell
unitarity-bootstrap method~\cite{Bootstrap} for computing complete
amplitudes in non-supersymmetric gauge theories.  It combines
unitarity~\cite{Neq4Oneloop, Neq1Oneloop} with on-shell
recursion~\cite{BCFRecurrence,BCFW}.  It thereby systematizes an
earlier unitarity-factorization bootstrap~\cite{ZFourPartons} used to
compute the one-loop amplitudes required for $Z \rightarrow 4$ jets
and $pp \rightarrow W,Z + 2$ jets.  In the combined approach,
cut-containing pieces are obtained via unitarity, while purely
rational terms are obtained via on-shell recursion.  The latter terms
had previously been the most difficult part of a one-loop QCD
calculation.  The use of an on-shell recursion reduces them to
tree-like calculations.

On-shell recursion relations rely on continuations of the amplitudes
to complex momenta~\cite{BCFW}.  As discussed in
refs.~\cite{OnShellRecurrenceI,Qpap,Bootstrap}, the complex
factorization properties of loop amplitudes are much more subtle than
those of tree-level ones.  In addition to the presence of branch cuts,
in loop amplitudes we encounter double and `unreal' poles, present
only when the momenta are complex.  Unlike the situation for real
momenta, there are as yet no general theorems describing how loop
amplitudes factorize for complex momenta.  For general helicity
configurations, some continuations of the amplitudes may require such
unknown factorizations.  On the other hand, in general it is not
possible to choose a continuation where all factorizations are known
without spoiling the vanishing of the contour integral at infinity
whose consideration gives rise to the recursion relation.  This
unhappy situation can come about because the continued amplitude may
not vanish as the continuation parameter becomes large.

How, then should we avoid channels with unknown factorization, without
spoiling the recursion? In this paper we provided a general strategy
for avoiding these difficulties.
We construct an auxiliary recursion
relation to fill in information required to re-establish
the contour integral and hence the primary recursion.  The
latter is designed to avoid channels with
non-standard factorizations, even at the price of bad behavior
as the continuation parameter becomes large.
The auxiliary recursion supplies the terms we need to subtract
from the contour integrand in order to obtain a well-defined
integral and thereby a proper recursion relation.
It may have channels with non-standard factorization,
but is designed so that these non-standard channels will not affect
the required subtraction terms.

As an illustration of these ideas, we obtained complete
$n$-gluon amplitudes with a scalar in the loop and three
nearest-neighbor negative helicities in the color ordering.  The
logarithmic parts of these amplitudes were obtained in
refs.~\cite{Neq1Oneloop,RecurCoeff}; here we constructed a recursive
expression for all the rational terms that appear in these amplitudes.
For the six-point scalar loop amplitude 
$A_{6;1}^{\NeqZero}(1^-,2^-,3^-,4^+,5^+,6^+)$ 
we found a compact representation for the amplitude.
Using the supersymmetric decomposition~\cite{GGGGG},
together with the previously computed $\NeqFour$ and $\NeqOne$
supersymmetric contributions~\cite{Neq1Oneloop, BCF7,
NeqFourSevenPoint,NeqOneNMHVSixPt, NeqFourNMHV, BBDPSQCD}, these
provide complete QCD amplitudes.  As an additional illustration, we
provided a recursive expression for the eight-point amplitude
with four-negative helicities, 
$A_{8;1}^{\NeqZero}(1^-,2^-,3^-,4^-,5^+,6^+,7^+,8^+)$.
We gave numerical values for the
amplitudes (in the supersymmetric decomposition) at specific points in
phase space, to use as a guide for constructing numerical programs in
the future.  In a companion paper we shall present all remaining
$n$-point MHV amplitudes~\cite{MHVQCDLoop}.  These calculations of
high-multiplicity and indeed all-$n$ amplitudes illustrate a crucial
feature of our approach: {\it the complexity of the calculation grows
only modestly as the number of external particles increases\/}.

The continuations typically used to derive recursion relations are
implemented by shifting a pair of
momenta associated with external legs.  We choose {\it pairs\/} of
such shifts.  Our strategy for choosing them is based on a several
empirical properties of amplitudes: that under a $\Shift{-}{+}$ shift,
gluon amplitudes are well behaved for large shift parameters, that the
large-parameter behavior is always polynomial in the shift parameter,
and that we can suppress unwanted channels with non-standard complex
factorization through appropriate choice of shifts.  As yet we cannot
offer formal proofs of these properties.  In lieu of such proofs, we
have checked the computed amplitudes by verifying the stringent
real-momentum factorization properties any amplitude must satisfy.
The correctness of the computed amplitudes may then be seen as further
evidence for the correctness of the empirical properties.

It would be useful to address some of these open formal issues, and to
develop a first-principles understanding of the complex factorization
properties and large-shift-parameter behavior of one-loop scattering
amplitudes.  Recent papers connecting tree-level on-shell recursion to
the gauge theory Lagrangian~\cite{VamanYao} provide one possible
avenue to deriving the analytic properties. Unitarity in $D$
dimensions may also assist in this formal
understanding~\cite{BernMorgan}, as it can be used to determine
rational parts of amplitudes, even though explicit computations are
often cumbersome.

The methods we have presented in this paper are systematic and thus
lend themselves to automation, most obviously in an analytic approach,
but also plausibly in a semi-numerical one.  The large number of
required subprocesses in many applications of phenomenological
interest makes such automation desirable.  The techniques of this
paper require as input the cut-containing parts of the amplitudes.
Efficient, semi-automated procedures for evaluating unitarity cuts are
therefore also necessary.  As remarked in the introduction, there have
been important developments in this direction in the past year or so,
which are applicable to generic helicity configurations and external
states in non-supersymmetric gauge theories, and which reduce the
problem to one of residue extraction~\cite{BBCFSQCD,BFM}.  In
particular, this approach has been used to complete the computation of all
cut-containing parts of six-gluon amplitudes.  These methods should
work well in tandem with the ones presented here.

The empirical properties of the amplitudes are robust enough, and
sufficiently broad, to allow the computation of arbitrary gluonic
amplitudes.  The same ideas should carry over to amplitudes with
external quarks, vector bosons or Higgs particles.  Beyond that, we
foresee that with suitable extensions the method will also work
for processes with massive particles propagating inside the loops.  We
are encouraged by the observation that the unitarity method applies to
massive loops~\cite{BernMorgan} and that at tree level, the on-shell
recursive approach carries over to massive theories without
difficulty~\cite{GloverMassive,Massive}.  This extension of the
on-shell bootstrap, to loop-level processes containing top quarks for example,
is an open problem.

Even before such extensions to amplitudes with massive internal
particles become available, however, the on-shell bootstrap method we
have described is ready to tackle many of the important multi-parton
next-to-leading order computations of phenomenological interest at the
Large Hadron Collider.

\section*{Acknowledgments}

L.D. thanks the Aspen Center for Physics, where part of this work
was performed, for hospitality.  We thank Academic Technology Services at
UCLA for computer support.


\appendix

\section{Previously Computed Amplitudes and Vertices}
\label{PreviousAmplitudesAppendix}

As discussed in the text, our bootstrap approach relies on using
previously computed amplitudes.  In this appendix
we list the amplitudes that enter into our
calculation. In the text, we also need the parity conjugates of the
listed amplitudes, which
are given by replacing $\spa{i}.{j} \leftrightarrow \spb{j}.{i}$,
\be
A_n(1^{h_1},2^{h_2},\ldots,n^{h_n}) = \left[
A_n(1^{-h_1},2^{-h_2},\ldots,n^{-h_n})
\right]_{\spa{i}.{j} \leftrightarrow \spb{j}.{i}} \, .
\ee

Some useful tree amplitudes are
\ba
A_3^\tree(1^-,2^-, 3^+) &=&
i {\spa1.2^4 \over \spa1.2 \spa2.3 \spa3.1}
\, ,
\label{A3treemmp}
\\
A_3^\tree(1^-,2^+, 3^+) &=&
-i {\spb2.3^4 \over \spb1.2 \spb2.3 \spb3.1}
\, , \label{A3treempp} \\
A_4^\tree(1^-,2^-, 3^+, 4^+) &=&
 i {\spa1.2^4 \over \spa1.2 \spa2.3 \spa3.4 \spa4.1}
\, ,  \label{A4treemmpp} \\
A^\tree_5(1^-,2^-,3^+,4^+,5^+) &=&
   i {\spa1.2^4\over\spa1.2\spa2.3\spa3.4\spa4.5\spa5.1}
\,,\label{A5treemmppp}\\
A_5^\tree(1^-,2^-, 3^-, 4^+, 5^+) &=&
        - i {\spb4.5^4 \over \spb1.2 \spb2.3 \spb3.4 \spb4.5 \spb5.1}
\, , \label{A5treemmmpp}
\\
A_6^\tree(1^-,2^-, 3^-, 4^+, 5^+,6^+) &=&
i {\sandmm1.{(2+3)}.4^3
              \over  s_{234} \spb2.3 \spb3.4 \spa5.6 \spa{6}.1
                                       \sandmm5.{(3+4)}.2}  \nn \\
&& \hskip 0.1 cm \null
+ i { \sandmm3.{(4+5)}.6^3
              \over s_{345}  \spb2.1 \spb1.6
                     \spa3.4 \spa4.5  \sandmm5.{(3+4)}.{2} }
\, .
\ea
(The three-point amplitudes are non-vanishing for complex momenta, even
though they vanish for real momenta.)
The $n$-point MHV amplitudes~\cite{ParkeTaylor,BGRecurrence},
and one sequence of NMHV amplitudes~\cite{TreeRecurResults},
are,
\ba
A_n^\tree(1^+,\ldots,m_1^-,\ldots,m_2^-,\ldots,n^+) & = &
i { {\spa{m_1}.{m_2}}^4 \over \spseq{1}.{n}}
\, , \label{MHVtree} \\
A_n^\tree(1^-, 2^-, 3^-, 4^+, \ldots, n^+) & = &
\label{adjNMHVtree} \\
& & \hspace*{-3.5cm}
-i \sum_{r=4}^{n-1}
 { \sandmp3.{\Ksl_{3\ldots r} \Ksl_{2\ldots r}}.1^3
                 \spa{r}.{(r+1)} \over
                 \sandpp2.{\Ksl_{3\ldots r}}.{r}
                 \sandpp2.{\Ksl_{3\ldots r}}.{(r+1)}
                  s_{2\ldots r}\, s_{3\ldots r}
                     \spseq{3}.{n}}
\, .
\nonumber
\ea
Tree amplitudes with all like-helicity gluons or only one gluon of
opposite helicity vanish.

We will also need certain classes of one-loop helicity amplitudes, or
rather, their rational parts $\Vertex_n$,
defined in \eqn{RationalDefinition}.
The following three-vertices with a scalar in the loop vanish,
\be
R_3(1^-,2^+, -\Ph_{12}^\pm ) = 0 \, ,
\label{R3mp}
\ee
related to the vanishing of the corresponding splitting amplitudes
for real momenta~\cite{Neq4Oneloop}.
In principle, there are also vertices with like-helicity external legs,
$R_3(1^-,2^-, -\Ph_{12}^\pm )$ and $R_3(1^+,2^+, -\Ph_{12}^\pm )$.
However, the associated complex factorization properties are not fully
understood as yet, so we avoid these channels, or suppress them in
a large-$z$ limit.

For multi-particle factorizations, factorization functions can appear,
as discussed in \sect{DerivationSection}.
For the scalar loop, the one-loop factorization function is,
\be
\Fact(K) = -\biggl({1\over 3\eps}
       + {1\over 3} \ln\biggl({\mu^2 \over -K^2} \biggr)
       + {8\over 9} \biggr) \,.
\label{FactFunction}
\ee
The two-point rational ``vertex'' associated with this is a constant,
\be
R_\Fact = -\biggl({1\over 3\eps} + {8\over 9} \biggr) \,.
\label{FactFunctionVertex}
\ee

As discussed in \sect{NotationSection}, it is convenient to apply a
supersymmetric decomposition of QCD loop amplitudes into $\NeqFour$,
$\NeqOne$, and $\NeqZero$ (scalar loop)
pieces~\cite{GGGGG,TwoQuarkThreeGluon}.
Just as the tree-level amplitudes with all gluons of like-helicity
or all but one gluons of like-helicity vanish, so do the corresponding
supersymmetric $\NeqFour$ and $\NeqOne$ amplitudes;
the scalar $\NeqZero$ amplitudes are nonvanishing, but 
purely rational and finite.  For the finite four-point helicity amplitudes
we have~\cite{BKStringBased,TwoQuarkThreeGluon},
\ba
A_{4;1}^{\NeqZero} (1^+, 2^+, 3^+, 4^+) & = &
- i\, {\cg \over 3} \, {\spb1.2 \spb3.4 \over \spa1.2 \spa3.4}
\, , \label{A4Neq0apppp}
\\
A_{4;1}^{\NeqZero} (1^-, 2^+, 3^+, 4^+) & = &
i\, {\cg \over 3} \, {\spa2.4 \spb2.4^3 \over \spb1.2 \spa2.3 \spa3.4 \spb4.1}
\,. \label{A4Neq0amppp}
\ea
The corresponding five-point amplitudes are~\cite{GGGGG},
\ba
A_{5;1}^{\NeqZero} (1^+, 2^+, 3^+, 4^+, 5^+) & = &
i {\cg \over 3} \, \Biggl[ { \spb1.2 \spb2.3
\over \spa3.4 \spa4.5 \spa5.1 }
+ {\spb4.5 \spb5.1 \over \spa1.2 \spa2.3 \spa3.4 }
+ {\spb2.5 \spb3.4 \over \spa1.2 \spa3.4 \spa5.1 }
\Biggr]
\, , \nonumber
\\
& & \\
A_{5;1}^{\NeqZero} (1^-, 2^+, 3^+, 4^+, 5^+) & = &
i\, {\cg \over 3} \,  {1\over \spa3.4^2}
\Biggl[-{\spb2.5^3 \over \spb1.2 \spb5.1}
       + {\spa1.4^3 \spb4.5 \spa3.5 \over \spa1.2 \spa2.3 \spa4.5^2}
       - {\spa1.3^3 \spb3.2 \spa4.2 \over \spa1.5 \spa5.4 \spa3.2^2}
     \Biggr]
\,. \nonumber \\
& &
\label{A5Neq0mpppp}
\ea
Expressions for all remaining one-loop finite amplitudes
may be found in refs.~\cite{AllPlus,Mahlon,OnShellRecurrenceI}.

We make use of the
following functions~\cite{GGGGG} to express the
other, cut-containing amplitudes at loop level,
\begin{eqnarray}
\Kz(s) & = & {1 \over \epsilon\,{(1-2\,\epsilon)}}
    \biggl({\mu^2\over -s}\biggr)^{\epsilon}
\; = \; {1\over \epsilon}\;+\;\ln\biggl({\mu^2\over -s}\biggr)\;+\;2
\;+\; {\cal{O}}(\epsilon)\,, \nn \\
\Ll_0(r) &=& {\ln(r)\over 1-r}\,, \nn \\
\Ll_1(r) &=& {\Ll_0(r)+1\over 1-r}\,, \nn \\
\Ll_2(r) &=& {\ln(r)-(r-1/r)/2\over (1-r)^3} \,,
\label{Lsdef}
\end{eqnarray}
in order to eliminate spurious singularities
for $r \rightarrow 1$ which are present in the pure cut terms.

The four-point amplitude $A_{4;1}(1^-,2^-, 3^+, 4^+)$
is given by,
\ba
A_{4;1}^{\NeqFour}(1^-,2^-, 3^+, 4^+) &=&
 \cg  A^\tree(1^-,2^-, 3^+, 4^+) \Biggl\{ -{2\over \e^2}
 \biggl[ \biggl({\mu^2\over -s_{12}}\biggr)^{\e}
  + \biggl({\mu^2\over -s_{23}}\biggr)^\e \biggr] \nn \\
&& \hskip 4 cm \null
       + \ln^2 \biggl( {-s_{12} \over -s_{23}} \biggr)
               + \pi^2  - {\delta_R \over 3}\Biggl\}
    \, , \\
A_{4;1}^{\NeqOne}(1^-,2^-, 3^+, 4^+) &=&
\cg  A^\tree(1^-,2^-, 3^+, 4^+) \biggl({1\over \eps}  +
       \ln\biggl({\mu^2 \over -s_{23}} \biggr) +2 \biggr)
  \, , \\
A_{4;1}^{\NeqZero}(1^-,2^-, 3^+, 4^+) &=&
\cg A^\tree(1^-,2^-, 3^+, 4^+) \biggl({1\over 3\eps}  +
 {1\over 3} \ln\biggl({\mu^2 \over -s_{23}} \biggr) + {8\over 9}
   \biggr)
\,, \label{A4N0mmpp}
\ea
where the regularization-scheme parameter $\delta_R$ was discussed in
\sect{NotationSection}.  The recursive vertex we need for the
$\NeqZero$ scalar loop contribution is therefore given by
\be
R_4(1^-,2^-, 3^+, 4^+) =
\biggl({1\over 3\eps}  + {8\over 9} \biggr) A^\tree(1^-,2^-, 3^+, 4^+)\, .
\label{R4mmpp}
\ee

The remaining five-point amplitude with adjacent like helicities 
that we use is,
\ba
A^{\NeqFour}_{5;1}(1^-,2^-,3^-,4^+,5^+) &=&
\cg A^\tree(1^-,2^-,3^-,4^+,5^+)
\nn\\
&&\outdenta\times \biggl[
-{1\over\e^2} \sum_{j=1}^5
  \biggl({\mu^2\over -s_{j,j+1}}\biggr)^\e
          + \sum_{j=1}^5
\ln \biggl({-s_{j,j+1}  \over -s_{j+1,j+2}}\biggr)\,
                   \ln \biggl({-s_{j+2,j-2}\over -s_{j-2,j-1}}\biggr)
+ {5 \over6}\pi^2 - {\delta_R\over3}\biggr]
\,, \label{A5NeqFourmmmpp} \\
A^{\NeqOne}_{5;1}(1^-,2^-,3^-,4^+,5^+) &=& \nn\\
&&\outdenta
  \cg A^\tree(1^-,2^-,3^-,4^+,5^+)\biggl[
   {1\over2 \e} \biggl\{  \biggl({\mu^2\over -s_{51}}\biggr)^{\e}
                         +\biggl({\mu^2\over -s_{34}}\biggr)^\e  \biggr\} + 2
                 \biggr]
                  \label{A5Neq1mmmpp} \\
&&\outdenta
  -i\, {\cg\over 2}
   {{\spb4.5}^2 \biggl(\spb5.1\spa1.2\spb2.4+\spb5.2\spa2.3\spb3.4\biggr) \over
    \spb1.2\spb2.3\spb3.4\spb5.1}
     {\Ll_0\biggl( {-s_{51}\over -s_{34}}\biggr)\over s_{34}}
 \,,\nn\\
A^{\NeqZero}_{5;1}(1^-,2^-,3^-,4^+,5^+) &=& \nn\\
&&\outdenta
  {1\over3} A^{\NeqOne}_{5;1}(1^-,2^-,3^-,4^+,5^+)
  +{2\over9} \cg A^\tree(1^-,2^-,3^-,4^+,5^+)
\label{A5Neq0mmmpp} \\
&&\outdenta\null
      + i \, {\cg\over 3}
   {\spa1.2\spb2.4\spb5.2\spa2.3
 \biggl(\spb5.1\spa1.2\spb2.4+\spb5.2\spa2.3\spb3.4 \biggr)
        \over\spb1.2\spb2.3}
     {\Ll_2\biggl( {-s_{51}\over -s_{34}} \biggr)\over s_{34}^3}
  + \cg \Remaining_5
\,,\nn
\ea
where
\be
\Remaining_5 =
 i\, \biggl[
 {1\over3}{\spb1.3\!\spa1.3^3\over\spb1.2\!\spb2.3\!\spa3.4\!\spa4.5\!\spa5.1}
     -{1\over3}{\spa1.3^2\spb4.5\over\spb1.2\!\spb2.3\!\spa3.4\!\spa5.1}
     +{1\over6}{\spa1.2\!\spa2.3\!\spb2.4\!\spb2.5\!\spb4.5\over
                  \spb1.2\!\spb2.3\! s_{34}s_{51}} \biggr]
\,. \label{Remaining5}
\ee
The vertex we need is given by setting the logarithms to zero in the
amplitude,
\be
\Rational_5(1^-, 2^-, 3^-, 4^+, 5^+)  =
{1 \over \cg} A^{\NeqZero}_{5;1}(1^-,2^-,3^-,4^+,5^+) \Bigr|_{\ln = 0} \, .
\label{R5mmmpp}
\ee

The $\NeqFour$ supersymmetric amplitude needed for
constructing the complete six-point QCD
amplitude with three color adjacent
negative helicities is given by~\cite{Neq1Oneloop},
\def\Wsix#1{W_6^{(#1)}}
\def\cc{\dagger}
\be
A_{6;1}^{\NeqFour}(1^-,2^-,3^-,4^+,5^+,6^+)=
 \cg\ [ B_1\,\Wsix1+B_2\,\Wsix2+B_3\,\Wsix3 ],
\ee
where
\ba
B_1 &=&
 i \, {(s_{123})^3
     \over \spb1.2\spb2.3\spa4.5\spa5.6
           \sandmm4.{(2+3)}.1 \sandmm6.{(1+2)}.3}
 \, , \\
B_2 &=&
i \, {\sandmm1.{(2+3)}.4^3 \over s_{234}
       \spb2.3\spb3.4\spa5.6\spa6.1 \sandmm5.{(3+4)}.2}
\nonumber \\
& & \null
  +  i \, {\spa2.3^3\spb5.6^3 \over s_{234} \spa3.4\spb6.1
                     \sandmm4.{(2+3)}.1   \sandmm2.{(3+4)}.5}
 \,,
                   \\
B_3 &=&
i \,
{\sandmm3.{(1+2)}.6^3  \over  s_{345}
               \spb6.1\spb1.2\spa3.4\spa4.5 \sandmm5.{(6+1)}.2}
\nonumber \\
& &  + i\, {\spa1.2^3 \spb4.5^3 \over s_{345}
               \spa6.1 \spb3.4 \sandmm6.{(1+2)}.3 \sandmm2.{(6+1)}.5}
\, .  \ea
We have taken the parity conjugate and somewhat simplified the results
compared to the expression of ref.~\cite{Neq1Oneloop}, and
used the four-dimensional helicity scheme where $\delta_R = 0$.
The combination of integral functions appearing in the amplitude is,
\ba
  \Wsix{i}
      &=& -{1\over2\e^2} \sum_{j=1}^6
         \left( { \mu^2 \over -s_{j(j+1)} } \right)^\e
  -  \ln\left({-s_{i \ldots (i+2)} \over -s_{i(i+1)}}\right)
      \ln\left({-s_{i \ldots (i+2)} \over -s_{(i+1)(i+2)}}\right) \nn \\
 &&
   - \ln\left({-s_{i\ldots (i+2)} \over -s_{(i+3)(i+4)}}\right)
      \ln\left({-s_{i\ldots (i+2)} \over -s_{(i+4)(i+5)}}\right)
  + \ln\left({-s_{i\ldots (i+2)} \over -s_{(i+2)(i+3)}}\right)
      \ln\left({-s_{i\ldots(i+2)} \over -s_{(i+5)i}}\right)\nn \\
 &&
  + {1\over 2}\ln\left({-s_{i(i+1)} \over -s_{(i+3)(i+4)}}\right)
         \ln\left({-s_{(i+1)(i+2)} \over -s_{(i+4)(i+5)}}\right)
  + {1\over 2}\ln\left({-s_{(i-1)i} \over -s_{i(i+1)}}\right)
         \ln\left({-s_{(i+1)(i+2)} \over -s_{(i+2)(i+3)}}\right)\nn \\
  &&
  + {1\over 2}\ln\left({-s_{(i+2)(i+3)} \over -s_{(i+3)(i+4)}}\right)
         \ln\left({-s_{(i+4)(i+5)} \over -s_{(i+5)i}}\right)
  +  {\pi^2\over3} \, .
\ea
%


\end{document}